\documentclass[a4paper,fleqn]{cas-dc}

\usepackage[numbers]{natbib}
\usepackage{tabularx}
\usepackage{bbding}
\usepackage{makecell}
\usepackage[figuresright]{rotating}
\usepackage[section]{placeins}%
\usepackage{array}
\usepackage{multirow}
\usepackage{hyperref}
\hypersetup{
    colorlinks=true,
    linkcolor=blue,
    urlcolor=blue,
    anchorcolor=blue,
    citecolor=blue
}

\usepackage{tikz}
\usetikzlibrary{backgrounds,mindmap}
\usepackage{xcolor}
\usepackage{pgfplots}
\usepackage{listings}
\usepackage{soul}
\usepackage{color, xcolor}  
\sethlcolor{yellow}

\begin{document}
\shorttitle{Watermarking Techniques for Large Language Models: A Survey} 
\shortauthors{Y. Liang \textit{et al.}}

%\begin{frontmatter}

\title[mode=title]{Watermarking Techniques for Large Language Models: A Survey}

\author[1]{Yuqing Liang\footnote{1}}
\ead{yqliang001@gmail.com}
\address[1]{College of Cyber Security, Jinan University, Guangzhou 510632, China}

\author[1]{Jiancheng Xiao\footnotemark[1]}
\ead{jcxiao001@gmail.com}
\fntext[equal]{Both authors contributed equally to this work.}
% \fntext[equal]{Both authors contributed equally to this work and share first authorship.}

\author[1]{Wensheng Gan}
\cortext[cor1]{Corresponding author}
\ead{wsgan001@gmail.com}
\cormark[1]

\author[2]{Philip S. Yu}
\ead{psyu@uic.edu}
\address[2]{Department of Computer Science, University of Illinois Chicago, Chicago, IL 60607, USA}

\begin{abstract}
   With the rapid advancement and extensive application of artificial intelligence technology, large language models (LLMs) are extensively used to enhance production, creativity, learning, and work efficiency across various domains. However, the abuse of LLMs also poses potential harm to human society, such as intellectual property rights issues, academic misconduct, false content, and hallucinations. Relevant research has proposed the use of LLM watermarking to achieve IP protection for LLMs and traceability of multimedia data output by LLMs. To our knowledge, this is the first thorough review that investigates and analyzes LLM watermarking technology in detail. This review begins by recounting the history of traditional watermarking technology, then analyzes the current state of LLM watermarking research, and thoroughly examines the inheritance and relevance of these techniques. By analyzing their inheritance and relevance, this review can provide research with ideas for applying traditional digital watermarking techniques to LLM watermarking, to promote the cross-integration and innovation of watermarking technology. In addition, this review examines the pros and cons of LLM watermarking. Considering the current multimodal development trend of LLMs, it provides a detailed analysis of emerging multimodal LLM watermarking, such as visual and audio data, to offer more reference ideas for relevant research. This review delves into the challenges and future prospects of current watermarking technologies, offering valuable insights for future LLM watermarking research and applications.
\end{abstract}

\begin{keywords}
     artificial intelligence  \\
     watermarking \\
     copyright protection \\
     multimodal data   \\
     large language model   \\ 
     watermark large model  \\
\end{keywords}

\makeatletter\def\Hy@Warning#1{}\makeatother
\maketitle

\section{Introduction}  \label{sec: introduction}

With the emergence of GPT-3.5 \cite{ouyang2022training} in late 2022, its outstanding capabilities have attracted worldwide attention. Numerous large language models (LLMs) \cite{zhao2023survey} with diverse functionalities have also been continuously developed and are being widely applied in various aspects of people's lives and production. So far, due to the abilities of model-as-a-service \cite{gan2023model}, LLMs have achieved preliminary intelligent applications in fields such as education \cite{gan2023large,kasneci2023chatgpt,xu2024large}, scientific research  \cite{beltagy2019scibert,sun2024scieval}, healthcare \cite{thirunavukarasu2023large,zheng2024large}, programming \cite{chen2021evaluating,xu2022systematic}, law \cite{cui2023chatlaw,lai2023large}, and robotics \cite{zeng2023large}, demonstrating remarkable performance that can efficiently assist or even replace some human tasks. Compared to traditional data mining \cite{gan2017data,gan2021survey}, natural language processing (NLP) \cite{chowdhary2020natural,nadkarni2011natural}, and AI systems \cite{mohseni2021multidisciplinary,wu2024mdgrl}, LLMs have exhibited unprecedented capabilities enhancements. In the future, LLMs may become as ubiquitous in human society as smart computers and smartphones \cite{kosinski2023theory}, and may even be deployed on every intelligent device. Besides, we should also be aware of the potential negative impacts and risks of LLMs \cite{wei2022emergent}. 1) LLMs may be misused, such as for automated hacking attacks \cite{wei2022emergent} or fraud bots \cite{weidinger2022taxonomy}. 2) The high training cost of LLMs means that their leakage could lead to significant profits for competitors. 3) LLMs can cause data pollution and be used to create inappropriate content, including pornography or extremist content \cite{adelani2020generating}. 4) LLMs can potentially leak user personal information or sensitive data during data processing \cite{liu2023summary}. 5) LLMs may lead to serious academic and intellectual property (IP) disputes, as the vast datasets used for their training may contain unauthorized data or infringe on IP rights, and the content generated by the models may also violate copyrights  \cite{pan2020privacy}. 6) The content generated by LLMs may involve academic misconduct \cite{weidinger2022taxonomy}, as the models have learned a large amount of online data during training, and their generated text may be highly similar or even identical to existing online content. To address the potential issues of LLMs, some research suggests that reviewing the data generated by these models could be a solution. However, in practice, this may raise more problems: 1) reviewing the generated data could significantly reduce the reaction speed of LLMs, affecting their efficiency, especially in time-sensitive scenarios; 2) it is difficult to establish relevant review standards, and this may overly restrict the creative freedom of LLMs; 3) the further introduction of review procedures or review agencies may lead to even more serious privacy data leakage \cite{kuvzina2021methods}; 4) the cost of the review process may lead to further increases in the costs of LLMs companies; 5) the high cost and relatively low efficiency of reviewing each generated data item, with only a small portion being used for malicious purposes.

Compared to reviewing the data generated by LLMs \cite{chang2023learning}, a feasible solution is to trace the source of the different forms of data (including text, images, videos, and sound) generated by each LLM. Through data tracing, the potential problems of LLMs can be effectively solved. We can trace the source of single or multiple data items generated by LLMs to identify the specific model that generated the data. Data tracing can effectively detect and quickly block automated hacking attacks and fraud bots, prevent the illegal use of LLMs by other companies, and find the source of privacy data leaks, quickly locating and repairing the problematic prompts that led to the generation of harmful information. The challenge is how to implement the data tracing of LLMs. In the past, digital watermarking \cite{cox2002digital} has been a common approach. A digital watermark is subtly integrated into conventional multimedia content that needs to be traced, with the watermark carrier including text, images, audio, video, documents, web pages, databases, and more. Digital watermarking can be used to achieve data tracing, digital rights authentication, and integrity verification. Research suggests using watermarking to safeguard the intellectual property of LLMs, to enable tracking or identification of the multimedia information generated by these models, and to prevent their misuse.

\begin{table*}[ht]
    \label{table:gaps}
    \scriptsize
    \centering
    \caption{Contributions and gaps of existing research.}
\begin{tabular}{|l|c|p{6.2cm}|c|c|c|} 
\hline
\multicolumn{1}{|c|}{\textbf{Study}} & \textbf{Year} & \multicolumn{1}{c|}{\textbf{Focus of survey}} & \textbf{Comparison} & \begin{tabular}[c]{@{}c@{}}\textbf{Model analysis}\end{tabular} & \begin{tabular}[c]{@{}c@{}}\textbf{Technology inheritance }\\\textbf{analysis}\end{tabular}  \\ 
\hline
Boenisch et al. \cite{boenisch2021systematic} & 2021 & Model watermarking for neural networks. & $\surd$ & Image & × \\ 
\hline
Liu et al. \cite{liu2023survey} & 2023 & A survey of text watermarking in the era of LLMs. & $\surd$ & Text & × \\ 
\hline
Hwang et al. \cite{hwang2023brief} & 2023 & \begin{tabular}[c]{@{}l@{}}Application status of generative AI watermarking \\ and related regulations.\end{tabular} & × & Image, text & × \\ 
\hline
Lalai et al. \cite{lalai2024intentions} & 2024 & Taxonomy and challenges in text watermarking for LLMs. & $\surd$ & \begin{tabular}[c]{@{}c@{}}Text\end{tabular} & $\surd$ \\ 
\hline

Our work & 2024 & A survey of watermarking techniques for LLMs. & $\surd$ & \begin{tabular}[c]{@{}c@{}}Image, text, audio,\\ multi-modal\end{tabular} & $\surd$ \\ 
\hline
\end{tabular}
\end{table*}

%Research Status
LLM watermarking can be divided into three categories based on functionality: 1) Data traceability function \cite{lin2021traceability}, which can efficiently detect the ownership of the invoked LLMs when the LLMs are used for illegal purposes; 2) Copyright protection function  \cite{jin2024proflingo}, which is used to realize the copyright recognition of multimedia data generated by LLMs; 3) Function to identify content generated by LLMs, i.e., to identify whether the content is generated by LLMs \cite{fu2024watermarking}. From the perspective of modality, it can be divided into four categories: 1) Text domain data  \cite{wang2023wasa,wei2024proving}: The text is the most common multimedia generation data in current LLMs, and is also the most common watermark embedding carrier in this research field; 2) Image domain data  \cite{wen2023tree,zhang2020model}: Images are common multimedia data in current LLMs, mainly manifested in text-to-image in AI-generated content (AIGC) \cite{wu2023ai}; 3) Video domain data: Except for the untested Sora, there are currently no stable and efficient video-generating LLMs, and there are no watermarks for video domain LLMs; 4) Audio domain data \cite{chen2023wavmark,cho2022attributable}: Audio LLMs are mainly focused on functions such as speech generation and music production. In the current research, the main problems encountered by LLM watermarking are as follows: 1) Watermark robustness problem: The current LLM watermarks are focused on text watermarks, and difficult to solve problems such as text clipping and paraphrasing attacks. 2) Semantic invariance problem: Embedding watermarks in text data may affect its semantics, affecting the model's answer effectiveness or making the text watermark less imperceptible, and may also lead to semantic shifts in generated multimedia data such as images/videos/audios. 3) Security vulnerability problem: LLM watermarking may require access to the inner structure and settings of the LLMs, which may pose security risks. 4) System consumption problem: LLM watermarking may lead to increased model system power consumption and computational complexity, thereby increasing model running costs.

\textbf{Research gaps}: Currently, there are many studies related to LLMs or watermarking, and details are summarized in Table \ref{table:gaps}. Boenisch et al. \cite{boenisch2021systematic} proposed an LLM watermarking classification framework to identify and analyze different types of LLM watermarking schemes, and introduced a unified LLM watermarking threat model for systematic analysis and evaluation of watermarking methods' effectiveness across different scenarios. It also introduced the security requirements for LLM watermarking and the types of attacks against them and conducted a survey and classification of the relevant literature based on this framework. Finally, it discussed the shortcomings and limitations of existing LLM watermarking methods and looked forward to future research directions of LLM watermarking. Liu et al. \cite{liu2023survey} discussed the role of text watermarking technology in the age of LLMs. Text watermarking technology is crucial for protecting the intellectual property rights of text material, mainly by inserting imperceptible watermark information within the text material to achieve copyright protection. With the development of LLMs technology, text watermarking technology has made significant progress in both technology and application scenarios (i.e., application in LLMs). It introduces in detail the definition, key attributes, classification, and evaluation indicators of LLM watermarking techniques, covering success rate, robustness, text quality, and resistance to forgery, explores the use of LLM watermarking in practical application scenarios, and emphasizes the ethical application of AI technology in LLMs. Hwang et al. \cite{hwang2023brief} introduce the adoption of watermarking technology in generative AI. The research discusses the significant potential of generative AI in text and image generation while highlighting the challenges that accompany advancements in AI technology, such as AI misuse and the creation of misinformation. This research states that various countries and regions are considering or have already implemented relevant AI regulations, using watermarking technology to identify AIGC \cite{wu2023ai}. It also analyzes how leading tech companies incorporate watermarking technology into their AI generation services, focusing on LLM watermarking technology and related laws and regulations. To summarize, we can observe the following limitations in the above reviews about LLM watermarking research: 1) Focusing mainly on LLMs text watermarking, lacking watermarking of other modalities, and lacking integration with the current development trends of LLMs; 2) Lacking consideration of technological inheritance, lacking analysis and discussion of traditional digital watermarking, not comparing current LLM watermarking with traditional digital watermarking, and not analyzing what current LLM watermarking can learn from traditional digital watermarking \cite{cai2024towards}. 

\textbf{Contributions}: The differences between this review and the existing studies are: 1) Not only exploring LLM watermarking, but also innovatively combining the current multi-modal trend of LLMs, and conducting a detailed analysis of the watermarking of emerging multi-modal LLMs such as image/audio; 2) Conducting a relatively detailed analysis of traditional digital watermarking and LLM watermarking technology, exploring the technological inheritance between LLM watermarking and traditional watermarking; 3) Analyzing in detail the distinctions between traditional digital watermarking and watermarking for LLMs in the same modality or the same application scenario. To fill these gaps, we provide a comprehensive investigation and analysis of LLM watermarking technology in detail. In summary, the main contributions of this review are as follows:

\begin{itemize}
    \item  Comprehensive analysis of the research status of watermarking for LLMs: This review offers an in-depth examination of the background, and current status, along with the technical development in watermarking research for LLMs, offering a comprehensive analysis and overview of this field.

    \item  Innovative analysis incorporating multimodal trends: This review incorporates the current multimodal development trends of LLMs, and provides a detailed analysis of watermarking for emerging multimodal LLMs, such as those for images and audio. This helps us understand the technologies and development of watermarking for multimodal LLMs, in line with the technological trends.

    \item  In-depth analysis of the relationship between traditional digital watermarking and watermarking for LLMs: Based on a detailed analysis of traditional digital watermarking, this review systematically analyzes and discusses the watermarking technologies for LLMs, delving into and analyzing the technical inheritance between them.

    \item  Analysis of challenges and future development directions: This review provides a thorough examination of the challenges facing watermarking for LLMs and their future development directions, and proposes some corresponding solutions.

    \item  Potential impact and significance: This review helps to understand the development trajectory of digital watermarking technology and watermarking for LLMs, and to apply traditional digital watermarking technology to related LLM watermarking, enabling effective utilization of the technical achievements of traditional digital watermarking research.
\end{itemize}

\begin{table}[H]
    \scriptsize
    \centering
    \caption{Important terms of acronyms and corresponding full form.}
    \label{table:acronyms}

\caption{Taxonomy of LLM watermarking techniques. }
\end{figure*}

\textbf{Organization}: The organization of this survey is structured as follows: Section \ref{sec:background} provides an overview of LLMs, traditional digital watermarking, and watermarking for LLMs. Section \ref{sec:Traditional} introduces traditional digital watermarking, providing detailed introductions to common digital watermarking based on different classification methods. Section \ref{sec:LLMs} introduces watermarking for LLMs, providing detailed introductions to different categories of LLM watermarking based on different classification methods, and analyzing the relationship and technical inheritance between traditional digital watermarking and LLM-enable watermarking. Section \ref{sec:adv&disadv} mainly analyzes the advantages and disadvantages of watermarking for LLMs. Section \ref{sec:Applications} analyzes the applications of watermarking for LLMs. Section \ref{sec:Challenges} analyzes the obstacles and future opportunities of watermarking for LLMs. Section \ref{sec:conclusion} is the conclusion. The detailed taxonomy of the state-of-the-art LLM watermarking techniques is presented in Figure \ref{fig:taxonomy}. Some important terms of acronyms and their corresponding full form are given in Table \ref{table:acronyms}.

\section{Relevant Background} \label{sec:background}
\subsection{Background of LLMs}

LLMs refer to machine learning models with large-scale parameters and complex computational structures. LLMs are mainly built on deep neural networks (DNN) \cite{sze2017efficient} and usually have a huge number of parameters. The main difference between LLMs and traditional small models lies in the number of parameters \cite{yao2024survey}. Small models \cite{shaikhina2015machine} usually refer to models with fewer neural network parameters, fewer layers, and smaller models. The advantage of small models is that they are small and suitable for scenarios with relatively small computational resources and data volume, and can be used for various lightweight applications such as embedded devices, IoT devices, connected vehicles, and mobile devices. LLMs typically have hundreds of billions or even trillions of parameters, which is a feature of having a huge number of parameters compared to general DNN.

From the perspective of the model scale, LLMs have experienced three stages \cite{han2021pre,zeng2023distributed}: pre-trained models, large pre-trained models, and super pre-trained models. The parameter scale increases at least tenfold per year, from hundreds of millions to tens of billions. LLMs with parameter scales in the hundreds of billions have become mainstream. As the training data, network parameters, and network layers of the model constantly expand, until a certain critical point is broken, the model will exhibit more complex and powerful capabilities, able to learn deeper features from the training data. This phenomenon is called an emergent phenomenon. Currently, AI models with emergent phenomena are considered LLMs, which is one of the main differences between large models and small models \cite{wei2022emergent}. The development of LLMs can be categorized into three stages: the germination period, the precipitation period, and the explosion period. 

From 1950 to 2005 was the germination period of the AI model. This period was represented by the convNet (CNN) \cite{xie2016theory}, signifying the emergence and evolution of traditional neural network models. In 1956, computer expert John McCarthy first proposed the notion of "artificial intelligence" (AI), initiating research in the AI field. In subsequent development, AI gradually evolved from the initial stage based on small-scale expert knowledge to the stage based on machine learning. In 1980, CNN provided an important tool for research in computer vision (CV) \cite{voulodimos2018deep} and NLP. In 1998, the basic structure of modern CNN, LeNet-5 \cite{lecun1998gradient}, was born, marking the transformation of machine learning methods from early models based on shallow machine learning to models based on deep learning. This stage laid the relevant mathematical foundations for the development of deep learning frameworks and LLMs, and had a far-reaching impact on research in CV and NLP.

From 2006 to 2018 was the precipitation period of LLMs, during which the brand-new neural network model represented by Transformer began to emerge. In 2013, the well-known Word2Vec \cite{mikolov2013efficient} was introduced, proposing the "word vector model" that converts words into vectors for the first time, providing a foundation for AI to better understand and process text data. In 2014, the generative adversarial network (GAN) \cite{goodfellow2014generative} was proposed, marking the entry of deep learning into a new stage of generative model research. In 2017, Google introduced the Transformer architecture \cite{vaswani2017attention} using a self-attention mechanism, establishing the basis for pre-training large models. In 2018, OpenAI and Google successively launched large pre-trained models like GPT-1 \cite{radford2018improving} and BERT \cite{devlin2019bert}, making pre-trained LLMs the mainstream in NLP. During the precipitation period, the brand new neural network architecture represented by Transformer laid a solid foundation for the development of large models, greatly improving the performance of LLMs.

From 2019 to the present, the explosion period of LLMs. OpenAI improved based on GPT-1 and released the GPT-2 model \cite{radford2019language}. In 2020, OpenAI unveiled GPT-3 \cite{brown2020language} featuring 175 billion parameters. In 2022, OpenAI released the groundbreaking ChatGPT (GPT-3.5) \cite{ouyang2022training}, triggering a new round of LLMs fever. 2022 was called the year of the element of LLMs, with Google releasing the PaLM \cite{chowdhery2023palm} and Microsoft releasing the BEIT-3 \cite{wang2023image}, marking the arrival of the era of multimodal LLMs \cite{wu2023multimodal}. In 2023, OpenAI released the more powerful GPT-4.0 \cite{achiam2023gpt} with support for multimodality and file reading; Google released PaLM2 \cite{anil2023palm} and LaMDA \cite{thoppilan2022lamda}; Meta released the LLaMA \cite{touvron2023llama} and LLaMA2 \cite{touvron2023llama}; Microsoft released the ChatGPT-supported Bing search engine. In 2024, OpenAI released the Sora \cite{peebles2023scalable}, which performed well in long-term video generation, and has excellent visual and multi-modal capabilities \cite{fei2022towards}. LLMs have penetrated various areas of daily life and become essential elements of everyday existence.

In terms of technical architecture, the Transformer architecture is the mainstream algorithmic foundation. Therefore, two main technical routes of GPT and BERT have formed \cite{chen2023universal}. Currently, GPT has gradually become the mainstream route of LLMs, and the vast majority of LLMs with parameters exceeding hundreds of billions adopt the GPT pattern. From the perspective of modality, LLMs can cover text, images, videos, audio, and other modalities. The modalities supported by LLMs are becoming more diversified, from unimodal to multimodal  \cite{wu2023multimodal}, so that LLMs can better perceive the world and learn various knowledge. From the perspective of task completion, LLMs can complete more and more tasks, from machine translation to article writing, from code writing to scientific computing, LLMs can efficiently complete various types of tasks. In terms of application areas, LLMs can be divided into general-purpose  \cite{chen2023universal} and industry-specific (expert)  \cite{wang2023cogvlm} types. General-purpose models have strong generalization capabilities and can complete tasks in various scenarios, including ChatGPT and BERT. Industry-specific (expert) models are fine-tuned based on industry knowledge to meet the needs of specific fields, such as BloombergGPT \cite{wu2023bloomberggpt} for finance and Med-PaLM \cite{singhal2023towards} for medicine.

\subsection{Background of Digital Watermarking}

Digital watermarking \cite{cox2002digital} is a method of embedding specific information into digital signals to verify the integrity and originality of digital material. The development of digital watermarking dates back to the mid-20th century. In 1954, the American firm Muzac filed a patent for "identification of sound and lide signals", which outlined a technique for embedding an imperceptible identification code into music to verify ownership. This is the earliest known digital watermarking technology. In 1994, Tirkel et al. \cite{van1994digital} formally introduced the idea of digital watermarking, but the proposed watermarking was implemented in the least significant bits of grayscale images, with relatively poor robustness. Since 1994, digital watermarking has shown importance in information security and economics. It has received active participation and investment support from research institutions, universities, and business groups around the world. In addition, companies such as IBM, Hitachi, NEC, Pioneer Electronics, and Sony have also announced joint research on electronic watermarking based on information hiding. In 1995, text digital watermarking \cite{brassil1995hiding} was introduced, and Macq et al. \cite{macq1995cryptology} initiated research on video watermarking. In 1996, Cox et al. \cite{van1994digital} introduced a digital watermarking method utilizing spread-spectrum communication, which improved the digital watermarking scheme's robustness. Cox et al. \cite{cox1996secure} embedded the watermark information in the DCT domain to enhance the watermark's robustness, but their watermark algorithm was a non-blind watermark extraction technique. It required the participation of the original image for watermark extraction, which restricted the application.

The international academic community has also published a large number of articles on digital watermarking and held many influential international conferences and special journals. For example, in 1998, a special session on digital watermarking was established by the International Conference on Image Processing (ICIP). In 1999, at the third international workshop on information hiding, 18 out of 33 works focused on digital watermarking research. Venkatesan et al. \cite{venkatesan2000image} introduced a technique for blind watermarking capable of extracting the watermark without the original image. In the late 1990s, some companies began to formally sell digital watermarking products. For example, the American company Digimarc was the first to launch a commercial digital watermarking software and integrated it into Adobe's Photoshop and Corel Draw image processing software. As a security product designed for printed documents, AlpVision's Safe Work is capable of hiding watermark information on the back of the document to verify its authenticity and identify its ownership. In 2002, the National Strategy for Securing Cyberspace, which focused on the use of digital watermarking, was released by the U.S. Department of Homeland Security. Today, digital watermarking has emerged as a crucial tool for protecting the IP rights of multimedia data and has become an indispensable part of the Internet. The functions of current digital watermarking are constantly expanding, from initial copyright protection to data integrity verification, data content authentication, data tampering detection and tracing, and data traceability.

\subsection{LLM Watermarking}

LLM watermarking \cite{christ2024undetectable,kirchenbauer2023watermark} and traditional digital watermarking have similarities and differences. Simply put, their similarity lies in the fact that LLM watermarking and digital watermarking serve similar functions, both achieving the purpose of copyright protection and ownership identification through embedding specific identifiers. The difference is that LLM watermarking is more similar to traditional neural network watermarking, while it differs greatly from traditional digital watermarking algorithms. The development of LLM watermarking is relatively recent. With the emergence of LLMs that have shown powerful generation and understanding capabilities, becoming a prominent research area in AI, research on LLM watermarking has gradually emerged. Because of the exceptionally high costs associated with training and deploying LLMs, safeguarding the IP rights of LLMs has become especially crucial, which has sparked research interest in LLM watermarking research. In 2023, Kirchenbauer et al.  \cite{kirchenbauer2023watermark} proposed an LLMs text watermarking approach that uses red-green token patterns to achieve embedding and detection of LLM watermarks. This innovation laid the foundation for the development of LLM watermarking and inspired academic and industrial interest in this research area.

LLM watermarking is similar to traditional neural network watermarking, mainly by embedding specific patterns or structures during the model training process and modifying model parameters or structures, so that the generated content or the model itself carries hidden identity information. This method can not only embed text watermarks, but can also be extended to image, video, and audio data. For example, LLM-based image watermarking can embed imperceptible watermark information in the generated images, LLM video watermarking can embed imperceptible watermark information in specific frames of the generated videos, and LLMs audio watermarking can embed watermark information that is inaudible to humans in the generated audio. These approaches seek to safeguard the models' IP and prevent unauthorized use and alterations. Currently, research has proposed LLM watermarking methods for images \cite{wen2023tree}, audio \cite{cho2022attributable}, and multimodal settings \cite{tang2023watermarking}.

As LLMs continue to develop, LLM watermarking has become a research hotspot to safeguard the IP of LLMs and prevent their misuse for unlawful activities. The research and application of LLM watermarking are rapidly evolving and have emerged as a key focus area in LLMs research. As the application of LLMs becomes increasingly widespread in various fields, protecting the IP of these LLMs has become increasingly important. LLM watermarking can not only be used for copyright protection but also for marking multimedia data generated by LLMs, preventing malicious tampering and forgery using LLMs. For example, in news and social networks, watermarking technology can ensure the source and authenticity of content, quickly identify false content generated by LLMs, and prevent the spread of false information generated by LLMs. However, the research on LLM watermarking also faces many problems and challenges \cite{boenisch2021systematic}, such as how to improve the robustness and undetectability of LLM watermarks, how to embed watermarks without affecting the performance of LLMs, and how to improve the embedding speed of LLM watermarks to reduce interaction latency with LLMs. These are important issues that need to be solved for the development of LLM watermarking.

\section{Traditional Digital Watermarking Techniques}
\label{sec:Traditional}
\subsection{Watermark Properties}
\subsubsection{Embedding Capacity-based Classification}

Embedding capacity-based classification: Zero watermarking \cite{rani2015zero} and multi-bit watermarking \cite{maiorana2016multi}. Traditional watermarking algorithms aim to simultaneously guarantee the watermark's invisibility and robustness. However, these two indicators are often contradictory. In traditional multi-bit watermarking, to improve watermarking's imperceptibility, it is essential to embed fewer watermark bits and minimize the effect of each bit on the original image. Considering the different needs of these two indicators, it is difficult to simultaneously satisfy the above two requirements when designing watermarking algorithms. To achieve the above goals, zero-watermarking  \cite{fierro2019robust,rani2015zero} was proposed, which preserves the original image and maintains its quality, with good imperceptibility. The watermark information is combined with the host media through encryption, and only users with the correct key and decryption algorithm can decrypt and extract the watermark data. Since the watermark data isn't directly inserted into the media, unauthorized users find it hard to identify the watermark's presence, which allows zero watermarking to effectively meet the requirements of imperceptibility and robustness.

\subsubsection{Perceptibility-based Classification}

Digital watermarks can be divided into two types based on perceptibility: visible and invisible watermarks \cite{craver1998resolving,dekel2017effectiveness}. Visible watermarks are often traditional physical watermarks, such as anticounterfeiting authentication on banknotes and checks, or station logos on TV channels. Due to the relatively poor imperceptibility of visible watermarks, they are more susceptible to attacks, resulting in the complete disappearance of the watermark information. The current main research direction is invisible watermarks, which, also known as blind watermarks or steganographic watermarks, have the characteristic of being imperceptible after being embedded. Imperceptibility refers to how undetectable it is to the human visual system (HVS). It indicates that the watermark remains largely invisible to the human eye once embedded and has the capacity to maintain the quality of the original image effectively, without significant damage to the image pixels and structure.

To achieve imperceptibility, research often designs watermarking algorithms based on HVS's characteristics \cite{gao2010image,roy2018hvs}, such as the human visual system is more responsive to low-frequency information, such as large brightness changes and gradual spatial changes than high-frequency information, such as details and sharp edges. This means that during watermark embedding certain high-frequency details can be compromised without greatly impacting the visible quality of the image. Among the three primary colors, the HVS is more sensitive to green, followed by red, and is least sensitive to blue. The HVS’s sensitivity to spatial frequency varies with the position in the visual field, being highest at the center and decreasing towards the periphery. Additionally, the HVS is less responsive to areas with high average brightness and regions of high texture complexity. These characteristics of the HVS allow us to make appropriate modifications to the corresponding parts of the image to embed watermarks without significantly impacting the visual appearance, thus enabling the design of relevant invisible watermarking algorithms.

In the 1990s, researchers began to explore invisible watermarking \cite{craver1998resolving,yeung1997invisible,yeung1998invisible}. We can provide several classic examples of invisible watermarking for images: 1). Spatial domain invisible watermarking, such as the classic LSB watermarking, which modifies the least significant bits of the image's pixel data (0-255) to embed the watermark. The process for embedding and extracting is described below. First, choose the LSB from each pixel in the original digital image; then, change the LSB of each pixel to 0, as this step of modifying the pixel values minimally (at most 1 bit) will not perceptibly affect the image's visual effect. Next, insert the encoded watermark data into the LSB of each pixel. The image with the watermark is subsequently created. When extracting the watermark information, the first 7 bits of the image pixels are set to 0, leaving only the LSB, which contains the extracted watermark information. Although the LSB algorithm operates in the spatial domain, it has good imperceptibility due to the minimal 1 bit change in the LSB. The LSB algorithm is relatively simple to implement, exhibiting low computational complexity and strong imperceptibility. However, it suffers from low robustness and is vulnerable to various attacks. Subsequent research \cite{chen2012robust,dehkordi2011robust,parthasarathy2005increased} has proposed various improvements to the LSB algorithm to improve its robustness. 2). Transform domain invisible watermarking \cite{choubisa2011permutation}, where operations in the transform domain eventually impact the spatial data. The difference between transform domain watermarking and direct spatial domain operation is that watermark insertion in the transform domain typically involves converting the image from the spatial domain to the transform domain, embedding the watermark, and then performing the inverse transformation to produce the watermarked image in the image domain. The classical domain of transformation watermarking mainly includes watermarking based on DFT, DCT, and DWT transforms.

\textbf{Discrete Fourier transform (DFT)} \cite{poljicak2011discrete,salah2021fourier} is a discrete version of the Fourier transform, representing the signal across time and frequency domains \cite{candan2000discrete}. The DFT changes a signal's time-domain samples into frequency-domain representations of its discrete-time Fourier transform (DTFT). The DFT can be used for spectrum analysis, data compression, differential equation solving, large integer and polynomial multiplication, and signal processing. We will introduce several invisible watermarking techniques based on the DFT. Cedillo et al. \cite{cedillo2021improving} introduced an enhanced watermarking algorithm based on the DFT that employs the particle swarm optimization (PSO) algorithm \cite{wang2018particle} to fine-tune key parameters, improving both the stealth and resilience of the image watermark. The initial step in the watermark embedding process involves transforming the RGB color image into the YCbCr color model and extracting the luminance component $Y$. Then, a zero-mean binary pseudo-random watermark $W$ is generated. The 2D DFT is performed on the luminance component $Y $ to derive the magnitude $M(u,v)$ and phase $P(u,v)$ information. A pair of radii $r1$ and $r2$ are defined to form a ring-shaped region in the DFT magnitude spectrum $M(u,v)$, covering the desired frequency band. The watermark $W$ is then encoded using direct sequence spread spectrum (DS-CDMA), forming the encoded watermark WDS-CDMA. The embedded watermark is incorporated additively into the upper half of the ring-shaped region in $M(u,v)$, and the frequency values in the lower half are adjusted according to the symmetry property of the DFT. The modified magnitude $M'(u,v)$ is then converted back to the spatial domain using the inverse discrete Fourier transform (IDFT) to obtain the watermarked luminance component $Y'(x,y)$. Finally, the luminance of the watermark $Y'(x,y)$ is combined with the original chrominance information to convert the YCbCr color model back to RGB, generating the final watermarked image $Iw$. %The procedure for extracting the watermark is as follows: The image with the watermark $Iw$ is first transformed from RGB to the YCbCr color space to retrieve the watermarked luminance $Y'(x,y)$. The 2D DFT is then computed on $Y'(x,y)$ to obtain the watermarked magnitude $M'(u,v)$. The ring-shaped region $A$ is determined using the predefined radii $r1$ and $r2$. The embedded watermark is extracted from the upper half of $M'(u,v)$ using the secret key $k3$, and the watermark data bits $W'$ are recovered using the sign function and rearranged using the secret key $k2$. 
To refine the essential settings of the watermarking technique, particularly the frequency band and the number of frequency parameters, along with the factor determining watermark strength $\alpha$, the research adopts the PSO algorithm to define an objective function that considers both the imperceptibility and resilience of the watermark. During the PSO training process, the watermark strength parameter $\alpha$ and the interval of radii $r1$ and $r2$ are adjusted to find the optimal parameter configuration for every image. This can determine the best watermark embedding parameters for each image, achieving improved robustness while maintaining image quality.

\textbf{Discrete cosine transform (DCT)} \cite{agrwal2016improved,yuan2020new} is a specific instance of the discrete Fourier transform, using only real numbers \cite{sundararajan2001discrete}. The DCT is extensively employed in signal and image processing due to its effective "energy compaction" characteristic, which ensures that most of the signal energy is concentrated in its low-frequency components. This property is beneficial for lossy compression of signals and images. The classic applications of DCT include the still image coding standard JPEG, the video coding standards MJPEG and MPEG, as well as some audio coding techniques. Several invisible watermarking techniques \cite{agrwal2016improved} based on DCT have also been proposed, such as an invisible watermark based on IWT+DCT, which reduces the fractional loss in the wavelet transform. Compared to DWT+DCT-based watermarking, this technique can improve imperceptibility and robustness. Choubisa et al. \cite{choubisa2011permutation} proposed a method for invisible digital watermarking utilizing the DCT transform and image permutation. It incorporates the watermark information by swapping the classical coefficients at specific positions (e.g., (3,3) and (4,4) pixels) in each 8 $\times$ 8 DCT coefficient block and enhances the watermark security through image rearrangement. The watermark extraction process employs a reverse algorithm to extract the watermark and recover the original image.

\textbf{Discrete wavelet transform (DWT)} \cite{mohammed2023blind,patil2013dwt} is a type of wavelet transform where the wavelets are sampled discretely \cite{shensa1992discrete}. Similar to other wavelet transforms, a major advantage of the DWT compared to the Fourier transform is its time resolution. DWT can capture both frequency and spatial information, and mainly utilized in signal and image analysis. Next, we introduce several invisible watermarking techniques based on DWT. Patil et al. \cite{patil2013dwt} developed an invisible digital image watermark utilizing DWT, which embeds watermark information in quantized DWT coefficients within the frequency spectrum. By inserting the watermark data into the sequential zero coefficients of the mid-high frequency elements within each reconstructed block of the 3-level 2-D DWT, a lossless digital watermark hiding scheme is achieved. This method primarily involves two steps: embedding and extraction. The watermark embedding process is invertible, allowing for perfect restoration of the original image once the watermark is removed from the invisibly watermarked version. This method seeks to enhance the quality of the invisibly watermarked image and boost the watermark capacity of the original, while ensuring the embedded watermark does not compromise the original's visual quality during extraction. A blind watermarking method \cite{mohammed2023blind} can embed and extract color image watermarks in digital images, utilizing DCT and DWT transforms. Initially, it employs an adaptive color selection strategy to choose either the blue or green channel of the source for embedding. Next, the chosen color channel is segmented into non-overlapping 4 $\times$ 4 blocks, and the DCT transform is applied to each block. Subsequently, it applies DWT to break down the DC coefficients of each block into four subbands, then further decomposes the low-frequency subband, and finally utilizes the LH1 subband for embedding the watermark. Experiments show that this approach can improve the watermark's imperceptibility while ensuring its robustness.

% This subsection discusses the classification of invisible watermarks and the importance of imperceptibility for digital watermarking. Digital watermarks can be categorized from the perceptibility perspective into visible and invisible types. Invisible watermarks have better imperceptibility, referring to both the invisibility to the HVS and the superior quality of image post-watermarking. To achieve imperceptibility, the watermark algorithm must account for the HVS's attributes, such as high sensitivity to low-frequency information and low sensitivity to regions with relatively uniform brightness. Classic invisible watermarking algorithms include spatial domain LSB watermarking and transform domain DCT, DFT, and DWT watermarking.

\subsubsection{Application Scenarios-based Classification}

Before the widespread use of LLMs, to protect the IP of related AI models, there was emerging relevant AI watermarking, especially AI watermarking based on classic neural networks \cite{li2021survey,wu2020watermarking,zhang2018protecting}. As shown in Table \ref{tab:watermarking_comparison}, different watermarking techniques can be classified into four categories: white-box, black-box, gray-box, and no-box watermarking. Details are presented below.

\begin{table*}[hb]
    \footnotesize
    \centering
    \caption{White-box, black-box, gray-box, and no-box watermarking.}
    \label{tab:watermarking_comparison}
    \begin{tabular}{|p{2cm}|p{7cm}|p{7cm}|} 
    \hline
    \centering \textbf{Classification} & \centering \textbf{Scenarios} & \centering \textbf{Common techniques} \arraybackslash \\    
    \hline
    
    White-box & The model is accessible internally & Based on internal weights, internal structure, and composite verification \\    
    \hline
    
    Black-box & The model is not accessible internally & (1) Constructing a trigger set by modifying the labels. (2) Incorporating information and altering labels. (3) Adding samples with embedded information. \\    
    \hline
    
    Gray-box & Combination of white-box and black-box watermarking & Combination of white-box and black-box watermarking \\    
    \hline
    
    no-box & The model itself is no longer required & Integrating frameworks to covertly embed watermarks \\ 
    \hline
    \end{tabular}
\end{table*}

\textbf{In a white-box model}, during the watermark embedding phase, the model owner integrates the watermark into the model's internal structure in the watermark extraction part. Because the network architecture and internal weights of the target model are accessible, it is possible to retrieve the watermark and restore the model's accuracy. White-box watermarking mainly includes the following types: \textbf{(1) White-box watermarking based on internal weights}: The parameters representing the connection strength between neurons in a neural network model are called weights. Weight-box watermarking involves inserting a watermark by altering the internal neural network parameters of the target model. Lim et al. \cite{lim2022protect} introduced white-box watermarking with two distinct embedding methods that integrate the secret key into the hidden memory state of recurrent neural network (RNN) \cite{graves2012long} to safeguard an image captioning model. They proposed two embedding operations: Element-wise addition and multiplication to integrate the key into the hidden state of the LSTM unit. Kuribayashi et al. \cite{kuribayashi2023white} proposed a new white-box watermarking approach, particularly targeting fully connected layers within DNN-based models. The method integrates the watermark to safeguard the trained DNN model, aiming to minimize its effect on the model's weights. The research assumed that the accuracy of the DNN model is insensitive to differences in local minimum points. During the fine-tuning process, they use secret key-based random sampling to initialize the DNN's weights, then iteratively update the DNN's weights and find the local minimum points. The watermark embedding process is carried out following each training cycle of DNN to confine the search area for local minima, thereby minimizing the watermark's effect on the model weights. This scheme has low model distortion and strong watermark robustness. \textbf{(2) White-box watermarking based on internal structure}: Structural-based white-box watermarking embeds the watermark by altering the internal architecture of the target neural network model. The white-box watermark is easy to discover or remove, causing the watermark to become invalid. To effectively resist such removal attacks, it is common to use neural network pruning to integrate the watermark. Neural network pruning can minimize a neural network's size while maintaining its basic performance by pruning the redundancies of the less important neural network. Zhao et al. \cite{zhao2021structural} introduced a structural watermarking method that incorporates channel pruning for embedding the watermark, thus protecting the IP of the neural network. Two techniques for channel pruning, called network slimming and L1 norm pruning, are introduced, which are simple and efficient for watermarking schemes. Initially, the watermark is split into several bit segments, with each segment used to define a pruning rate, and then the derived pruning rates are allocated to the convolutional layers for channel pruning. In the IP verification process, the watermark can be accurately retrieved and restored by analyzing the channel pruning rates. Deepmark \cite{xie2021deepmark} leverages the pruning of neural network models. First, by copying the original model and setting 50\% of its parameters to 0, and then fine-tuning, Deepmark identifies redundant parameters that do not affect the model's performance. It then generates a pruning mask using the owner's signature, which determines the watermark embedding locations through hash processing and redundant parameter information. In the model pruning stage, the forward propagation function of the model is modified based on the pruning mask, which involves using the element-wise product of the mask and the model parameters to decide which neural network connections to keep or discard, thereby integrating the owner's information into the model. Finally, during watermark extraction, ownership can be confirmed by comparing the model parameters with the pruning mask. This approach integrates the watermark while keeping the performance impact of the neural network model to a minimum. (\textbf{3) White-box watermarking based on composite verification}: This type of watermark is similar to zero watermarking. However, unlike zero watermarking, which constructs a watermark that avoids embedding into the watermark carrier based on its structural properties, this approach is used to verify whether the carrier has been modified. White-box watermarking based on composite verification divides the watermark into two parts: one is incorporated into the target model, while the other is kept by the model owner. When verifying the validity of the watermark, the two parts are merged for validation.

\textbf{In a black-box model}, the model owner is unable to access the internal parameters and architecture of the potentially compromised model (i.e., the suspected stolen model). In this case, the owner can only use the model API to obtain the output to verify the copyright, usually by constructing a trigger set (i.e., a dataset used to achieve a specific output) to implement copyright verification. We specifically introduce a black-box watermarking method based on a multi-classification task: (1) Construct a trigger set by modifying the labels. Label modification refers to changing the correct label corresponding to the original sample to a label that does not match the original sample content. This belongs to the zero-watermarking method. (2) Create a trigger set by incorporating information and altering labels within the original training data. (3) Construct a trigger set by adding samples with embedded information. It creates distinct trigger sets from unique user signatures, and then fine-tunes the embedded watermark, and distributes them to the corresponding users to achieve traceability of the neural network model. Next, we introduce several classic black-box watermarking methods. Plmmark \cite{li2023plmmark} is a secure and resilient black-box watermarking framework specifically designed to protect the intellectual property (IP) of PLM. This framework mainly includes three main stages. Firstly, it encodes a watermark with the owner's identity information, establishes a strong association between the digital signature and the related trigger set using the original vocabulary of the PLM, and ensures the security of the scheme through public-key cryptography \cite{hellman1978overview}. To embed a robust, task-agnostic, and transferable digital watermark in the PLM, it introduces a supervised contrastive loss to make the output of the trigger set deviate from the result from the original clean sample, allowing the watermarked model to respond to the anomalous output of the trigger set and identify the ownership, while avoiding affecting the result from the clean sample. Finally, to ensure reliable verification of model ownership, it utilizes a dual verification method to prevent forgery. Blackmarks \cite{chen2019blackmarks} is an end-to-end multi-bit black-box watermarking framework for protecting DNN's IP. BlackMarks accepts a pre-trained, unlabeled model and the owner's binary signature as input, producing the corresponding marked model along with a set of watermark keys. The framework initially develops a model-specific encoding scheme to assign all possible classes to binary '0' and '1'. Following this, utilizing the watermark key provided by the owner (already mapped to a binary string), it uses a target adversarial attack to design a set of trigger images and labels. Subsequently, it fine-tunes the model using the generated watermark trigger set to integrate the watermark into the prediction patterns of the target DNN. During the watermark extraction phase, the framework employs trigger images to interact with the remote model and extracts the owner's signature from the resulting predictions.

\textbf{In a gray-box model}: the watermarking approach integrates elements from both white-box and black-box watermarking techniques. It utilizes the white-box method to insert data into the model's internal structure, while employing the black-box method to validate the model's ownership. However, unlike the black-box watermarking method, which achieves watermark embedding through the creation of a trigger set, the gray-box watermarking technique directly embeds information within the model itself to implement watermark embedding. HufuNet \cite{lv2021hufunet} is a gray-box watermarking method specifically designed for deep neural networks. HufuNet divides a small-scale DNN into two parts: one part (the left EPH slice) is incorporated into the target DNN model to serve as the watermark, and the other part (the right SPH slice) is retained by the model owner for verifying model ownership. This method offers significant resilience to model fine-tuning and pruning. HufuNet maintains the core performance of the DNN model while also safeguarding its IP. 

\textbf{In a no-box model}, the watermarking approach is distinct from white-box, black-box, and gray-box methods. It verifies model ownership without embedding a watermark in the model or using a trigger set to generate specific input-output pairs, i.e., the model itself is no longer required. Meanwhile, the input content, after passing through the model, will carry the watermark information, and the model ownership can be confirmed by retrieving watermark information from the output data. Zhang et al. \cite{zhang2021deep} introduced a no-box watermarking framework designed to safeguard DNN models. This task-agnostic framework, when integrated into the target model, embeds an invisible watermark into the model's output. If an attacker uses the input-output pairs of the target model to train a substitute model, the hidden watermark will be incorporated, extracted, and subsequently utilized for copyright verification. Wu et al. \cite{wu2020watermarking} additionally introduced a no-box watermarking framework for DNN models, where the original neural network and the watermark extraction network are jointly trained. By optimizing a combined loss function, the trained neural network embeds the watermark into the output image while performing the original task, which stands apart from earlier watermarking schemes that alter neural network weights or the trigger set's classification labels, by directly identifying the watermark within the output image. It determines the ownership of the original neural network and verifies if the image is generated by a specific neural network. Abdelnabi et al. \cite{abdelnabi2021adversarial} proposed a new no-box text watermarking technique called adversarial watermarking transformer (AWT) to trace and verify the text source. AWT utilizes adversarial training and learns how to hide and decode binary information in text without significantly changing the semantics and correctness of the text. This model can effectively hide information while maintaining the text's usefulness and is resistant to a series of potential neural network attacks.

% Conclusion: In white-box watermarking, the watermark is directly embedded in the model and the watermark is embedded primarily by altering the model's parameters and structure. In black-box watermarking, the watermark authenticates the model's copyright by constructing a specific data set or input-output pair. Gray box watermarking combines the characteristics of white-box and black-box watermarking. It not only integrates the watermark within the model but also verifies copyright through external verification methods. Boxless watermarking eliminates the need for model modification and instead verifies the model's copyright using input and output images. These methods aim to protect the IP rights of AI models, prevent unauthorized usage, and protect the IP associated with related research.

\subsubsection{Robustness-based Classification}

The robustness of a digital watermark denotes its capacity to be extracted or recovered despite undergoing different types of attacks, which reflects the system's resilience. Digital watermarks are inevitably subjected to unintentional or intentional attacks during the transmission process. For digital watermarks employed in copyright protection, ensuring robustness is a crucial concern, as the embedded watermark may suffer from compression, noise, cropping, filtering, and rotation attacks, among others. To ensure that the watermark can effectively carry out its copyright protection task, research has conducted in-depth studies on how to improve watermark resilience. Watermarks can be categorized into three types according to their level of robustness: robust watermarks, semi-fragile watermarks, and fragile watermarks \cite{zhao2011robust}. Robust watermarks \cite{tao2014robust} refer to those that can resist various signal processing and attacks, or have relatively strong resistance to specific attacks. As mentioned earlier, this type of watermark is primarily utilized for copyright protection and offers substantial resistance to attacks. Fragile watermarks \cite{rakhmawati2019recent} are those that are destroyed by any change to the image and cannot resist any attacks, being sensitive to any attack. This type of watermark is used primarily to verify data integrity, such as in the financial industry. Semi-fragile watermarks \cite{feng2020novel} can resist certain unintentional attacks and have robustness against unintentional attacks, but are sensitive to malicious attacks. Semi-fragile watermarks are applied mainly in the authentication of common multimedia information, such as applying very small JPEG compression (1\%) to the image during transmission, which is nearly undetectable by the human eye. In this case, if a fragile watermark is used for detection, it will indicate that the watermarked image has been altered. Semi-fragile watermarks with robustness against unintentional attacks are more suitable for detecting malicious attacks in such environments. In the following, we summarize robust watermarks, semi-fragile watermarks, and fragile watermarks.

\textbf{Robust watermarks} \cite{barni1997robust,praun1999robust,wagner2000robust}. In recent years, with advancements in deep learning and coding theory, robust watermarking has been further developed. Here are some examples. Singh et al. \cite{singh2021optimized} proposed an optimized robust watermarking technique, combining the application of the chaotic k-best gravitational search algorithm (CKGSA) within the domains of DCT and SVD, to address issues of false positives (FPP) in traditional SVD watermarking techniques and attain a compromise between invisibility and durability. This watermarking technique initially segments the original image into $8 \times 8$ blocks, applies the Arnold transform (AT) and DCT transform to each block, and then performs SVD to extract the principal components from these blocks. The watermarking image is also subjected to Arnold chaotic mapping to enhance its robustness, and then the embedding factor is optimized through CKGSA to integrate the key components of the watermark image with chaotic mapping into the original image. The watermark extraction process is similar to the embedding process; however, the host image is not required for extraction. only the watermarked image needs to be processed similarly. CKGSA is utilized to identify the optimal embedding factor by maximizing the correlation between the original watermark and the extracted watermark, as well as the structural similarity between the original image and its watermarked version, to calculate the fitness value. Xu et al. \cite{xu2021compact} introduced a streamlined deep learning algorithm utilizing the INN architecture, called the invertible watermarking network (IWN). This method does not rely on AT or SVD decomposition, but directly embeds and extracts watermarks within the reversible neural network to improve the information loss problem in traditional deep learning watermarking techniques. IWN employs a single reversible neural network to handle watermark embedding and extraction as inverse operations, achieving a one-time solution. In the embedding stage, this method first performs block-wise processing on the image and then forward propagation through the INN. It incorporates a message normalization component and a noise addition layer to improve robustness while embedding watermark information into the image's DCT coefficients. In the extraction stage, this watermarking method uses the INN backpropagation process to restore the embedded watermark information, and the entire process shares the same set of network parameters. Fang et al. \cite{fang2022encoded} introduced a three-phase watermarking framework for enhancing the resilience of digital watermarks to distortions. The initial phase involves noise-free training to set up the encoder for watermark insertion and train the decoder for watermark retrieval. The second stage implements a mask-guided frequency enhancement approach to refine the encoded features, informed by the analysis of these features and associated distortions, ensuring that these features are retained during the subsequent watermark distortion process. The third stage further strengthens the decoder through adversarial training to enable accurate watermark extraction from distorted images.

\textbf{Fragile watermarking}  \cite{dittmann1999content,wolfgang1999fragile,yeung1998fragile}. Nejati et al. \cite{nejati2022fragile} introduced a fragile watermarking method for verifying image authenticity. Initially, the Fourier transform \cite{bracewell1989fourier} is applied to shift the host image from the spatial domain to the frequency domain, thereby enhancing the visual fidelity of the marked image. Next, QR decomposition is applied to both the transformed original image and the watermark image. During watermark embedding, a coefficient $\alpha$ from the R matrix of the watermark image is added to the R matrix of the host image, incorporating the watermark information and creating a new upper triangular matrix. Finally, it multiplies the Q matrix of the original image with the newly formed upper triangular matrix to obtain the image matrix containing the watermarked information. During verification, if the watermark can be successfully extracted from the image, it indicates that the host image has not been attacked; if not, it means the image may have been tampered with. This method is highly sensitive to minor watermark attacks and is suitable for applications with strict requirements for image integrity. Huang et al. \cite{huang2022self} introduced a robust fragile watermarking method that operates within the encrypted domain. This approach involves creating and embedding two distinct types of watermarks: one for detecting tampering and another for image recovery. First, the original image is segmented into 2 $\times$ 2 non-overlapping blocks, and the average value for each block is computed. The top 5 bits of these averages are then utilized as the recovery watermark. Next, the original image undergoes perturbation through a logistic-logistic system to create a modified image. This modified image is then segmented into 2 $\times$ 2 non-overlapping blocks, where the average value of each block is computed and converted to binary form. Combined with the identification watermark and restoration watermark, a second-level watermark is generated using a specific algorithm. This approach employs a three-tier tampering detection system. The initial detection occurs in the cloud, while the recovery phase utilizes a "block-level detection and pixel-level restoration" method to accurately locate and restore tampered areas of the image across different scenarios. Su et al. \cite{su2021hybrid} introduced a fragile watermarking approach utilizing a hybrid Sudoku framework for detecting and authenticating tampered digital images, which employs a dual-perspective system to conceal the watermark data. Initially, the watermark is digitally integrated into each pixel pair within the first Sudoku perspective, producing provisional coordinate data. Subsequently, in the second Sudoku perspective, this temporary coordinate information is hidden in the original image, achieving the watermark embedding. The watermark generation is completed using a pseudo-random number generator and a pre-set key, and the generated binary sequence is then converted to a 9-ary numeral system representation as the watermark S. During the data embedding phase, the original image is segmented into non-overlapping pixel pairs, which are then encoded based on the 9-ary watermark S. Through the data hiding process based on hybrid Sudoku, the watermark is embedded into the pixel pairs, resulting in the creation of a watermarked image. During the data extraction phase, the recipient retrieves the watermark S' from the potentially tampered image using the same method and compares it with the original watermark. This comparison allows for verification of image integrity and identification of tampered areas with pixel pair precision. The approach enhances both the security and invisibility of the watermark, while preserving the high quality of the watermarked image.

\textbf{Semi-fragile watermarking} \cite{hassan2006semi,ho2004semi,lin2007semi}. Molina et al. \cite{molina2020effective} introduced a semi-fragile watermarking approach that integrates IWT and DCT methods for detecting and recovering from image tampering. This technique generates two types of watermarks: an authentication watermark and a restoration watermark. The authentication watermark is created with IWT and integrated into the original image's coefficients to produce the watermarked image. The restoration watermark is generated by partitioning the watermarked image and applying DCT, with each 2$\times$2 block producing a 10-bit watermark. These watermarks are combined into a restoration label and sent with the watermarked image. Upon receipt, authenticity is verified using tampering detection methods, and the restoration watermark is utilized to restore altered areas based on the recovery label. Rhayma et al. \cite{rhayma2021semi} introduced a novel semi-fragile watermarking method designed to improve the security and integrity verification of JPEG2000 images throughout the compression process. The central concept of this watermarking technique is to utilize a perceptual hash function (PHF) to generate the watermark, which can resist compression attacks from the JPEG 2000 encoder while maintaining sensitivity to image content changes. Specifically, the host image undergoes a transformation via DWT, followed by processing with PHF to derive the DWT coefficients. The PHF extracts essential visual features from the image to produce a distinctive hash value. This hash value is insensitive to minor image changes but can detect substantial content alterations. The watermark is embedded in the image with minimal visual impact and can be detected if the image is altered. During watermark embedding, index-modulation quantization directly integrates the watermark into the fifth-level wavelet decomposition's approximate subband coefficients within the JPEG2000 compression framework, allowing the watermark to be effectively extracted during image decoding without any additional files or data. Ouyang et al. \cite{ouyang2024semi} introduced a new semi-fragile reversible watermarking approach designed for detecting and locating image tampering, which overcomes the limitation of conventional semi-fragile watermarking techniques, which cannot restore the original data carrier, and enhances the precision of tampering detection and localization and embeds the watermark into the mid-frequency quaternion discrete Fourier transform (QDFT) \cite{sangwine1997discrete} coefficients to ensure semi-fragility, while also incorporating reversible compensation data into the watermarked image to enable recovery. Experimental results indicate that it can successfully withstand various common watermarking attacks, precisely identify tampered areas, and restore the original image without loss in the absence of attacks. Yuan et al. \cite{yuan2024semi} introduced an innovative semi-fragile watermarking technique using neural networks aimed at content authentication and tampering detection. This approach creates a collection of semi-fragile samples without altering the parameters or structure of the neural network model. By examining the model's responses to these samples, it can swiftly detect any malicious tampering and authenticate the content of the neural network model. The scheme is not only applicable to specific DNNs but also exhibits significant transferability. Waterlo \cite{beuve2023waterlo} is an innovative image protection approach designed to defend against Deepfake attacks. This scheme uses a local semi-fragile watermarking technique. If the image is altered, the semi-fragile watermark in the affected area will be compromised, allowing the detector to identify the image tampering and accurately locate the position. A compression module is incorporated into the training process to improve the watermark's resilience to JPEG compression.

% So far, we introduced three types of digital watermarks generated for the robustness requirements of different attack types and analyzed their characteristics and application scenarios, respectively. When analyzing different watermarking schemes, we listed some specific technical methods and their advantages, including robust watermarking technology that combines deep learning and coding, fragile watermarking schemes based on encrypted domains, and IWT and DCT. Semi-fragile watermarking technology, etc. These approaches are tailored to address the security requirements of digital content security across various domains.

\subsection{Extraction-based Methods}
\subsubsection{Content-based Classification}

\begin{table*}[hb]
\scriptsize
\centering
\caption{Comparison of Spatial and Transform Domain Watermarking}
\label{tab:watermarking_comparison_1}
\begin{tabular}{|p{3.8cm}|p{5.8cm}|p{5.8cm}|}  
\hline
\centering \textbf{Classification} & \centering \textbf{Advantages} & \centering \textbf{Disadvantages} \arraybackslash \\ 
\hline
\begin{tabular}[c]{@{}l@{}}Spatial domain\\watermarking\end{tabular} & 
\begin{tabular}{@{\labelitemi\hspace{\dimexpr\labelsep+0.5\tabcolsep}}l@{}}Easy implementation.\\Low computational complexity.\end{tabular} & 
\begin{tabular}{@{\labelitemi\hspace{\dimexpr\labelsep+0.5\tabcolsep}}l@{}}Poor invisibility and robustness.\\Limited embedding capacity.\end{tabular} \\ 
\hline
\begin{tabular}[c]{@{}l@{}}Transform domain\\watermarking\end{tabular} & 
\begin{tabular}{@{\labelitemi\hspace{\dimexpr\labelsep+0.5\tabcolsep}}l@{}}Good invisibility and robustness.\\Large embedding capacity.\end{tabular} & 
\begin{tabular}{@{\labelitemi\hspace{\dimexpr\labelsep+0.5\tabcolsep}}l@{}}Complex implementation.\\High computational complexity.\end{tabular} \\ 
\hline
\end{tabular}
\end{table*}

Content-based classification includes spatial domain watermarking and transform domain watermarking. As shown in Table \ref{tab:watermarking_comparison_1}, spatial domain watermarking and transform domain watermarking have distinct advantages and disadvantages. Spatial domain techniques are easier to implement but lack robustness, while transform domain techniques offer better security and robustness at the cost of increased computational complexity.

\textbf{Spatial domain watermarking} \cite{su2018robust,yuan2020fast}. It refers to the direct operation on the pixels (i.e., the spatial domain of the image) to perform watermark embedding. This type of watermarking mainly includes several operation methods: (1) Pixel intensity modification: Watermark data can be embedded by altering pixel intensity values. Techniques include embedding information through changes in pixel brightness, contrast, or hue, or by modifying the relative brightness and contrast between adjacent pixels to enable differential embedding. (2) Pixel expansion: Extending certain pixels in the original image to other regions to embed the watermark information. For instance, duplicating a pixel’s value to multiple locations within the image can extend the watermark information's impact. (3) Pixel rearrangement: Watermark information can be embedded by altering the pixel values in the image. For example, substituting original pixel values with a defined pattern or sequence achieves the embedding of watermark data. Several new spatial domain watermarking methods are introduced as follows. Su et al. \cite{su2018robust} introduced a spatial domain blind watermarking technique that embeds a binary watermark by directly altering the pixel values of the blue channel in the RGB image. To insert the watermark, the generation principle and distribution characteristics of the DC coefficients are utilized. The DC coefficients are quantized using a secure key, then four distinct sub-watermarks are inserted into various areas of the original image. During watermark extraction, the sub-watermarks are blindly extracted based on the DC coefficients of the watermarked image and the key. Because this algorithm operates directly in the spatial domain instead of the transform domain, it benefits from both the simplicity and speed of spatial domain processing and the robustness typically associated with the DCT domain. Yuan et al. \cite{yuan2020fast} introduced an efficient and resilient blind watermarking technique for color images. This algorithm achieves good imperceptibility and strong robustness, combining the DCT applied to the frequency domain to solve the copyright protection problem of color images. This approach takes advantage of watermarking techniques in both the spatial and frequency domains. It achieves watermark embedding and blind extraction by directly modifying the DC coefficients of pixel blocks in the spatial domain, eliminating the need to process the actual DCT domain. The method applies distinct quantization processes to the red, green, and blue channels of the image, enhancing the correlation among the R, G, and B layers. It also leverages the correlation between the DC coefficients of neighboring pixel blocks to improve both the watermark's invisibility and robustness. Compared to conventional DCT-based methods, this approach offers quicker embedding and extraction, supports color watermark images, and provides greater embedding capacity. Abraham et al. \cite{abraham2019imperceptible} introduced an innovative watermarking technique for the spatial domain for color images that gradually disperses the watermark information in the pixels of a certain region, similar to transform domain techniques. This approach aims to achieve the two fundamental features of a watermarking system: high imperceptibility and high robustness. Through various watermark image quality measurements and experiments of common watermark attacks, the findings demonstrate that this approach offers excellent imperceptibility and strong resistance to attacks. Nana et al. \cite{nana2016watermarking} introduced a blind digital watermarking method for spatial domain images, employing SVD theory and its properties. This method first rearranges the digital watermark image using a chaotic sequence, then segments the original image into blocks, and applies SVD to each block to derive small singular value matrices. Then it embeds the quantized singular values with the watermark. Tests involving JPEG compression, cropping, noise, rotation, and filtering attacks demonstrate that this approach exhibits both high robustness and excellent imperceptibility.

\textbf{Transform domain watermarking} \cite{fares2020robust,khare2021reliable} refers to the process of embedding a watermark by first applying a reversible mathematical transformation to the spatial-domain data, then modifying the transformed coefficients using a watermark embedding algorithm, and then applying the inverse transformation to retrieve the watermarked multimedia data. The transformations utilized consist of orthogonal techniques such as DCT, DWT, and DFT. Transform-domain watermarking algorithms implement the watermark signal superposition in the frequency domain and leverage techniques from spread-spectrum communications to effectively encode the watermark signal, thereby improving its imperceptibility and robustness. Next, we introduce several recent transform domain watermarking techniques. Fares et al. \cite{fares2020robust} introduced an innovative digital watermarking method for color images utilizing the Fourier transform domain. This method involves dividing the image into components R, G, and B and applying different Fourier transforms (e.g., DFT, FFT, and QDFT) to each component. The digital watermark is concealed by leveraging the parity of the middle-frequency band coefficients after transformation. This technique ensures that the watermarked images retain high quality while offering excellent imperceptibility and robust resistance to common attacks. Khare et al. \cite{khare2021reliable} introduced an advanced image watermarking method that integrates DWT, homomorphic transformation (HT), SVD, and AT. The approach starts by applying a single-level DWT to the original image to extract various subbands. It then focuses on the HL subband and applies HT to break it down into illumination and reflectance components. The watermark undergoes AT for increased security before being embedded in the singular values of the reflectance component. This component, which holds the primary features of the image, offers improved resistance to attacks when used for watermark embedding. Furthermore, rapid variations in the reflectance component contribute to better imperceptibility. The tests demonstrate that this method provides strong robustness. Khare et al. \cite{khare2021reliable} introduced a robust and secure watermarking method that integrates DWT and SVD within the FRFT domain. The algorithm begins by applying FRFT to both the original and the watermark images to extract their amplitude components. It then performs a two-level DWT on the amplitude of the original image. Following this, a singular value decomposition (SVD) \cite{stewart1993early} is performed in the low-frequency subband of the original image and the watermark amplitude, creating a new matrix with singular values to embed the watermark. The optimal FRFT transformation order is then determined numerically to balance imperceptibility, robustness, and security. This approach demonstrates enhanced performance in imperceptibility and resistance to conventional signal processing and geometric attacks, including image rotation, cropping, average filtering, median filtering, and Gaussian filtering. Zheng et al. \cite{zheng2020robust} introduced a robust watermarking method that integrates DWT, DCT, and SVD to enhance resistance to common watermarking attacks. The approach starts with decomposing the original image using DWT, followed by applying DCT to the low-frequency subband. SVD is then used to extract rotation-invariant features for watermark insertion. To mitigate false positives often associated with SVD-based methods, the scheme incorporates a signature-based authentication system to validate the watermarked image's integrity. This method excels in maintaining visual quality and robustness, especially in demonstrating strong resistance to rotation attacks. Kishore et al. \cite{kishore2020novel} developed a new blind watermarking technique utilizing DCT. It ensures security, invisibility, durability, and simplicity. The watermark is embedded by altering DCT coefficients at two distinct positions within the image. Notably, the extraction process does not depend on the original or watermark images, and the method demonstrates excellent invisibility, minimal computational demands, and high resilience.

% This subsection provides an overview of the concepts of spatial domain watermarking and transformed domain watermarking, as well as some new watermarking methods. Currently, there are several spatial domain watermarking algorithms, including modifying the blue component of RGB images, using DWT and HT techniques to implement color image watermarking, and using SVD and chaotic sequences to implement watermark embedding. Recent advances in spatial domain watermarking focus on leveraging transformed domain features to enable faster watermark embedding. Transformed domain watermarking involves performing a reversible mathematical transformation and then embedding the watermark. Existing research has proposed transformed domain watermarking algorithms based on the Fourier transform, DWT, and FRFT. These methods have seen improvements in imperceptibility and robustness, and have better resistance to common attacks. The more recent transformed domain watermarking emphasizes the use of multiple transformed domain features to achieve joint watermark embedding, thereby improving the overall performance of the watermark.

\subsubsection{Bitstream-based Classification}

\textbf{Moving picture experts group-2 (MPEG-2)} \cite{iso13818} is a standard for digital video and audio compression, created by the ISO in 1994. It is a further development of the MPEG-1 standard, focusing on applications in broadcast television, DVD video, and digital broadcasting. MPEG-2 employs efficient compression algorithms to reduce the transmission and storage costs of video and audio data \cite{tudor1995mpeg}, while maintaining relatively high audiovisual quality. This standard is primarily used in digital television, satellite broadcasting, cable television, DVD, and other broadcast and storage media. The MPEG-2 system stream typically includes two basic elements: video data/audio data and timestamps. Their roles and relationships in the MPEG-2 system are as follows:

\textbf{1. Video data + timestamps}: The video data in the MPEG-2 system stream contains the compressed video signal. This data has been processed by the MPEG-2 compression algorithm to reduce the data volume, making it more suitable for transmission and storage. The video data contains the information of the images, and through decoding, the images can be restored to a visible form. The timestamps are associated with the video data to identify the time position of each video frame. The timestamps ensure that the video frames can be presented in the correct sequence and time interval during playback or transmission. The timestamp information is crucial for achieving synchronized playback, especially in multimedia streams, where the video and audio timestamps are used to maintain the synchronization between the two.

\textbf{2. Audio data + timestamps}: The audio data in the MPEG-2 system stream contains the compressed audio signal. Similar to video data, the compression of the audio data helps to reduce the data volume, making it more suitable for transmission and storage. The audio data can be restored to audible audio through decoding. The audio data also carry timestamps to identify the time position of each audio frame. Like video, audio timestamps ensure that audio frames can be played in the correct temporal order and synchronized with video frames during playback or transmission.

Various digital watermarking methods utilizing MPEG-2 bitstreams have been introduced: Choi et al. \cite{choi2010blind} introduced a novel blind watermarking technique aimed at enhancing the resilience of MPEG-2 video watermarks against attacks from camcorder recordings. The method takes advantage of the stability of low-frequency DCT coefficients under geometric transformations and embeds watermark data by altering the mean of these coefficients over time. To prevent the watermark from shifting across frames, the approach restricts watermark embedding to B frames in MPEG-2 video, which reduces the need for partial decoding and enhances the efficiency of watermark extraction. Lu et al. \cite{Lu2011mpeg2} introduced a blind watermarking algorithm for MPEG-2 video. Using DCT, it identifies and groups certain mid-frequency DCT coefficients from the video frames, embedding the watermark by modifying these mid-frequency coefficients within each group, which has a large watermark capacity, certain robustness against video compression attacks, and minimal impact on video quality. Cruz et al. \cite{cruz2010blind} developed a blind watermarking algorithm for MPEG-2 video that achieves fully blind watermarking without requiring the original video data, original watermark, or any other related information. This method embeds two-dimensional visually recognizable images, such as company logos and owner icons, into the DWT domain of video frames for copyright protection. To improve security, the watermark data is converted to noise-like data using a chaos-based mixing method with two numeric keys. This scheme offers several benefits, including fully blind detection, resilience to common video attacks and combined attacks, and low complexity. Jiang et al. \cite{jiang2011video} proposed an MPEG-2-based video watermarking scheme for copyright protection, which embeds watermark information into the DC coefficients of block images in instantaneous decoding reference frames during the differential pulse code modulation process in the MPEG-2 encoder. A synchronization code is also designed to address watermark synchronization issues, allowing blind extraction without the original image or full decoding. The watermark is periodically embedded and undergoes random testing, providing strong robustness against common video attacks, including low-bitrate MPEG-2 compression and synchronization attacks.

\textbf{Moving picture experts group-4 (MPEG-4)} \cite{mpeg4_overview} is a collection of standards for compressing and encoding audio and video information, created by the MPEG, which is part of the International Organization for Standardization (ISO) and the International Electrotechnical Commission (IEC). The standard was first approved in October 1998 and the second version was approved in December 1999. The main application areas of the MPEG-4 standard include network streaming, optical disc storage, voice transmission (e.g., video telephony), and television broadcasting. MPEG-4 integrates most of the functionalities of MPEG-1 and MPEG-2, and further extends support for the virtual reality modeling language (VRML) \cite{brutzman1998virtual}. It introduces object-based composition, encompassing audio, video, and VRML objects, along with digital rights management (DRM) \cite{subramanya2006digital} and various interactive features. Compared to MPEG-2, MPEG-4 no longer uses macroblocks for image analysis, but instead records the changes in individual objects within the image, avoiding the blocky artifacts that can occur when the image changes rapidly and the bitrate is insufficient. MPEG-4 has higher interactivity and flexibility compared to MPEG-1 and MPEG-2. Additionally, MPEG-4 offers the advantage of a high compression ratio (up to 4000:1). It achieves this by discarding redundant elements and only processing the different elements between image frames, significantly reducing the file size of the multimedia composition. Due to its low bitrate, small core program space, strong computational capabilities, and excellent communication application integration, MPEG-4 has become the most important standard format in the digital video and audio industry.

Here are several digital watermarking schemes based on MPEG-4 bitstreams. Rajpal et al. \cite{rajpal2019novel} introduced an innovative video watermarking method for real-time MPEG-4 video frame watermarking. The method integrates extreme learning machine (ELM) \cite{ding2014extreme}, fuzzy logic, and PSO for embedding and extraction of watermarks in video frames. It utilizes ELM to choose appropriate video frames for watermarking and applies the fuzzy-PSO technique \cite{abdelbar2005fuzzy} within the DWT-SVD domain for embedding. Barni et al. \cite{barni2005watermarking} introduced a watermarking algorithm for MPEG-4 video elements. In the macroblocks of the video object plane, it selects pre-defined DCT coefficient pairs and adds specific relationships between these coefficient pairs to integrate the watermark data. The watermark embedding considers mid-frequency quantized DCT coefficients, balancing imperceptibility and robustness. The method can efficiently embed watermarks without significantly degrading video quality and is reasonably robust against common video attacks. Essaouabi et al. \cite{essaouabi2010wavelet} introduced a watermarking system for MPEG-4 video authentication that utilizes the DWT. This system employs a shape-adaptive discrete wavelet transform (SA-DWT) for watermark embedding and adjusts the watermark strength using a visual model based on HVS principles. This approach ensures a balance between visibility and robustness by embedding the watermark in the average of wavelet blocks, rather than in individual wavelet coefficients, making the system resilient to distortions in some wavelet coefficients. This watermarking scheme is visually imperceptible and can resist various attacks, such as lossy compression in MPEG-1, MPEG-2, MPEG-4, and H.264. Moreover, the scheme allows for watermark detection without requiring the original video. Yesilyurt et al. \cite{yesilyurt2015robust} introduced a peak-based watermarking method for MPEG-4 videos to enhance the resilience of compressed video files. The method involves two primary steps: First, it transforms the intraframes (I-frames) from RGB to YCbCr color space and determines the peak value of each luminance (Y) component to set a threshold peak value for the video. Second, based on this threshold, it embeds the watermark into the mid-frequency DCT coefficients of the Y component, placing each binary watermark bit in a distinct DCT block. Experimental results demonstrate that this approach offers excellent imperceptibility and robustness during binary watermark embedding.

We have introduced digital watermarking based on bitstream classification, using MPEG-2 and MPEG-4 digital video and audio compression standards as examples. For MPEG-2, several watermarking algorithms are introduced for MPEG-2 video, leveraging low-frequency DCT coefficients along with timestamps to enhance the watermark's resilience during video transmission. For MPEG-4, we introduced some novel watermarking techniques. These methods maintain video quality while providing high imperceptibility and robustness, which are applicable to various video processing operations and attack scenarios.

%\subsection{Extraction-based: Blind Watermark Extraction, Semi-blind Watermark Extraction, Nonblind Watermark Extraction}

\subsection{Extraction-based Watermark Extraction}

Based on the extraction method, digital watermarks can be divided into three types: blind watermark extraction, semi-blind watermark extraction, and non-blind watermark extraction. Blind watermarking: the original unwatermarked media is not required during watermark extraction. Non-blind watermarking requires access to the original, unwatermarked media for extracting the watermark. Semi-blind watermarking does not depend on the original host media for watermark detection and extraction; instead, it utilizes the watermarked signal itself for the process.

\textbf{Blind watermark extraction}. Hong et al. \cite{hong2001blind} developed a blind watermarking method using wavelet transform that allows for copyright verification without requiring the original image. The process begins with applying a wavelet transform to the image and then generating a signature with a secret key. This key is utilized both for creating the initial signature and for extracting it during the watermarking process. Najafi et al. \cite{najafi2017robust} developed a robust image watermarking algorithm utilizing the DWT. Logo watermark elements are inserted directly into the subbands of the three-level DWT decomposition. This scheme is entirely blind for both the host image and the watermark, meaning no original image or embedded watermark information is needed during detection. It ensures imperceptibility, blindness, and robust resistance to various geometric and non-geometric attacks. Yuan et al. \cite{yuan2020dct} introduced a blind watermarking technique for color digital images using two-dimensional DCT and variable step-size quantization. It addresses the issue of copyright protection for color digital images. It initially splits the host and watermark images into R, G, and B color components, followed by dividing them into fixed-size blocks. It then selects specific DCT coefficients from these blocks, processes them in a particular scanning order, and applies varying quantization step sizes to the DCT coefficients in different positions for watermark embedding and blind extraction. This technique enhances watermark imperceptibility and robustness against various attacks, and is ideal for protecting the copyright of color digital images.

\textbf{Semi-blind watermark extraction}. Mohammed et al.  \cite{mohammed2020imperceptible} introduced a semi-blind image watermarking scheme utilizing DWT-SVD combined with an efficient embedding method. This scheme ensures high robustness and imperceptibility for digital image copyright protection. During embedding, the watermark data is first transformed into the frequency domain via the DWT algorithm; subsequently, the SVD algorithm is applied to the LL subband values, and the diagonal matrix is integrated into the original image. The original image is decomposed into two levels using DWT. Watermark bits are then embedded into the HL2 and HH2 subbands of the original image using the Zigzag technique to enhance imperceptibility. For watermark extraction, only the watermarked image and the U and V values of the SVD transformation are needed. Palani et al. \cite{palani2024semi} introduced a semi-blind watermarking scheme for detecting and recovering tampered medical images. This technique uses the invariant watermark features generated by the CNN, combined with the turtle shell matrix data hiding (TSDH) \cite{liu2017data} algorithm and the DWT-SVD, to perform dual-layer embedding, enhancing the watermark security and enabling tamper localization and recovery of the generated image. This scheme not only improves the imperceptibility of the watermark and the accuracy of tamper detection but also demonstrates robust tamper detection and image recovery performance against various common malicious watermark attacks. Similarly, to protect medical images, Rezayatmand et al. \cite{rezayatmand2022robust} developed a resilient semi-blind watermarking technique utilizing DWT and SVD. This method applies a two-level Haar wavelet transform (HWT) \cite{stankovic2003haar} to the original image, followed by single-level SVD on its low-frequency components, which are then combined with the watermark coefficients. An additional SVD layer is used to integrate the watermark, boosting its robustness. The approach enables semi-blind watermark extraction, allowing retrieval without the original image, thus enhancing both the imperceptibility and durability of the watermark.

\textbf{Non-blind watermark extraction}. Dappuri et al. \cite{dappuri2020non} introduced a non-blind color image watermarking method that utilizes the SVD algorithm within the translation-invariant wavelet (TIW) \cite{liang1996translation} domain, further refining the SVD-TIW approach with the Enhanced Grey Wolf Optimizer (E-GWO) \cite{nadimi2021improved}. A non-blind image watermarking method \cite{kartikadarma2021comparison} evaluates the efficacy of DCT and HWT algorithms in image processing, and the results showed that HWT performs better, especially in blind watermarking. A pseudo-zero watermarking approach \cite{thanh2022pseudo} leverages non-blind watermarking and visual secret sharing (VSS) \cite{feng2008visual} for protecting digital image copyrights. This method divides the copyright information into n shares using a k-out-of-n distribution technique, which resembles a $(k,n)$ threshold secret sharing scheme and is also referred to as a $(k,n)$ threshold secret sharing scheme and is also referred to as $n$-1 shares are filed with the copyright office. To verify the watermark, the user extracts it from the watermarked image and decodes it using any $k$-1 of the $n$-1 shares to retrieve the copyright information.

% In summary, there are three types of digital watermarking based on the extraction method: blind extraction, semi-blind extraction, and non-blind extraction. Blind watermark extraction uses algorithms such as wavelet transform, DWT, and variable-step quantization to verify the watermark without the original multimedia data information. Semi-blind watermark extraction employs methods like DWT and SVD to enhance robustness and imperceptibility. Non-blind watermark extraction methods utilize techniques like SVD and gray wolf optimization to retrieve the watermark, given that the original multimedia data is accessible. These schemes are applied in different fields, such as digital image copyright protection and medical image tampering detection.

%\subsection{Based on Application Purpose: Copyright Protection Watermark, Authentication Watermark, Covert Communication Watermark, Annotation Watermark}
\subsection{Application-based Purpose}

Based on different application purposes, digital watermarks can be classified into the following types: copyright protection watermark, authentication watermark, covert communication watermark, and annotation watermark. Copyright protection watermark: mainly used to protect the copyright of media content and the IP of related digital media. Authentication watermark: Primarily used to ensure the authenticity and integrity of content, especially in sectors such as finance. Covert communication watermark: Using watermark technology to convey hidden information in media for covert communication. Annotation watermark: Watermarks are used to add additional information, such as creator annotations, copyright notices, etc.

\textbf{Copyright protection watermark}. Sinhal et al. \cite{sinhal2021machine} introduced a blind color image watermarking approach leveraging machine learning for copyright protection. This method uses the YCbCr color space and applies IWT and DCT for embedding watermarks in color images. In the watermark embedding process, the Y channel is segmented into fixed-size blocks, and a random number generator utilizing the Mersenne Twister selects the blocks for embedding. This generator uses a key, enhancing the security of the watermarking scheme. To minimize computational complexity, an ANN architecture was developed for watermark embedding. Experimental results indicate that this scheme excels in imperceptibility and robustness. The ANN framework provides faster embedding speed for the watermarking scheme, which can be used for robust watermarking applications with low computational time or time-sensitive requirements, such as watermarking multimedia data for generative AI. Ahmadi et al. \cite{ahmadi2021intelligent} introduced an intelligent dual-color image watermarking scheme for both copyright protection and image authentication. The scheme uses the blue channel in the RGB color space to embed an invisible, robust watermark for copyright protection. Additionally, it introduces a novel method for manipulating diagonal singular values, embedding a fragile watermark in each RGB channel to ensure image authentication. By integrating robust and fragile watermarks, this scheme effectively safeguards color images. Darwish et al. \cite{darwish2020dual} introduced a novel dual watermarking method for safeguarding color image copyrights. The scheme combines continuous watermarking and segmented watermarking techniques, and optimizes the watermark embedding location and scaling factor using a genetic algorithm to balance imperceptibility and robustness. The scheme utilizes DWT to capture the original image's prominent features and applies Walsh-Hadamard Transform (WHT) \cite{ahmed1975walsh} to encrypt the watermark, enhancing the security of the watermarking process. This method demonstrates strong robustness against typical watermark attacks.

\textbf{Watermarking for authentication}. Guan et al. \cite{guan2020reversible} introduced a reversible watermarking method aimed at verifying the integrity of CNNs. The approach leverages model compression techniques based on pruning theory to develop the host sequence used for watermarking and employs the histogram shifting method to embed the watermark information. Experimental results indicated that the reversible watermark affects classification performance by less than ±0.5\%, a minimal impact, and allows for the complete recovery of model parameters post-watermark extraction. Additionally, the integrity of the neural network model can be confirmed using the reversible watermark. Gong et al. \cite{gong2020secure} introduced an image authentication method that uses dual fragile watermarks to ensure image integrity and identify tampered areas,which features two types of sensitive fragile watermarks: the diffusion watermark and the authentication watermark. To strengthen the security of these watermarks, both types employ secret keys to create random sequences, which are then embedded into the two LSB layers of the original image. Experimental results and security analysis demonstrate that the scheme can resist selected image attacks. Swain et al. \cite{swain2022effective} introduced an image watermarking method that employs block truncation coding (BTC) \cite{delp1979image}  and SVD for image verification and tampering recovery. This approach involves segmenting the image into non-overlapping 4 $\times$ 4 blocks and embedding the watermark data for both authentication and recovery within these blocks. The authentication watermark for each block is created using SVD combined with XOR operations, whereas the recovery watermark is derived from BTC. The watermark data is embedded into the two least significant bits of every pixel within each block.

\textbf{Covert communication watermarking}. RoSteALS \cite{bui2023rosteals}, is a novel steganography method, which utilizes a pre-trained autoencoder to hide watermark data. RoSteALS achieves robust watermarking with limited training and no content specialization by directly injecting the secret information into the pre-trained autoencoder. The technique has a lightweight secret encoder with only about 300k parameters, is easy to deploy and train, and demonstrates good secret recovery performance and strong image quality in three benchmark tests. Gladwin et al. \cite{gladwin2020combined} proposed a watermarking algorithm that integrates elliptic curve cryptography (ECC) \cite{koblitz2000state} with Hill cipher to bolster image security and confidentiality. Initially, an ECC-based key generation mechanism is employed to create a key, which is then used to produce ciphertext via the Hill cipher algorithm. This ciphertext, along with the image's DCT coefficients, is embedded into the least significant bits of the original image. By merging steganography and cryptography, this approach ensures a higher level of security and privacy for covert data transmission, rendering it challenging to retrieve the hidden data and images without the correct key. Hossen et al. \cite{hossen2020new} introduced an image steganography scheme that utilizes AES and RC5 encryption systems for data hiding. The scheme first encrypts the secret information using the AES and RC5 algorithms, and then utilizes digital watermarking techniques to hide the encrypted data in the image. Through this scheme, even if the data is intercepted by an attacker, it cannot be read without the correct key.

\textbf{Annotation watermark}. Korus et al. \cite{korus2014new} introduced a high-capacity annotation watermarking method utilizing digital fountain coding. Existing watermarking schemes are severely limited in effective capacity due to the strict requirement of watermark transparency. This approach employs the fountain coding model and establishes an efficient watermark communication system akin to conventional data packet networks. The research primarily concentrates on achieving annotation with high capacity, and it is assumed that the fidelity requirement for the watermarked image is relatively low. The scheme demonstrates strong resistance to JPEG compression and cropping attacks. He et al. \cite{he2008high} introduced a high-fidelity image watermarking method for image annotation, designed to offer moderate robustness against image distortion. This visual perception model quantifies an image's noise tolerance, combining outputs from an entropy filter and a differential standard deviation filter. Two watermark types were created: guide watermark and information watermark. The guide watermark indicates the presence of the watermark, while the information watermark conveys tens of bits of data. This technique aims to embed 32-bit data in a single image and shows strong robustness against JPEG compression and cropping attacks. Applied to photography and medical images, the scheme demonstrated the visual model's effectiveness and the practicality of the proposed annotation technique. Vielhauer et al. \cite{vielhauer2006image} introduced an image annotation watermark that embeds hierarchical data linked to user-selected regions within the image. Unlike earlier methods, this research focuses on robustness against cropping attacks while preserving hierarchical object relationships after watermark recovery. Two relationship categories were defined: visual-functional and visual-spatial. A new encoding scheme was developed for the visual-functional relationships. The system features an ontology-based interactive editor and an image watermarking scheme, with an extension of block-based image watermarking to meet annotation watermarking requirements.

% This section detailed several applications scenarios and related technologies of digital watermarking. Based on different application purposes, digital watermarking can be categorized into copyright protection watermarking, authentication watermarking, steganographic communication watermarking, and annotation watermarking, which were then introduced in detail. In copyright protection watermarking schemes, research uses techniques such as machine learning, color space transformation, and wavelet transformation to achieve copyright protection of media content. Authentication watermarking is primarily employed to validate the authenticity and integrity of content, utilizing methods like deep learning and dual fragile watermarking. Steganographic communication watermarking is used to hide information transmission, using technologies such as autoencoder and ECC. Annotation watermarking is used to add additional information such as author annotations and copyright statements, using technologies such as high-capacity annotation and visual perception models. These digital watermarking technologies have shown good prospects in different application fields.

%\subsection{Based on Algorithm Characteristics: Adaptive Watermarking and Non-Adaptive Watermarking}
\subsection{Algorithm-based Characteristics}

Based on the classification of algorithm characteristics, digital watermarking can be categorized into adaptive and non-adaptive types. Adaptive watermarking: the watermarking algorithm can adjust the watermark embedding strategy according to the characteristics of the media content. Non-adaptive watermarking: The watermark embedding algorithm is independent of the specific properties of the media content. The specific characteristics of the algorithm are detailed below.

\textbf{Adaptive watermarking}. Bhinder et al. \cite{bhinder2020improved} introduced an enhanced robust image adaptive watermarking method that embeds two watermarks into high entropy 8×8 blocks of the image. Initially, DWT decomposes these blocks into four subband coefficients. The approximation and vertical frequency coefficients are then modeled with a Gaussian distribution. Watermark embedding is adjusted using an adjustable strength factor (ASF) based on the fourth statistical moment, kurtosis. The watermark also transmits limited side information, including the location of high entropy blocks and Gaussian distribution parameters. During watermark extraction, a statistical-based maximum likelihood decoder is employed to retrieve the watermark. Su et al. \cite{su2022robust} introduced a resilient and adaptive blind watermarking technique specifically for color images, designed to resist geometric distortions. This method leverages the frequency domain characteristics of the oblique transform, identifies the highest energy coefficients within pixel blocks in the spatial domain, and embeds watermark information by quantifying these coefficients. The approach also incorporates several adaptive strategies, including adjustments to image dimensions, quantization parameters, watermark encoding, and embedding locations, to enhance the robustness of the watermarking process. Additionally, it leverages the image's geometric features to recover from distortions. Altay et al. \cite{altay2021self} introduced a resilient digital image adaptive watermarking method that leverages DWT and SVD for copyright protection. Initially, DWT decomposes the original image into various subbands. The method then identifies non-overlapping blocks within the low-frequency subbands based on their standard deviation values for watermark embedding. These selected blocks are subjected to SVD. An adaptive step firefly algorithm is employed to determine the embedding scaling factor, optimizing the balance between robustness and imperceptibility. To further secure the watermark, a Fibonacci-Lucas transform is applied to scramble the binary watermark.

\textbf{Nonadaptive watermarking}. Loan et al. \cite{loan2018secure} introduced a secure and robust non-adaptive watermarking approach for digital images, integrating coefficient difference techniques with chaotic encryption. The process begins by dividing the image into 8$\times$8 pixel blocks and applying DCT. Specific DCT coefficients from adjacent blocks are then chosen for differential operations to embed the watermark. The method conceals the watermark by altering the difference between selected coefficients, which enhances its resistance to various signal processing and geometric attacks. For enhanced security, chaotic mapping and AT are utilized to encrypt the watermark. This technique supports embedding into grayscale images, the luminance component of color images, or each RGB channel, thereby increasing watermark storage capacity. Parah et al. \cite{parah2016robust} introduced a robust non-adaptive watermarking technique for digital images, known as the block-based inter-coefficient difference method within the DCT domain. This technique involves partitioning the image into non-overlapping 8 $\times$ 8 pixel blocks and applying the DCT to each. During watermark embedding, the differences between DCT coefficients at corresponding positions in adjacent blocks are computed and adjusted based on the watermark bit, which helps ensure the watermark's invisibility and robustness. By regulating the amount of change applied to the DCT coefficients, the method leverages the mid-frequency subband coefficients, thus preserving image quality while improving the watermark's resilience against various types of attacks.

% This section presents the two primary types of digital watermarking techniques categorized by their algorithmic characteristics. Adaptive watermark technology modifies the watermark embedding strategy according to the media content's features, enhancing the watermark's robustness and imperceptibility. We introduce several recent adaptive watermarking techniques, including an improved robust image-adaptive watermarking technique, a resilient adaptive blind color image watermarking method, and an adaptive digital image watermarking technique utilizing DWT and SVD. In contrast, non-adaptive watermarking techniques do not depend on the specific characteristics of the media content. We also present several recent non-adaptive watermarking techniques, such as a robust and secure method using coefficient difference and chaotic encryption, and a resilient non-adaptive digital image watermarking approach utilizing inter-block coefficient difference. These techniques are crucial in their respective application areas.

% This subsection presents the two primary types of digital watermarking techniques, categorized by their algorithmic characteristics. Adaptive watermark technology modifies the watermark embedding strategy according to the media content's features, enhancing the watermark's robustness and imperceptibility. In contrast, non-adaptive watermarking techniques do not depend on the specific characteristics of the media content.

%\subsection{Based on Content Classification: Meaningful Watermark, Meaningless Watermark}
\subsection{Content-based Classification}

Digital watermarking can be classified into meaningful watermarks and meaningless watermarks based on the content. A meaningful watermark is one where the watermark encodes a digital image (e.g., a logo) or a digital audio fragment (e.g., a music clip). A meaningless watermark corresponds only to a sequence number, such as a hash value. The advantage of a meaningful watermark is that if the watermarked carrier is attacked or the decoded watermark is damaged for other reasons, people can still visually confirm whether there is a watermark. For a meaningless watermark, if there are several bit errors in the decoded watermark sequence, one can only determine whether the signal contains a watermark through statistical decision-making.

\textbf{Meaningful watermark.} Yan et al. \cite{yan2011watermarking} introduced a digital watermarking scheme utilizing visual cryptography, generating meaningful watermark images. These images are no longer meaningless noise but have a certain understandability, which helps to avoid detection by attackers. Compared with previous visual cryptography schemes, this scheme can be applied not only to binary black-and-white images but also to grayscale and color images, without changing the pixel expansion. The scheme is easy to implement and highly feasible, and can flexibly select appropriate watermark embedding algorithms and watermark image sizes to provide efficient and secure information protection in practical operations. Liu et al. \cite{liu2005design} introduced a significant digital image watermarking algorithm. This method employs a radial basis function neural network (RBFNN) \cite{bors1996median} to adaptively calculate the highest watermark embedding strength that the DCT coefficients can handle, simulating human visual characteristics. The watermark used is a meaningful binary image processed through Arnold confusion, which allows for maximum information embedding without compromising the original image quality and offers robust resistance to various common attacks.

\textbf{Meaningless watermark.} Zhang et al. \cite{zhang2023m} introduced a highly robust audio meaningless watermarking technique designed to withstand large-scale cropping attacks, known as the m-sequence and sliding window combined audio watermarking scheme (m-SW-LSC). Initially, the method applies DWT, gabor-based transform (GBT), and SVD operations on the host audio signal to produce transform coefficients. It then uses a spread spectrum to embed the encrypted chaotic watermark information into these coefficients. The scheme merges the m-sequence and encrypted watermark to form the watermark key, leveraging the periodic properties of the m-sequence to theoretically ensure that the watermark can self-recover even if cropped. The scheme discards the synchronization mechanism, implementing an effective sliding window strategy to extract residual watermark fragments from the DWT-GBT-SVD coefficients and recover the full watermark using the watermark key. Experimental results demonstrate that the m-SW-LSC scheme withstands various common attacks, such as large-scale cropping and desynchronization attacks, significantly enhancing the robustness of the audio watermark. Artiles et al. \cite{artiles2022robust} introduced a meaningless image watermarking algorithm employing chaotic sequences and error-correcting coding to bolster the watermark's robustness. The algorithm's essence lies in merging the watermark information with a bit sequence produced by a key-controlled chaotic mapping, forming an error-corrected watermark information sequence, and then embedding the error-correction check bits into specific frequencies of the image using the DCT method. This approach enables authorized users to retrieve the watermark information using intact check bits and error-correction capabilities, even in the presence of image attacks.

% This section introduces meaningful watermarks and meaningless watermarks. Meaningful watermarks refer to digital watermarks that have certain meanings, such as trademark images or coded music fragments, as well as technologies that use visual cryptography to generate shareable images with understandable content. These meaningful watermarking techniques provide high robustness and information protection while maintaining image quality. In contrast, meaningless watermarks only correspond to a sequence number, such as a hash value, whose advantage is to provide higher security, but may require statistical decision-making operations during decoding to determine whether the signal contains a watermark.

\section{Common LLM Watermarking Techniques}
\label{sec:LLMs}

%\begin{figure}
%    \centering
%    \includegraphics[width=0.96\linewidth]{figs/classification.pdf}
%    \caption{Classification of LLM watermarking techniques. }
%    \label{fig:enter-label}
%\end{figure}

Over the past decades, digital watermarking has been widely employed to secure the ownership of text, images, audio, and video content. However, extending watermarking techniques to LLMs to achieve reliable protection of LLMs IP is still in its early stages. Given the exponential growth in training costs for LLMs compared to traditional neural network models, research on digital watermarking for LLMs for IP protection is of great urgency. Although LLM watermarking borrows fundamental ideas and techniques from traditional watermarking, embedding watermarks into LLMs and recovering them have both similarities and differences compared to traditional watermarking.

Compared to traditional watermarking, the main characteristics of LLM watermarking are as follows: 1) Similar to the neural network watermarking in traditional watermarking, LLM watermarking can adopt white-box watermarking or black-box watermarking approaches for watermark embedding. Due to the "large" nature of LLMs, LLMs have a larger watermark embedding capacity compared to traditional neural network models, and the watermark can be embedded at the most suitable embedding layer. Moreover, many LLMs have a multimodal nature, making it easier to construct less detectable trigger sets in black-box neural network watermarking, increasing the stealthiness of the watermark. 2) The output end of LLMs can adopt traditional watermarking methods for watermark embedding, i.e., direct watermark embedding in the output text/image/video/audio data of the LLMs. Alternatively, similar to traditional neural networks, intentional watermark embedding modifications can be made to the training data set.

The application scenarios of LLMs are very broad, including NLP tasks. In addition to processing text data, LLMs can also be applied in the image and audio domains, learning multimodal data and performing tasks such as image-to-text, text-to-image, speech-to-text, and text-to-speech. Similarly to the division of traditional digital watermarking, this section mainly divides LLM watermarking into three aspects according to different application fields: the text domain, the image domain, and the audio domain. Finally, this section also briefly discusses multimodal watermarking and dynamic watermarking. The comparison of LLM watermarking techniques is shown in Table \ref{table:LLM1}, Table \ref{table:LLM2}.

\subsection{Text Domain Watermarking for LLMs}

The emergence of LLMs has made the research of text domain watermarking for LLMs increasingly important. The most mainstream applications of LLMs include generating natural language text, machine translation, and text classification tasks. LLMs are trained on a large amount of text data to generate high-quality text, making it difficult to distinguish them from human-written text. This increases the potential for inappropriate use of text produced automatically. For example, malicious users can generate fake news articles and spread them on social networks. As LLMS have demonstrated great potential for the text they produce to be used for commercial purposes, concerns have also been raised about the IP of LLM data. Text domain watermarking technology for LLMs inserts distinctive signs or data into models that process text tasks. These watermarks can be implemented through statistical differences, text-based, model-based, and other methods to help verify whether the text is produced by a particular model, protect the IP of the model, track and identify the source of the model, and prevent unauthorized use and abuse. Classification of text domain LLM watermarking techniques with their advantages and disadvantages are shown in Table \ref{table:advdisadv}. Here are several common text domain watermarking for LLMs:

\subsubsection{Embedding Watermarks into Text}

Text-based watermarking techniques \cite{hou2023semstamp,kirchenbauer2023watermark,wang2023wasa,yang2023watermarking} utilize the unique linguistic features of each language model to embed watermarks.  These features can include word frequency, character modifications, word substitution, or sentence reordering.  For example, by using specific vocabulary or phrases, or using specific syntactic structures in the generated text, watermarks can be embedded.  Such watermarks may be inconspicuous and difficult to remove or modify.  These techniques can be categorized by when the watermark is inserted: watermark embedding in the text pre-processing stage, generation stage, and post-processing stage. In traditional neural network-based text digital watermarking, watermarks can also be embedded in different stages.  First, in the text preprocessing stage, watermarks can be embedded in the training data by modifying specific text attributes. Second, in the generation stage, watermarks can be directly embedded in the text generation model by adjusting the parameters or structure of the neural network model.  Finally, in the post-processing stage, additional processing can be performed on the generated text, such as adding specific text vocabulary, text phrases, or text syntactic structures, to embed watermarks.  In traditional text digital watermarking, watermarks are embedded by modifying specific text attributes, such as text word frequency and text sentence structure.  In the following sections, we will discuss the inheritance and development of text domain watermarking for LLMs, making adjustments to accommodate the inherent characteristics of LLMs.

\textbf{1) Pre-processing before text generation}. The watermarking technique in the pre-processing stage involves embedding predefined watermark information into the training data to achieve watermark insertion. It first maps watermark information of different lengths to the watermark, assigns a unique watermark to each data provider, and then embeds the watermark into sentences that represent the unique characteristics of the data provider. This allows the LLMs to learn the precise mapping of the corresponding watermark from the text of different data providers. In the generated sentences, the watermark is embedded in random positions, making it more difficult to remove or modify. The output from LLMs could potentially violate the IP rights of the data employed in their training. To address this issue, it is crucial to be able to trace the contributors of the data used to train the models. By having the LLMs generate synthetic text with embedded watermarks, the source attribution and provenance can be established. Watermark-augmented source attribution (WASA)  \cite{wang2023wasa} uses Unicode characters that are imperceptible to the human eye to construct the watermark. Each watermark consists of 10 characters, each selected from 6 invisible Unicode characters. The six Unicode characters include: zero-width space, zero-width non-joiner, zero-width joiner, invisible time, invisible separator, and invisible plus. Wei et al. \cite{wei2024proving} designed two types of watermarks: one is a watermark with an inserted random sequence, and the other is a watermark with random replacement of similar Unicode characters. The inserted watermark is a random character sequence appended to the end of the text. The replacement watermark involves curating a replacement list similar to Unicode for uppercase and lowercase ASCII letters. These methods have two limitations: they require pre-embedding a fixed watermark into the training data, which is costly for large-scale datasets; and they can only embed a single, fixed message as a watermark unless the model is retrained.

\begin{figure}[b]
    \centering
    \includegraphics[width=1\linewidth]{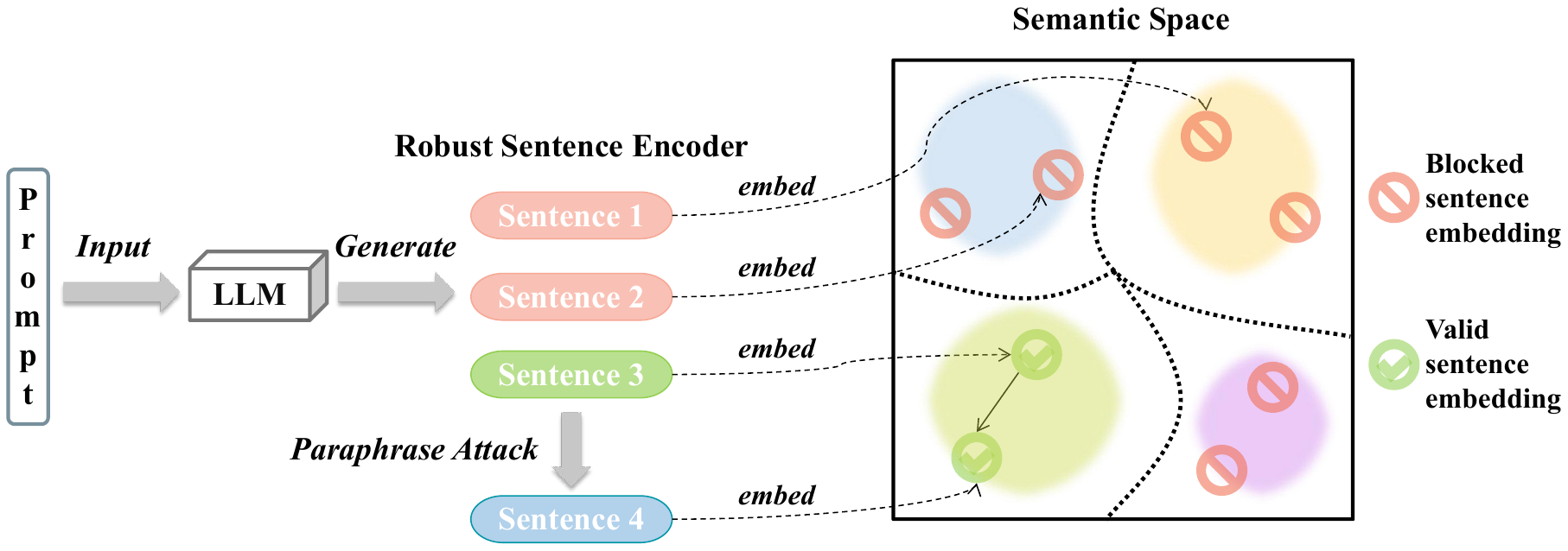}
    \caption{Modifying the model generation process: sentence level \cite{hou2023semstamp,hou2024k}.}
    \label{fig:sentence}
\end{figure}

\begin{figure}[t]
    \centering
    \includegraphics[width=0.9\linewidth]{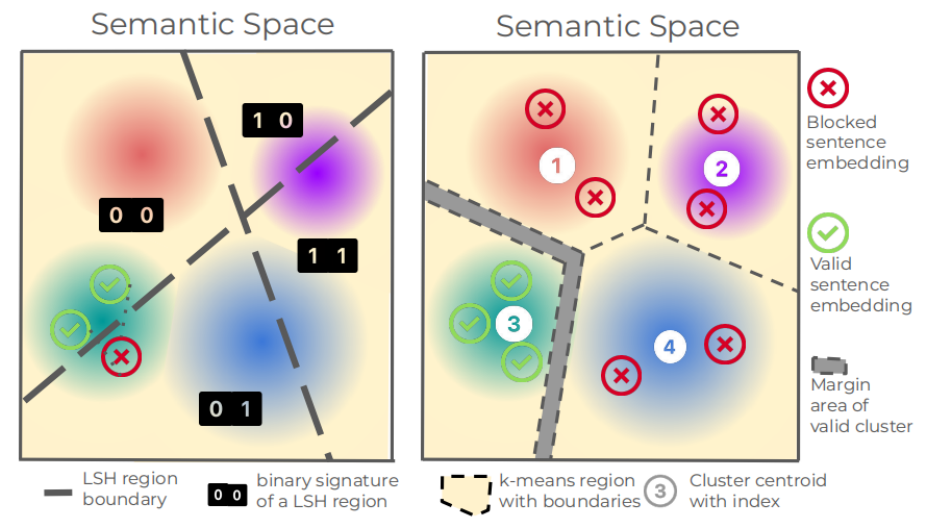}
    \caption{SEMSTAMP vs. $k$-SEMSTAMP \cite{hou2024k}.}
    \label{fig:semstampcompare}
\end{figure}

\textbf{2) Modifying the generation process of the model at the sentence level}. Watermarking algorithms that modify the model's generation process are mostly applied to white-box text models, although some can perform watermark verification in black-box scenarios. Since this watermarking method requires retraining the model from scratch and intervening in the model, the watermarked text must be generated under white-box conditions. This method is suitable for model owners who can access the model's output probability distribution and potentially interfere with the sampling process. However, third-party users of black-box text model services cannot access the internal structure of the model or alter the model’s vocabulary probability distribution. This feature makes this watermarking algorithm infeasible for black-box text model use cases. This watermarking algorithm modifies the output probability distribution during the text generation process to embed the watermark or selects text that meets the conditions during sampling for output. The model generation process is illustrated in Figure \ref{fig:sentence}. SEMSTAMP  \cite{hou2023semstamp} embeds semantic watermarks at the sentence level, utilizing $d$-dimensional locality-sensitive hashing (LSH) \cite{charikar2002similarity,indyk1998approximate}, which can produce a $d$-bit binary signature. The method employs LSH for segmenting the semantic realm of expressions, processes encoded options produced by LLMs, and conducts sentence-based rejection sampling until a selected sentence falls within a designated watermark section within the semantic embedding zone. The sentence-level semantic watermarking algorithm can well maintain the quality of the generated text. However, SEMSTAMP has limitations, since the arbitrary plane might split two sentences with comparable semantics into distinct areas. Rewriting sentences near the region boundaries may transfer the sentence embeddings to nearby regions, leading to suboptimal watermark strength. Subsequently, $k$-SEMSTAMP \cite{hou2024k} was proposed, an enhanced version of SEMSTAMP. The approach adopts $k$-means clustering \cite{lloyd1982least}, an alternative to LSH, to segment the semantic space while maintaining semantic coherence.  To more effectively counteract paraphrasing attacks, $k$-SEMSTAMP enforces cluster boundary constraints, ensuring that sentence embeddings are adequately distant from these boundaries, which significantly enhances robustness and sampling efficiency. The comparison between SEMSTAMP and $k$-SEMSTAMP is shown in Figure \ref{fig:semstampcompare}. Since the watermark information of the sentence-level semantic watermarking algorithm is associated with the semantic content of the sentence, the watermark information can still be intact and has stronger robustness even if the sentence is replaced with synonyms, rewritten, and other operations. However, because this kind of watermarking algorithm uses the rejection sampling method, the speed of generating text will be slow.

\begin{figure}[b]
    \centering
    \includegraphics[width=0.8\linewidth]{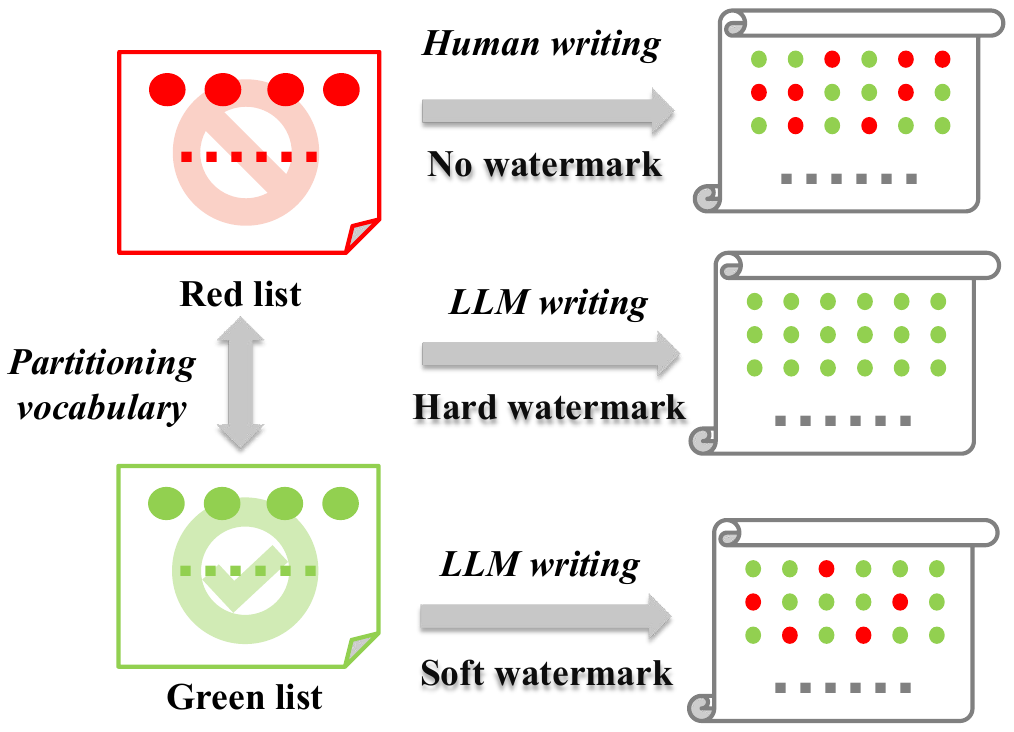}
    \caption{(1) No watermark: About half the words in the text are in the green list. (2) Hard watermark: The words of the text are all green lists. (3) Soft watermark: The offset increases the probability of generating a green word.}
    \label{fig:comparison}
\end{figure}

\textbf{3) Modifying the generation process of the model at the token level}. This watermarking method chooses to modify the output probability distribution or sampling at a smaller unit, from sentences to words. The watermarking method randomly separates the vocabulary into two distinct groups, referred to as the green list and the red list. The red list generator uses the previous token as a seed and can regenerate the red list using a hash function without accessing the entire generation sequence. If the text is human-written, the probability of a word belonging to the green list or the red list is random, approximately 50/50.

i) Soft/Hard watermarking: For hard watermarking, if the text is generated by LLMs, the entire text must be composed of words from the green list and cannot contain any words that are on the red list. However, such generated text has lower quality and cannot generate some common phrases. To alleviate this issue,  soft watermarking (KGW) 
\cite{kirchenbauer2023watermark} does not strictly require the text to be entirely composed of words from the green list, but relaxes the hard watermarking constraints by introducing a green list offset to increase the probability of generating green words. The model generation process is illustrated in Figure \ref{fig:token}. The main feature of this watermarking is that the expected score per token of the watermarked text is different from that of natural text. For unwatermarked text, the final statistics are consistent with the standard Gaussian distribution. In contrast, for watermarked text, the final statistics deviate, usually presenting a significantly higher value. The generated text will then consist of mostly green words and a small portion of red words. In the watermark extraction phase, a third party can use the red list generator to regenerate the red list for each token and calculate the number of rule violations. A null hypothesis test is then used to indicate that a watermark is detected if the number of violations exceeds a set threshold. This method has a relatively simple embedding and extraction process, and the watermark can be verified in a black-box manner without the need to decode the watermark using the original model. The comparison of no watermark, hard watermark and soft watermark is shown in Figure \ref{fig:comparison}. The reliability of watermark detection largely depends on the hashing scheme, i.e., the number of tokens used for hashing and generating the green list. To color the token at location $t$, KGW  \cite{kirchenbauer2023watermark} uses only the single token at position $t$-1, with the green list depending on a single token. KGW-reliability  \cite{kirchenbauer2023reliability} uses the tokens at positions $t$-1 and $t$ to color the token at position $t$, with the green list depending on two tokens, making it more difficult to brute-force the watermarking rules. In contrast, token-level-based algorithms are less robust than sentence-level ones. Because they rely on lexical level matching, even simple editing and rewriting of the text, such as an attacker replacing words in the green list with synonyms, may lead to loss or change of watermark information.

\begin{figure}[t]
    \centering
    \includegraphics[width=1\linewidth]{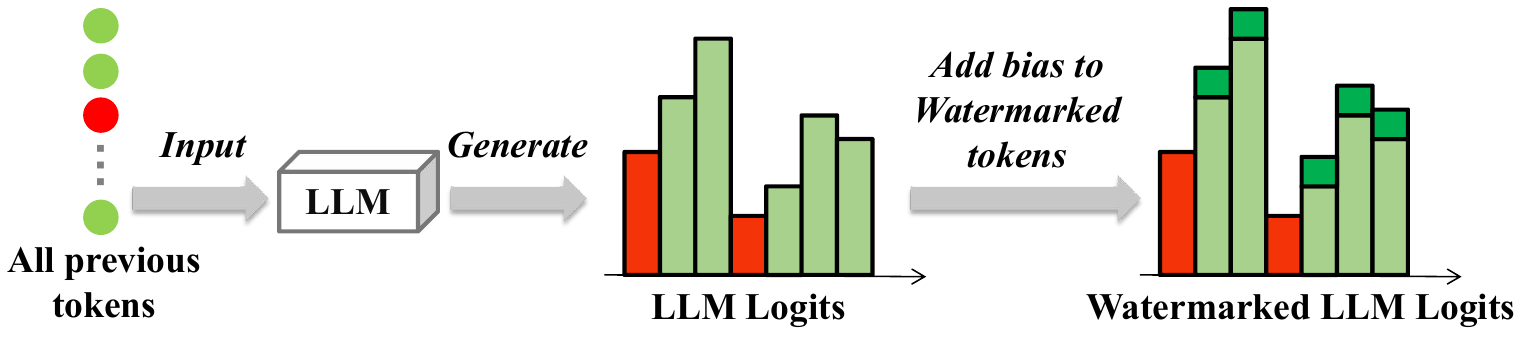}
    \caption{Modifying the model generation process: token level \cite{liu2023private,liu2023semantic}.}
    \label{fig:token}
\end{figure}

ii) Improvements and diversity issues of soft watermarking. One of the drawbacks of soft watermarking is that if there is insufficient occurrence of green tokens in low-entropy scenarios, it can lead to increased difficulty in detection. Building upon the foundation of soft watermarking  \cite{kirchenbauer2023watermark}, SWEET  \cite{lee2023wrote} improves soft watermarking based on information entropy. It selectively applies watermarking to tokens with sufficiently high entropy above a defined threshold. In the generation process, the green/red list rules are not applied to every token, but only to the tokens with entropy high enough to meet the given threshold, ensuring that the watermark can still be detected in low-entropy scenarios. The tendency of soft watermarking to produce the same result for the same prompt has a significant negative influence on both the user experience and the diversity of model output. To enhance diversity, GumbelSoft  \cite{fu2024gumbelsoft}, a variant of soft watermarking, uses two strategies to introduce variability in the decoding function, as well as a strategy to introduce a pseudo-random function (PRF). For example, directly combining Gumbel noise into the logit vector influences the sampling of the next token, increasing the uncertainty of the output. While the GumbelSoft watermarking technique achieved diversity, the efficiency of the watermark is lower, requiring more tokens to achieve accurate detection. Duwak  \cite{zhu2024duwak} is a dual watermarking algorithm, which embeds two independent secret modes into the token probability distribution and sampling scheme. Duwak enhances text diversity while minimizing the number of tokens required for detection, using 70\% fewer tokens than existing methods. As the scale and user base of LLMs grows, the efficiency and practicality of existing watermarking methods are limited, as it becomes impractical to generate a unique green list and red list for each text generation. Additionally, model owners need to remember the list corresponding to each generated text sequence to distinguish LLM-generated and human-generated text using the red/green lists, resulting in a significant memory burden. To address the limitations of the existing watermarking algorithms, a topic-based LLM watermarking algorithm 
\cite{nemecek2024topic}, generates a pair of red/green lists based on the text topic to adjust the token weights of the watermarked LLMs. This watermarking algorithm has higher practicality, as it avoids the need to generate a large number of seed lists for each text input, and does not require remembering the corresponding red/green lists, significantly reducing the computational load.

\begin{figure}[b]
    \centering
    \includegraphics[width=1\linewidth]{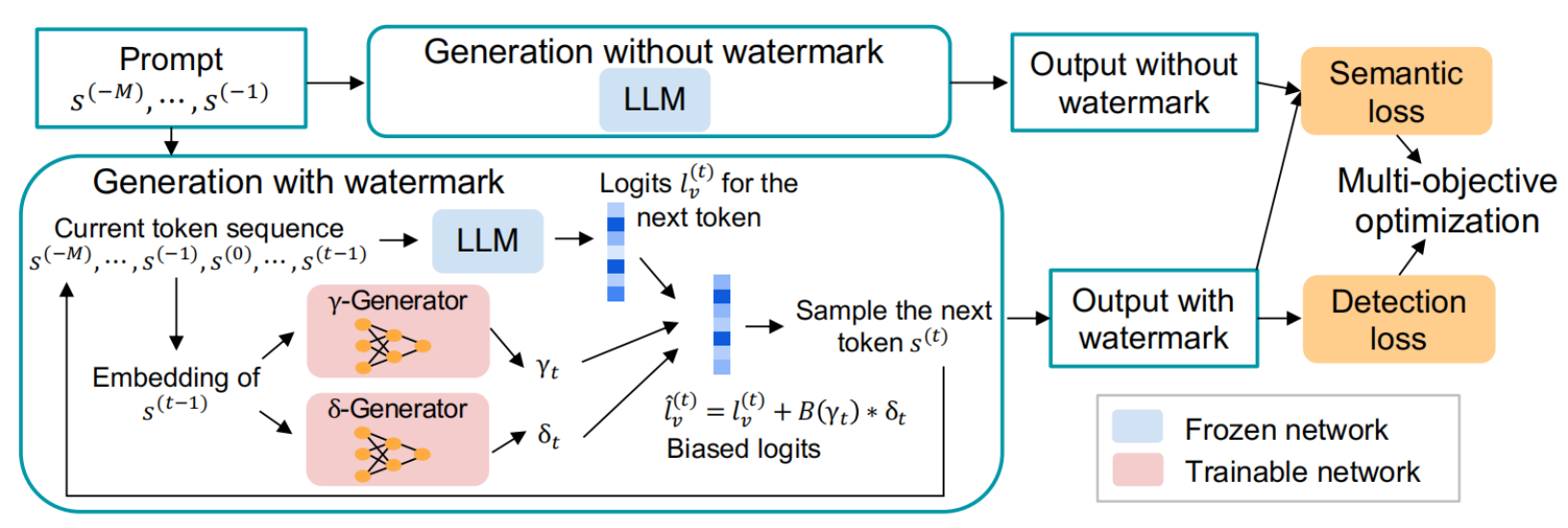}
    \caption{Watermark detectability and semantic integrity are achieved based on multi-objective optimization problem  \cite{huo2024token}.}
    \label{fig:huo2024token}
\end{figure}

iii) Balancing the watermarking metrics. Although soft watermarking has relaxed the constraints on generated text, the quality of the generated text is still not ideal. The higher the offset value, the higher the distortion, and the greater the possibility of reducing the quality of the text. Besides, due to the rather rigid text, the low stealth of the watermark makes it easier for attackers to locate the watermark and perform text attack operations, reducing the robustness. Subsequent research has attempted to find a balance between text quality and watermark recognizability to enhance the robustness, stealth, and improve the text quality of the watermark. NS-Watermark  \cite{takezawa2023necessary} was proposed, which can generate more natural text with the minimum constraints to distinguish whether the text is generated by  LLMs. Wouters et al. \cite{wouters2023optimizing} converted the trade-off between recognizability and text generation quality into a multi-objective optimization problem, aiming to maximize the watermark test quality and minimize the text quality degradation, and determined the relevant Pareto-optimal watermark solutions. Huo et al. \cite{huo2024token} based on the multi-objective optimization problem, found a Pareto-optimal solution to achieve detectability and semantic integrity. This method aims to optimize two lightweight networks simultaneously, responsible for handling the representation of the preceding token, deciding the ideal partition ratio for the subsequent token, and applying the appropriate watermark logit. More details are shown in Figure \ref{fig:huo2024token}.

Some research has attempted to improve upon soft watermarking  \cite{kirchenbauer2023watermark}, making the watermarking algorithm better at maintaining generation quality and ensuring more coherence between the contexts. DiPmark  \cite{wu2023dipmark} introduces a special permutation-based distribution-preserving reweighting strategy. This dual method improves the utilization of the green list by reweighting the log probability of words and ensures that the distributions of the samples from the original language model and the watermarked language model are similar, thus enhancing the imperceptibility of the watermark. A simple and effective semantic-based watermarking algorithm  \cite{fu2024watermarking} considers the connection between watermarked text generation and the input context. To ensure the quality of the text that is generated, it calculates the word vectors of the input context and prioritizes selecting the most semantically relevant words from the green list as the watermark. Neural networks can also be used to preserve the quality of text generation. A semantic-invariant watermarking algorithm \cite{liu2023semantic} uses another embedded LLM to create semantic embeddings for each prior token, and adds a small watermark logit to the logit of the next generated token, which is then transformed into a watermark logic through a trained non-linear neural network.

In addition to the trade-off between the quality and robustness of the watermarking technique, there is also research exploring the trade-off between the robustness and complexity of the watermark. WaterMax  \cite{giboulot2024watermax} first defines a watermark detection algorithm, where the $p$-value is uniformly distributed on [0, 1] if the text does not contain a watermark, and is very small otherwise. WaterMax innovatively adopts a block-wise text generation approach to reduce the computational cost. WaterMax generates multiple text candidate segments and selects the one with the lowest $p$-value to embed the watermark, improving the detection effect. WaterMax does not modify any components of the LLMs, keeping the overall structure unchanged, thereby balancing the robustness and complexity.

iv) Improving the robustness of watermarks: The current text watermarking techniques do not have cross-language consistency, meaning that the text watermark loses its effectiveness after being translated into other languages. X-SIR \cite{he2024can} is a defense strategy against cross-language watermark removal attacks (CWRA). The primary elements influencing the uniformity of text watermarks across different languages are the cross-language semantic clustering of the vocabulary and the vocabulary partition. X-SIR watermark meets the above two factors, ensuring that semantically similar markers or prefixes must be in the same partition, either in the green list or the red list. Additionally, the public watermark algorithm means that the key from the watermark generation stage is needed for the watermark detection stage. Because of this, it is simple for attackers to delete and fake text watermarks using these public keys. Liu et al. \cite{liu2023private} proposed a private watermark technique. It is asymmetric because it uses two distinct neural networks for watermark generation and detection rather than one key for both stages. Using a separate detection neural network to detect the watermark efficiently ensures the algorithm's confidentiality and possesses strong anti-decryption capabilities, in contrast to earlier detection approaches that require the watermark key for detection. The detection network can efficiently obtain a very high accuracy rate since the token embedding parameters are shared by both the generation network and the detection network.

\begin{figure}[b]
    \centering
    \includegraphics[width=1\linewidth]{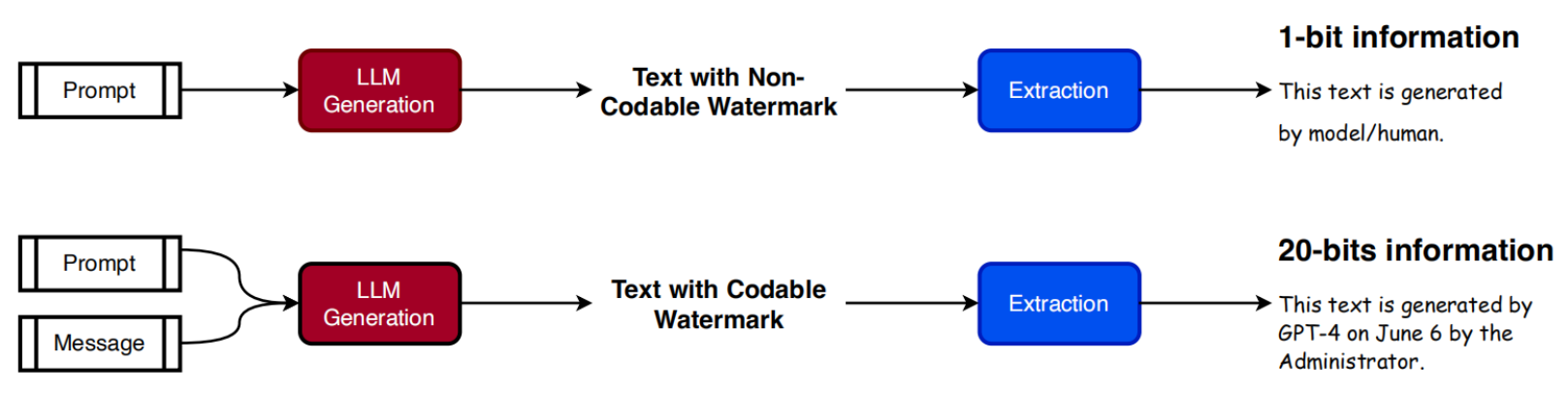}
    \caption{Zero-bit watermarks vs. multi-bit watermarks  \cite{wang2023towards}.}
    \label{fig:bitcomparison}
\end{figure}

v) Zero-bit watermarks \cite{fu2024gumbelsoft,he2024can,kirchenbauer2023watermark,takezawa2023necessary} and multi-bit watermarks \cite{qu2024provably,wang2023towards,yoo2023advancing}: The watermarking methods mentioned above are all zero-bit watermarks, meaning they can only complete the task of distinguishing whether the text is generated by LLMs. However, zero-bit watermarks are unable to adapt to the various information encoding requirements of different LLM application scenarios. Therefore, subsequent research has proposed multi-bit watermarking, which, in addition to distinguishing whether the text is written by LLMs or a human, can also obtain model-related information through the watermark, such as the generating time, user ID, and model version, etc., to track and identify the source of the model. The source of the content can be quickly discovered if it is used maliciously for things like disseminating fake information or plagiarizing in academic writing. The comparison between zero-bit watermarks and multi-bit watermarks is shown in Figure \ref{fig:bitcomparison}. Codable text watermarking for LLMs (CTWL) \cite{wang2023towards} can encode richer information into the watermark compared to previous watermarking methods. CTWL is designed based on ensuring that the probabilities of the usable and unusable vocabulary lists are approximately equal and proposes the lMproxy-AIDED balanced vocabulary partitioning method, which can greatly enhance the quality of the generated text and automatically filter out the low-entropy parts of the text. To further increase the watermark bit capacity, in the "multi-bit watermarking through position allocation" (MPAC) \cite{yoo2023advancing}, the vocabulary can be divided into $r$ color lists instead of just the green list/red list, thereby expanding the number of different color lists. The binary form of the message is encoded by converting it to base $r$. In this way, each token can encode $r$ states, not just the binary 0 or 1 states. The conversion of the message to base $r$ determines which color list the token at the corresponding position is generated from. While MPAC 
 \cite{yoo2023advancing} realizes the embedding of multi-bit watermarks, it cannot accurately extract the embedded bit strings. A watermarking algorithm  \cite{qu2024provably} based on error-correcting codes solves this challenge, choosing the BCH coding scheme \cite{bose1960class} to encode the message, which, although introducing additional bits, provides a certain tolerance for the message due to its error-correcting properties. It proves that under bounded adversarial attacks, this watermarking method can correctly extract the watermark, significantly improving its robustness and detectability.

\textbf{4) Post-processing of generated text}. To enable third-party language model providers to autonomously embed watermarks in their black-box models, some research has proposed a post-processing watermarking technique. This involves inserting invisible markers into the generated text, which can be done through character modifications, word replacements, or sentence reordering. Compared to embedding the watermark during the LLMs text generation process, which requires lengthy training times, the post-processing watermarking technique does not modify the original model, thus bypassing complex operations and becoming more efficient. However, the drawback is that, unlike the earlier approaches, the post-processing watermarking technique cannot leverage the generative capabilities of the LLMs and can only work with individual tokens. As a result, the created sequence cannot be adaptively changed. This limits the freedom of the watermarked text.

\begin{figure}[b]
    \centering
    \includegraphics[width=1\linewidth]{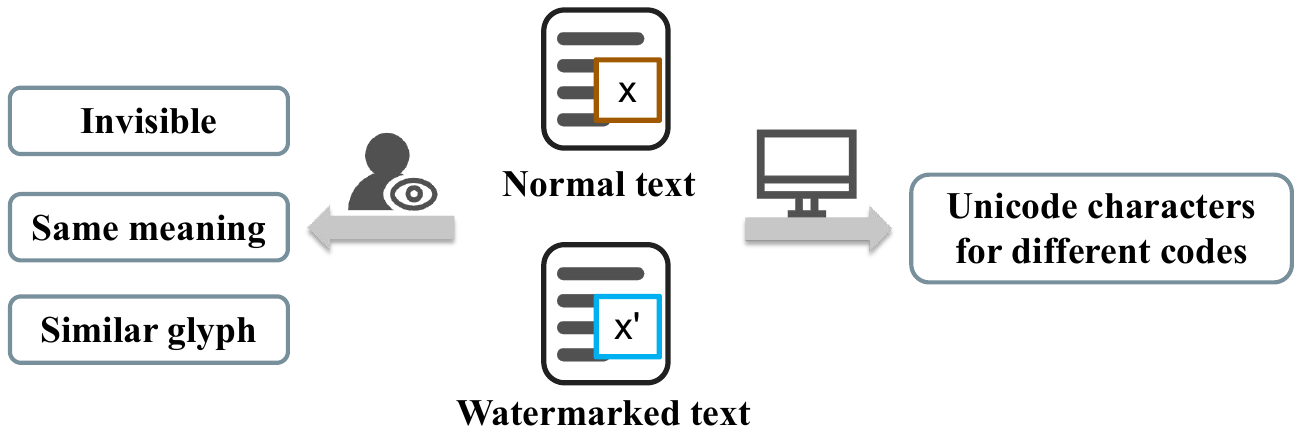}
    \caption{Character encoding-based watermarking.}
    \label{fig:characterencoding}
\end{figure}

i) Character encoding-based watermarking: The aforementioned hard watermarking and soft watermarking techniques may compromise the quality of the generated text, as the LLMs are forced to generate less text to accommodate a sufficient number of green words. The character encoding-based watermarking does not reduce the text's quality, and the watermarked text looks the same as the original text. This technique is also very easy to implement, the amount of code is small and does not require modifying the decoding program like hard/soft watermarking. This method takes advantage of the numerous Unicode characters that share visual similarities or are identical and replaces some symbols with letters and punctuation marks that have different Unicode representations but look the same, thereby embedding invisible watermarks into the text without changing its meaning. For example, this can be done by inserting specific hidden character sequences or embedding markers in the letters of certain words within the generated text. The principle of character encoding-based watermarking is shown in Figure \ref{fig:characterencoding}. EASYMARK  \cite{sato2023embarrassingly} designs three variants: WHITEMARK, VARIANTMARK, and PRINTMARK, each suitable for different scenarios. WHITEMARK is the simplest, applicable to numeric text, where it replaces spaces with different space code points. Through statistical examination of the space code points, the existence of the watermark is identified. VARIANTMARK is suitable for Chinese text, using Unicode variant selectors to replace Chinese characters and embed secret information. PRINTMARK is for printed text, where the watermark is embedded through ligatures between two characters. This watermarking technique does not compromise text quality at all, and it does not require knowledge of the LLM's internal structure, allowing users to implement the watermark injection on the user-side, making it a black-box application.

\begin{figure}[b]
    \centering
    \includegraphics[width=0.8\linewidth]{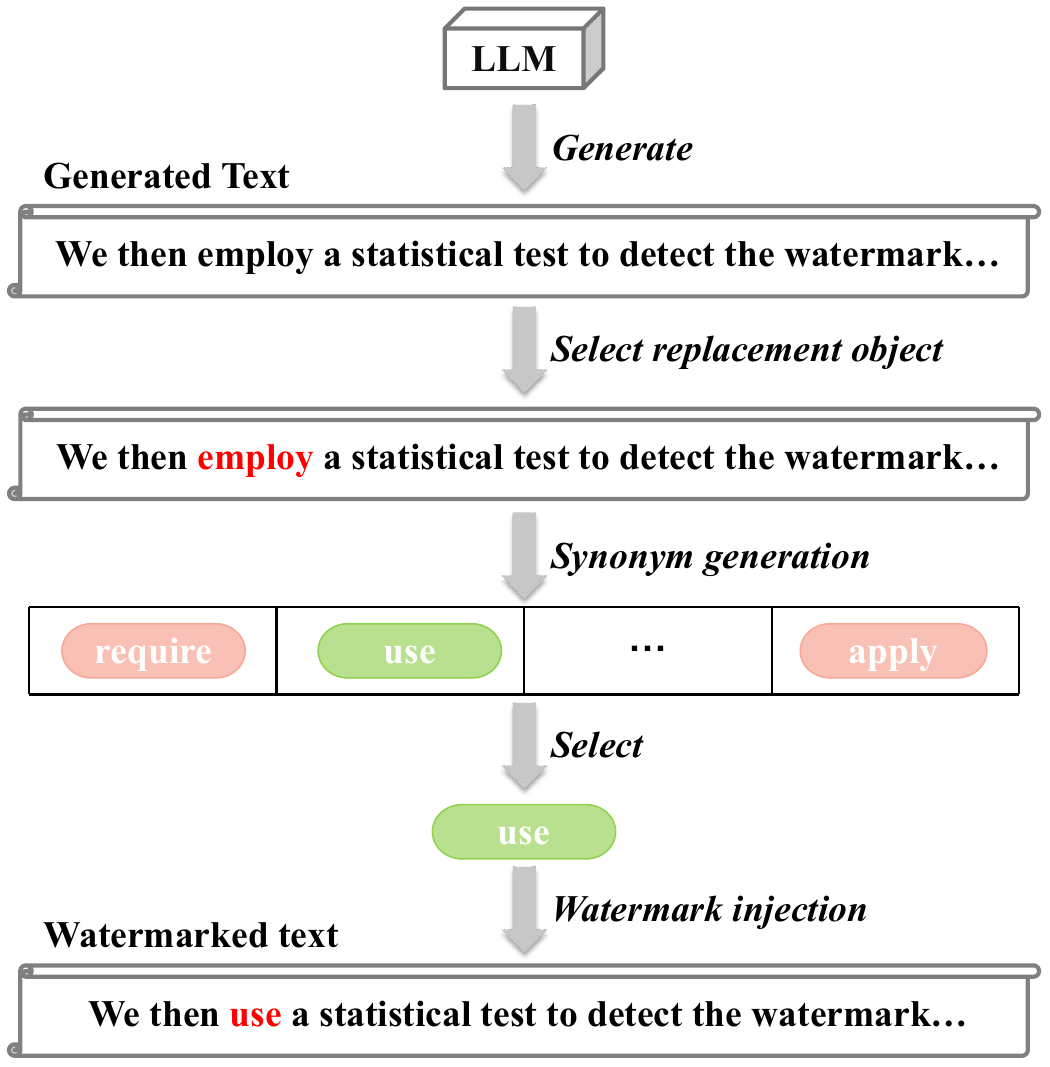}
    \caption{Watermarking based on synonym substitution \cite{yang2023watermarking}.}
    \label{fig:synonymsubstitution}
\end{figure}

ii) Based on synonym substitution. The watermarking method based on character encoding can be easily implemented, but this method is fragile because attackers can easily remove the watermark without modifying the meaning of the text. The watermarking technique based on synonym substitution \cite{munyer2023deeptextmark,yang2023watermarking} usually first generates multiple synonym candidates based on the context, then filters the candidates based on certain criteria, and finally replaces the original words or sentences to embed the watermark. It does not compromise the original semantics, thus achieving the watermarked text's imperceptibility and improving its robustness. The framework of synonym substitution-based watermarking is shown in Figure \ref{fig:synonymsubstitution}. Yang et al. \cite{yang2023watermarking} designed a context-based synonym generation algorithm and a watermark-driven synonym sampling algorithm. First, they define a binary encoding function to calculate the random binary encoding corresponding to a word. In non-watermarked text, the number of words representing bit-0 and bit-1 is almost balanced. However, for watermarked text, the probability of words representing bit-1 is far greater than 0.5. To inject the watermark, for each selected word, they generate its context-based synonym candidates, encode the synonym set using the binary encoding function, and replace the word with the one with the highest word-level similarity score and the encoding result as bit-1. The injected watermark will produce a relatively higher proportion of words representing bit-1. Consequently, statistical analyses can be employed to identify the existence of a watermark. DeepTextMark 
\cite{munyer2023deeptextmark} involves word substitution. It first uses Word2Vec \cite{mikolov2013distributed} to convert each candidate word into an embedding vector and searches for the $n$ semantically closest replacement words. Then, it uses a universal sentence encoder \cite{cer2018universal} to score the quality of each sentence, calculates how similar the suggested sentence embedding is to the original sentence embedding, and keeps the suggested sentence with the highest similarity score as the watermarked sentence. For watermark detection, a transformer classification block can be used to classify the embeddings. But this watermarking method is also vulnerable to simple attack methods such as random synonym substitution.

iii) Based on deep learning. This technique automatically embeds and extracts watermarks using an additional neural network, which typically has two components: a message encoder and a corresponding message decoder. The model is jointly trained to achieve the goal of encoding the message with minimal loss, successfully decoding the message, and simultaneously deceiving the adversary. The framework of deep learning-based watermarking is shown in Figure \ref{fig:deeplearning}. An adversarial watermark transformer (AWT)  \cite{abdelnabi2021adversarial} uses adversarial training and other smooth auxiliary losses to encode fixed-length watermark information in English text in an imperceptible way. The network learns to replace inconspicuous words (e.g., prepositions, conjunctions, and punctuation) to encode information, and the watermark can be extracted even if some words are changed, as long as the inconspicuous words remain intact. However, the downside is that the capacity for embedded signatures is limited. Subsequently, REMARK-LLM  \cite{zhang2023remark} was proposed, which can embed signature sequences up to 2 times longer into the same content without affecting the coherence of the text compared to the AWT. It consists of three components: message encoding, reparameterization, and message decoding. The message encoding uses a sequence-to-sequence (Seq2Seq) model \cite{sutskever2014sequence} to insert the invisible watermark into the text generated by the LLMs. The reparameterization uses Gumbel-Softmax \cite{jang2016categorical} to convert the watermarked distribution generated by the Seq2Seq model into a sparser watermarked text token distribution. The message decoding first uses a mapping network to map the reparameterized distribution to their respective embedding representations and then uses a transformer-based decoder to retrieve the hidden message from these embeddings.

\begin{figure}[t]
    \centering
    \includegraphics[width=0.99\linewidth]{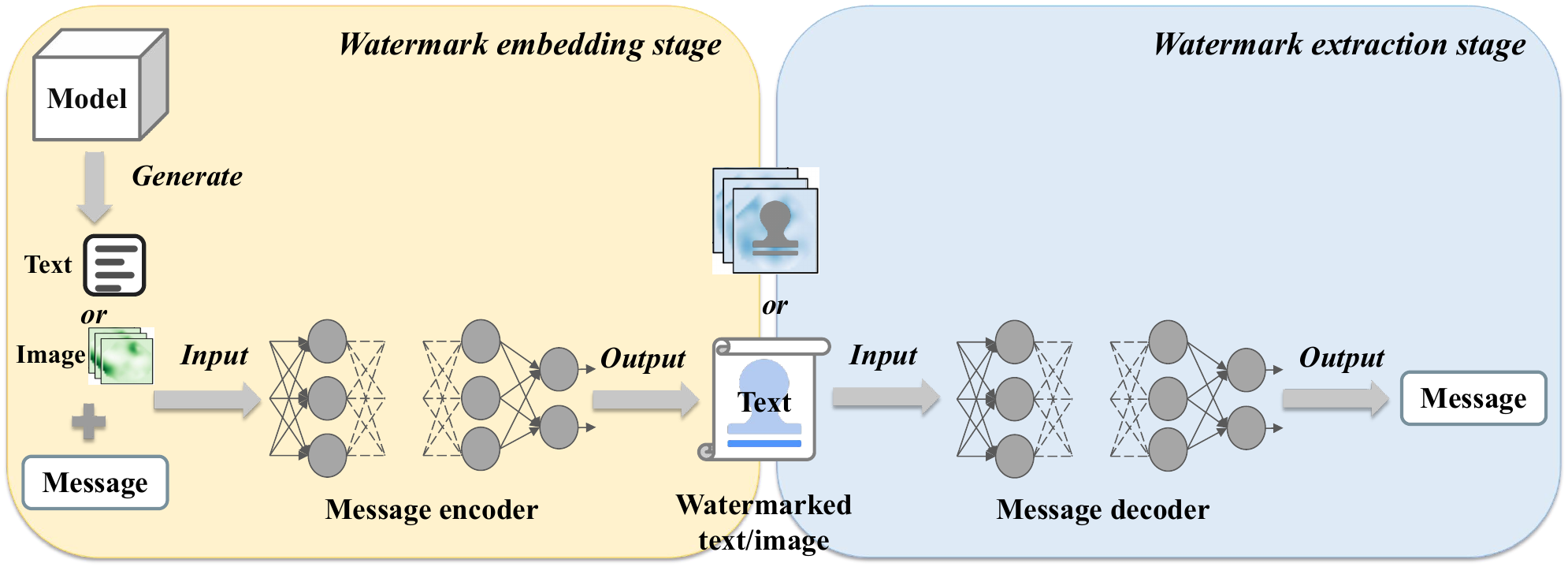}
    \caption{Framework of watermarking based on deep learning.}
    \label{fig:deeplearning}
\end{figure}

\subsubsection{Embedding Watermarks into Models}

Model-based watermarking technology usually considers embedding the watermark into the parameters of the model, various network architectures, or backdoors, which is similar to the traditional neural network watermarking, to achieve the authentication of the model, copyright protection, and other purposes. Research has extensively borrowed from the traditional neural network watermarking models to implement LLM watermarking models and has made improvements based on the characteristics of LLMs. Especially for neural network black-box watermarking, as LLM models are often only callable in the form of an interface, it meets the applicable conditions of black-box watermarking. \textbf{Parameter watermarking} \cite{fernandez2024functional,li2023watermarking,zhang2024emmark}: Integrating watermark data into the model's parameters is a popular technique for watermarking. By adjusting the parameters throughout the training phase, the watermark information can be embedded within the weights, biases, or other model parameters. \textbf{Backdoor watermarking} \cite{liu2023watermarking,peng2023you,shetty2024warden}: Backdoor watermarking, primarily targeting black-box watermarking, pre-defines certain specific trigger conditions in the model so that the backdoor can be triggered unauthorized, and the watermark information can be extracted. Usually, specific input-output patterns are added to the model during the training process or a special trigger set is set to activate the backdoor.

\begin{figure}[b]
    \centering
    \includegraphics[width=0.8\linewidth]{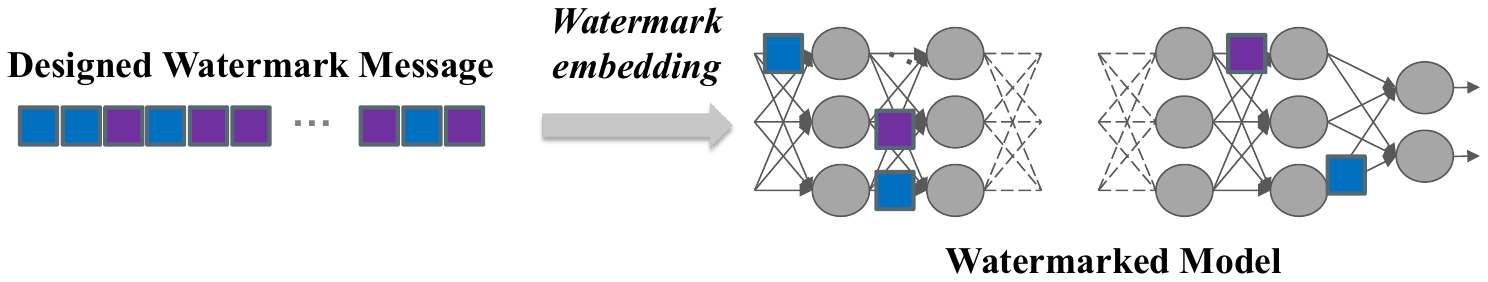}
    \caption{Embedding watermarks into model parameters \cite{zhang2024emmark}.}
    \label{fig:modelparameters}
\end{figure}

\textbf{1) Embedding watermarks into model parameters.} To embed a watermark, a simple approach is to modify the network parameters to embed the watermark. The watermark is embedded into the target model's parameter space using these techniques.  By contrasting the parameter watermark taken from the suspicious model with the watermark of the owner's model, the owner can confirm the identity of the model. The framework of embedding watermarks into model parameters is shown in Figure \ref{fig:modelparameters}. A new watermarking algorithm \cite{li2023watermarking} based on model quantization uses the gap between quantization and full precision to embed the watermark into the model weights. When the model is utilized in FP32 mode, the watermark is effective, and when the model is quantized to int8, it stays concealed. It adopts a rollback strategy and interval optimization strategy. Fernandez et al. \cite{fernandez2024functional} utilized the invariance of Transformers \cite{vaswani2017attention} to implement lightweight watermark addition with almost no computing cost, suitable for non-blind white-box scenarios. Two permutations are chosen for embedding by this method: one for the attention block and one for the FFN block. The above methods do not have a fixed standard for selecting parameters. If the selected parameters are inappropriate, it could impact the model's efficacy or diminish the robustness of the watermark. The EmMark \cite{zhang2024emmark} employs a strategic selection of watermark weight parameters to ensure robustness and preserve model quality. Two aspects were used to evaluate the parameters, and the best performance weight parameters were selected to encode the watermark signature. One is that the weight parameter should be less sensitive to the watermark after insertion to ensure the model's quality. The other is that the weight parameters should play a crucial role in achieving watermark robustness. To remove the watermark, attackers must disturb a large proportion of the significant weights, leading to a decrease in LLMs performance. However, parameter embedding techniques require that the model parameters should be available during watermark detection, which restricts their application to white-box scenarios. A drawback of this method is that the watermark can be easily removed through minimal re-training of the watermarked model, which is often infeasible in real-world attacks.

\begin{figure}[b]
    \centering
    \includegraphics[width=0.9\linewidth]{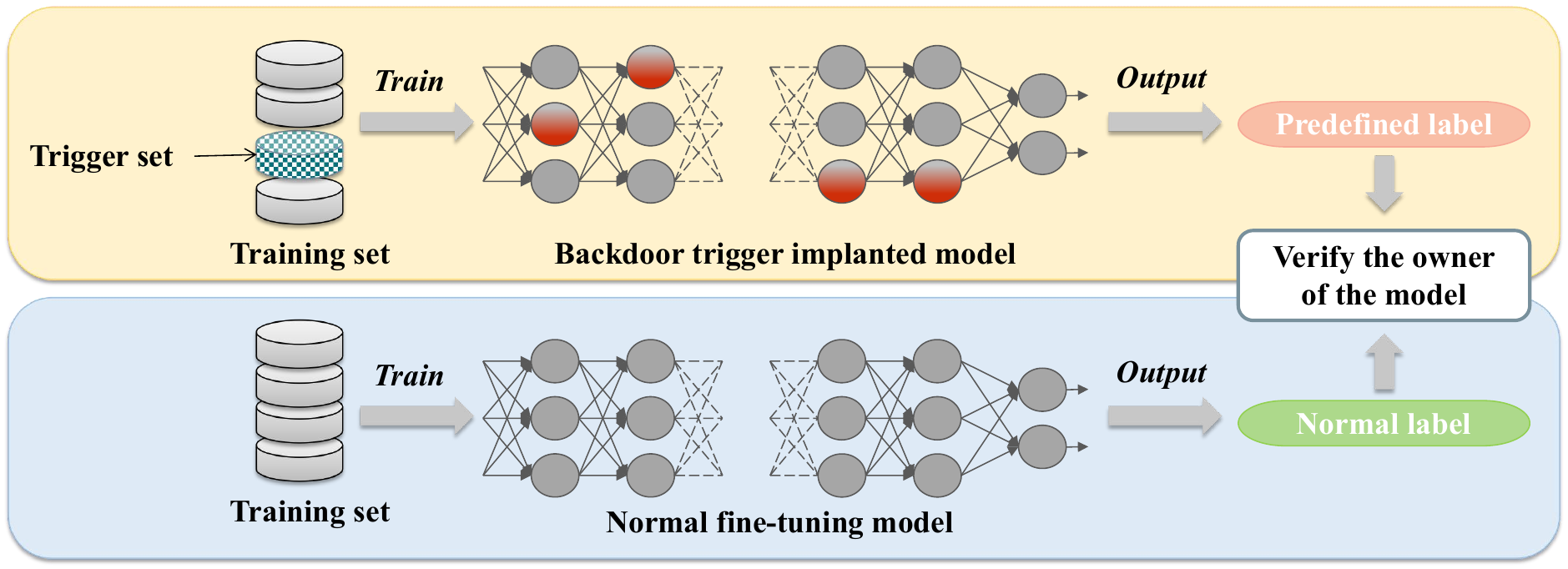}
    \caption{Embedding watermarks into model backdoors \cite{liu2023watermarking,peng2023you}.}
    \label{fig:modelbackdoors}
\end{figure}

\begin{figure}[t]
    \centering
    \includegraphics[width=0.9\linewidth]{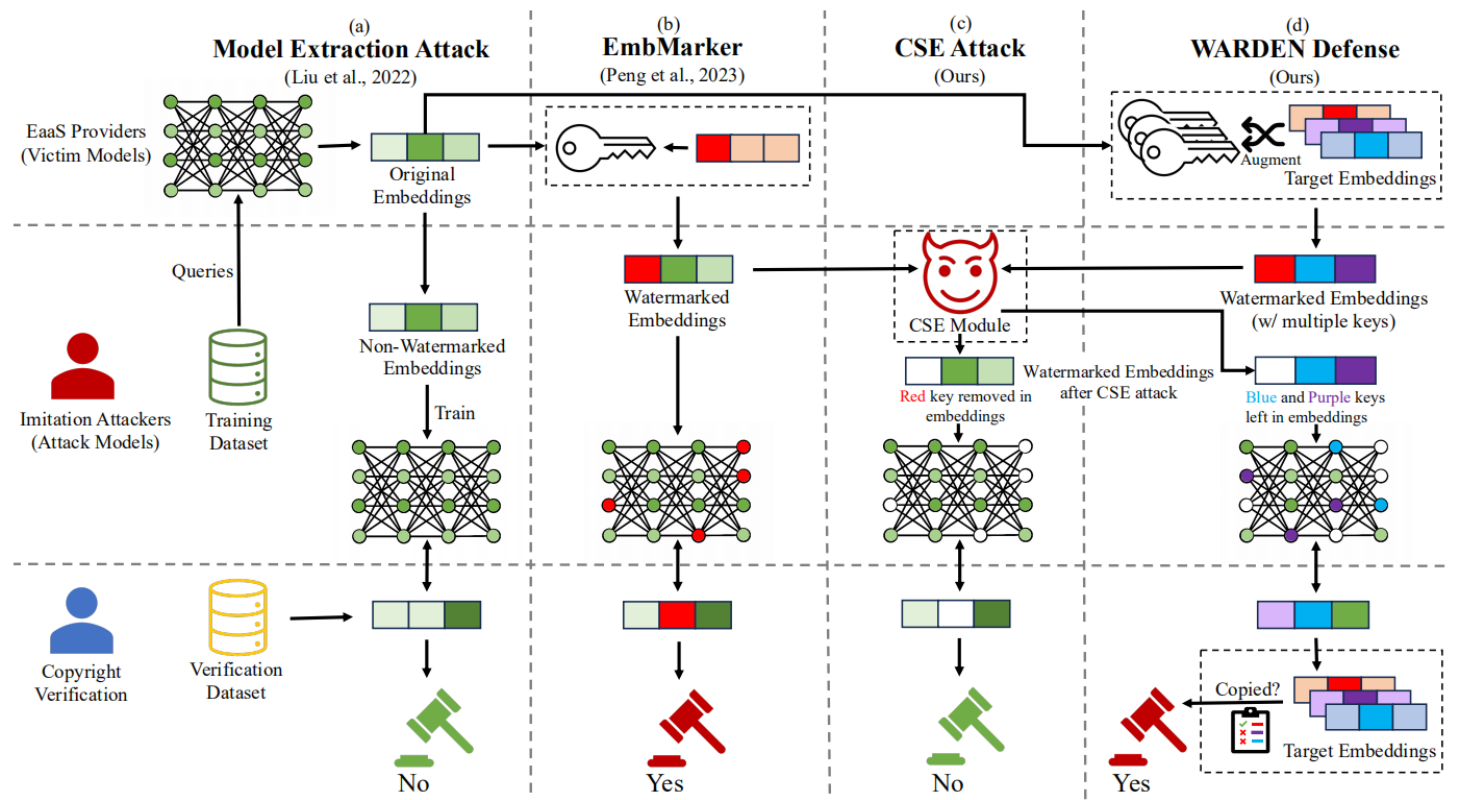}
    \caption{EmbMarker vs. WARDEN  \cite{shetty2024warden}.}
    \label{fig:embmarkvswarden}
\end{figure}

\textbf{2) Embedding watermarks into model backdoors.} In real-world scenarios, model parameters and network architectures in commercial services are usually kept confidential, so the aforementioned approaches of embedding watermarks into model parameters and network layers are often infeasible. Backdoor-based techniques are more suitable for black-box settings and are more widely adopted in the real world. The watermark verification process can remotely extract the watermark trigger set without accessing the model weights. Models possess numerous parameters, a significant portion of which are not essential for the completion of the model's tasks. These extraneous parameters can carry information other than what is required for the model's task. This property is exploited by backdoors, where the model is intentionally made to output erroneous predictions for some selected backdoor trigger sets. They can be triggered by certain specific elements, and in text LLMs, the trigger sets are often some meaningless words in the sentences. It fine-tunes the pre-trained model by using a set of triggers assigned to one or more secret target classes, embedding the backdoor triggers into the model. Traditional neural network black-box watermarking also often adopts trigger-set-based designs. Due to the multimodal nature of LLMs, the trigger sets can be more flexibly designed to make the watermarks more stealthy. Model owners can train a watermarked model with a trigger set and a large training dataset with correct labels, and then provide it to their clients. They can validate their ownership by feeding the trigger set into the suspected model to identify unauthorized or improperly used variants of their models. The framework of embedding watermarks into model backdoors is shown in Figure \ref{fig:modelbackdoors}. EmbMarker \cite{peng2023you} chooses to embed a backdoor on the model. First, it selects a set of medium-frequency words from a general text corpus to form the trigger set. Then, it embeds the target embedding into the text, with the inserted weights proportionate to the text's trigger word count. For copyright verification, the text with the backdoor trigger is used to query a suspicious embedding as a service (EaaS) API, and hypothesis testing is used to calculate the probability that the output embedding is the target embedding. However, recent CSE (clustering, selection, elimination) attacks can successfully remove the EmbMarker watermark by comparing the victim model's embeddings with those of a standard model, identifying the embedding pairs with significant distance changes as suspicious samples, and then neutralizing the watermark effect in the embeddings. To address the CSE attack, a multi-directional watermark enhancement defense algorithm called WARDEN \cite{shetty2024warden} was proposed. While EmbMarker adds a single watermark to the text, WARDEN adds multiple watermark embeddings to improve the defense capability by combining multiple possible watermark directions. WARDEN demonstrates greater robustness to CSE attacks and exhibits superior stealthiness compared to EmbMarker. The comparison between EmbMarker and WARDEN is shown in Figure \ref{fig:embmarkvswarden}. TextMarker \cite{liu2023watermarking} is based on the backdoor membership inference method, which only requires the data owner to label a small number of samples under black-box access for data copyright protection. Each user can design their backdoor text. They can use backdoor techniques to add triggers to their private text, creating marked text, and then using their triggers to verify if their private text has been used without authorization to train NLP models. Although backdoor-based watermarking techniques are more suitable for black-box scenarios, the watermark capacity is generally lower than embedding watermarks in model parameters or network architectures. Most backdoor-based watermarking techniques are zero-bit watermarks, meaning the owner can only verify whether the model contains their watermark, and cannot embed more useful information into the model.

\subsubsection{Watermarking Techniques Based on Cryptography}

In traditional cryptography-based digital watermarking techniques \cite{li2023plmmark,liu2023watermarking}, research often uses cryptographic algorithms to embed watermarks to achieve purposes such as copyright protection and digital product content verification. The key features of these cryptography-based digital watermarking techniques are the stealth and undetectability of the watermark, meaning that an attacker lacking the key cannot distinguish the watermark from background noise. The watermark can only be successfully detected when the exact key is known. Similarly, LLM watermarking techniques based on cryptography can also apply cryptographic algorithms to incorporate undetectable watermarks:  watermarks that are only detectable with knowledge of the key. It is computationally challenging to identify the watermarked model output in the absence of the key. This property enhances the stealth of the watermark and prevents unauthorized parties from discovering and removing the watermark. Based on the idea of one-way functions, Christ et al. \cite{christ2024undetectable} constructed an undetectable watermark that makes the distribution of the generated text identical to the original text. They achieves zero-bit lossless watermarking. It defines empirical entropy to measure the amount of entropy involved in the process of generating a specific consecutive substring. It samples from the model until enough empirical entropy is collected. Once enough empirical entropy is collected in the text block $r$, the watermark is embedded using $r$ as a seed and then embedded in the next subsequence of the token, guaranteeing that the watermark will be detected with a high probability. Based on this, the multi-bit lossless DISC  \cite{boroujeny2024multi} was proposed, which opens the way for many applications in authentication and communication security. This watermarking employs a distortion-free mapping rule, associating the binary token values with PRF's values to achieve watermark embedding. The comparison of mapping rules for Christ et al. \cite{christ2024undetectable} and DISC is shown in Figure \ref{fig:mappingrules}.

\begin{figure}[t]
    \centering
    \includegraphics[width=1\linewidth]{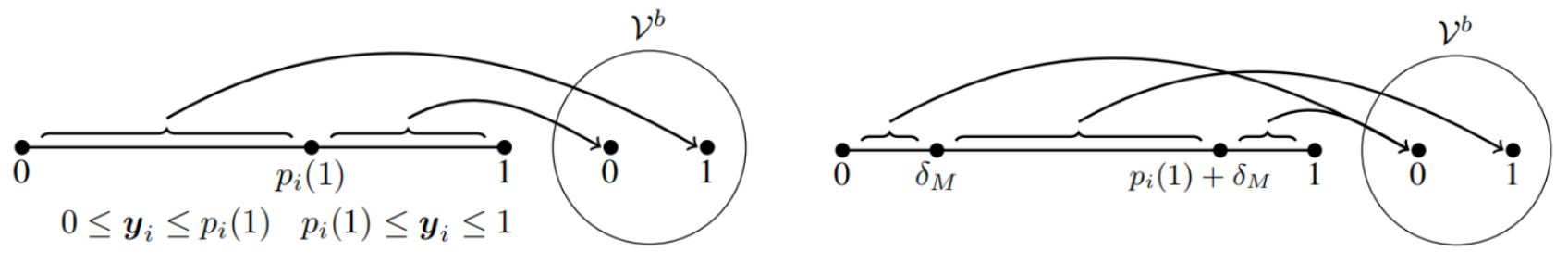}
    \caption{Watermark mapping rules:
    Christ et al.  \cite{christ2024undetectable} vs. DISC \cite{boroujeny2024multi}.}
    \label{fig:mappingrules}
\end{figure}

\begin{table*}
\scriptsize
\centering
\caption{Classification of text domain LLM watermarking techniques and their advantages and disadvantages.}
\label{table:advdisadv}
\begin{tabular}{|l|l|l|l|l|} 
\hline
\multicolumn{3}{|c|}{\begin{tabular}[c]{@{}c@{}}\textbf{Classification of watermarking}\\\textbf{~techniques}\end{tabular}}                                                                                                                     & \multicolumn{1}{c|}{\textbf{Advantages}}                                                                                                                                          & \multicolumn{1}{c|}{\textbf{Disadvantages}}                                                                                                                                                                                                                                                                                                                                     \\ 
\hline
\multirow{6}{*}{\begin{tabular}[c]{@{}l@{}}Embedding \\into text\end{tabular}}   & \multicolumn{2}{l|}{Pre-processing stage}                                                                                                                    & \begin{tabular}{@{\labelitemi\hspace{\dimexpr\labelsep+0.5\tabcolsep}}l@{}}Easy to implement.\\Low computational complexity.\end{tabular}                                         & \begin{tabular}[c]{@{}l@{}}\begin{tabular}{@{\labelitemi\hspace{\dimexpr\labelsep+0.5\tabcolsep}}l@{}}The cost is high for large-scale data sets.\end{tabular}\\\begin{tabular}{@{\labelitemi\hspace{\dimexpr\labelsep+0.5\tabcolsep}}l@{}}The watermark information is fixed and~cannot\end{tabular}\\adapt to different text contents.\end{tabular}  \\ 
\cline{2-5}
                                                                                 & \multirow{2}{*}{\begin{tabular}[c]{@{}l@{}}Modifying the\\generation\\process\end{tabular}} & \begin{tabular}[c]{@{}l@{}}Sentence \\level\end{tabular}      & \begin{tabular}{@{\labelitemi\hspace{\dimexpr\labelsep+0.5\tabcolsep}}l@{}}The generated text is high quality.\\Strong robustness.\end{tabular}               & \begin{tabular}{@{\labelitemi\hspace{\dimexpr\labelsep+0.5\tabcolsep}}l@{}}Generating text is slower.\end{tabular}                                                                                                                                                                                                                                                      \\ 
\cline{3-5}
                                                                                 &                                                                                              & \begin{tabular}[c]{@{}l@{}}Token \\level\end{tabular}         & \begin{tabular}{@{\labelitemi\hspace{\dimexpr\labelsep+0.5\tabcolsep}}l@{}}It is faster to generate text.\end{tabular}                                                                        & \begin{tabular}{@{\labelitemi\hspace{\dimexpr\labelsep+0.5\tabcolsep}}l@{}}The robustness is relatively weak.\\Reducing text quality.\end{tabular}                                                                                                                                                                                                                                \\ 
\cline{2-5}
                                                                                 & \multirow{3}{*}{\begin{tabular}[c]{@{}l@{}}Post-\\processing\\stage\end{tabular}}            & \begin{tabular}[c]{@{}l@{}}Character\\encoding\end{tabular}   & \begin{tabular}{@{\labelitemi\hspace{\dimexpr\labelsep+0.5\tabcolsep}}l@{}}Text quality is not affected.\\Easy to implement.\\Suitable for black-box scenarios.\end{tabular}      & \begin{tabular}{@{\labelitemi\hspace{\dimexpr\labelsep+0.5\tabcolsep}}l@{}}The watermark is fragile and~can be easily deleted.\end{tabular}                                                                                                                                                                                                                                     \\ 
\cline{3-5}
                                                                                 &                                                                                              & \begin{tabular}[c]{@{}l@{}}Synonym\\substitution\end{tabular} & \begin{tabular}{@{\labelitemi\hspace{\dimexpr\labelsep+0.5\tabcolsep}}l@{}}Text quality is not affected.\\The robustness is high.\\Suitable for black-box scenarios.\end{tabular} & \begin{tabular}{@{\labelitemi\hspace{\dimexpr\labelsep+0.5\tabcolsep}}l@{}}Vulnerable to random synonym substitution attack.\end{tabular}                                                                                                                                                                                                                                       \\ 
\cline{3-5}
                                                                                 &                                                                                              & \begin{tabular}[c]{@{}l@{}}Deep\\learning\end{tabular}        & \begin{tabular}{@{\labelitemi\hspace{\dimexpr\labelsep+0.5\tabcolsep}}l@{}}Strong robustness.\\Suitable for black-box scenarios.\end{tabular}                                     & \begin{tabular}[c]{@{}l@{}}\begin{tabular}{@{\labelitemi\hspace{\dimexpr\labelsep+0.5\tabcolsep}}l@{}}The computational complexity is high, which~affects\end{tabular}\\the model performance to some extent.\end{tabular}                                                                                                                                                      \\ 
\hline
\multirow{2}{*}{\begin{tabular}[c]{@{}l@{}}Embedding \\into models\end{tabular}} & \multicolumn{2}{l|}{Parameters}                                                                                                                              & \begin{tabular}{@{\labelitemi\hspace{\dimexpr\labelsep+0.5\tabcolsep}}l@{}}Large watermark capacity.\\The robustness is strong.\end{tabular}                                      & \begin{tabular}{@{\labelitemi\hspace{\dimexpr\labelsep+0.5\tabcolsep}}l@{}}Only for white-box scenarios.\end{tabular}                                                                                                                                                                                                                                                           \\ 
\cline{2-5}
                                                                                 & \multicolumn{2}{l|}{Backdoors}                                                                                                                               & \begin{tabular}{@{\labelitemi\hspace{\dimexpr\labelsep+0.5\tabcolsep}}l@{}}Suitable for black-box scenarios.\end{tabular}                                                         & \begin{tabular}[c]{@{}l@{}}\begin{tabular}{@{\labelitemi\hspace{\dimexpr\labelsep+0.5\tabcolsep}}l@{}}The watermark has a limited capacity.\\The robustness is affected by~the choice of trigger set.~~\end{tabular}.\end{tabular}                                                                                                                                          \\
\hline
\end{tabular}
\end{table*}

\subsection{ Watermarking in Image Domain for LLMs}

In recent years, LLMs have demonstrated remarkable capabilities in image learning and generation, significantly enhancing the flexibility and diversity of AI art. While LLMs were initially designed for text generation, their powerful generative and learning abilities have enabled them to expand into other domains. LLMs can be used for image description generation. The model can produce natural language descriptions of an image's content when it receives an image as input. For tasks like image annotation, image search, and visual assistance, this is incredibly helpful. Furthermore, LLMs can be used for image generation, in which the model receives a textual description as input and outputs an image that corresponds to the description. In addition to generation tasks, LLMs can also be used for image understanding and analysis. By inputting an image into the model, the model can extract features from the image, which can be utilized for tasks such as image classification and object detection. It costs a lot of money to train an image processing model as it requires substantial data, hiring machine learning specialists, and providing high computational resources. Protecting image processing models from infringement is a crucial task. Besides, improper use can lead to issues of deep forgery and image infringement. Deep generative models can synthesize forged images, such as generating fake facial images, making it difficult for the naked eye to distinguish whether the image is real or false, eroding people's trust in digital media. Unlike traditional techniques, watermarking in the image domain for LLMs aims to leverage deep learning techniques to embed watermarks that are adaptive to the model, so that the generated images or the model itself carry tracing information. Three categories of image domain watermarking approaches are image-based, model-based, and output distribution-based watermarking. By embedding identifiable watermark codes into the models used for image processing tasks and then extracting them, it is feasible to differentiate between real images and those generated by a model, thereby safeguarding the IP and commercial rights of the model.

\subsubsection{Embedding Watermarks in Images}
Similar to text domain watermarking techniques, although these watermarking techniques all involve adding watermarks to the images, they can be categorized into three types based on the timing of watermark embedding: pre-processing stage, image generation stage, and post-processing stage. In the pre-processing stage, the image dataset is first modified to embed the watermark information, and then the watermarked training data is used to train the model so that the watermark is transferred from the input to the output, and the watermark can be detected in the generated image. In the image generation stage, by redefining the loss function of the generator, the watermark is embedded into the output image while generating the image. In the post-processing stage, the generated image is processed directly, the encoder inserts the watermark into the image, and the decoder extracts the watermark information from the watermarked image.

\textbf{1) Pre-processing before image generation} \cite{cui2023diffusionshield,ditria2023hey,yu2021artificial}. The watermarking technique in the pre-processing stage refers to embedding pre-defined watermark information into the training data to achieve watermark embedding. It encodes the information into the parameters of the generator so that all generated images are associated with the watermark. The advantage of this method is that the pre-embedded information can be effectively retained during the generation process, and the quality of the generated image is less affected. The limitations are the same as the pre-processing watermarking algorithm before text generation, that is, the cost is high for large-scale data sets and the watermark is single and fixed. Specifically, it can be divided into the following four stages (as shown in Figure \ref{fig:preprocessing}):

\begin{itemize}
    \item Joint training of watermark encoder and watermark decoder: The encoder's goal is to embed the watermark information into the image by optimizing the loss function. It minimizes the reconstruction error between the watermarked image and the non-watermarked image. The decoder's goal is to recover the pre-defined watermark information from the watermarked image.
    
    \item Model training: Use the watermark encoder to embed the pre-defined watermark information into the training set, obtaining a training set with watermarks. Train the image generation model using the watermarked training set, so that the watermark is transferred from the input to the output, allowing the generated images to be watermarked.

    \item Model usage: Input prompts or images, use the trained model to generate fake images, and the fake images will have watermarks.

    \item Ownership verification: To confirm ownership, use the watermark decoder to decode the watermarked data from the watermarked fake images.
\end{itemize}

\begin{figure}
    \centering
    \includegraphics[width=0.9\linewidth]{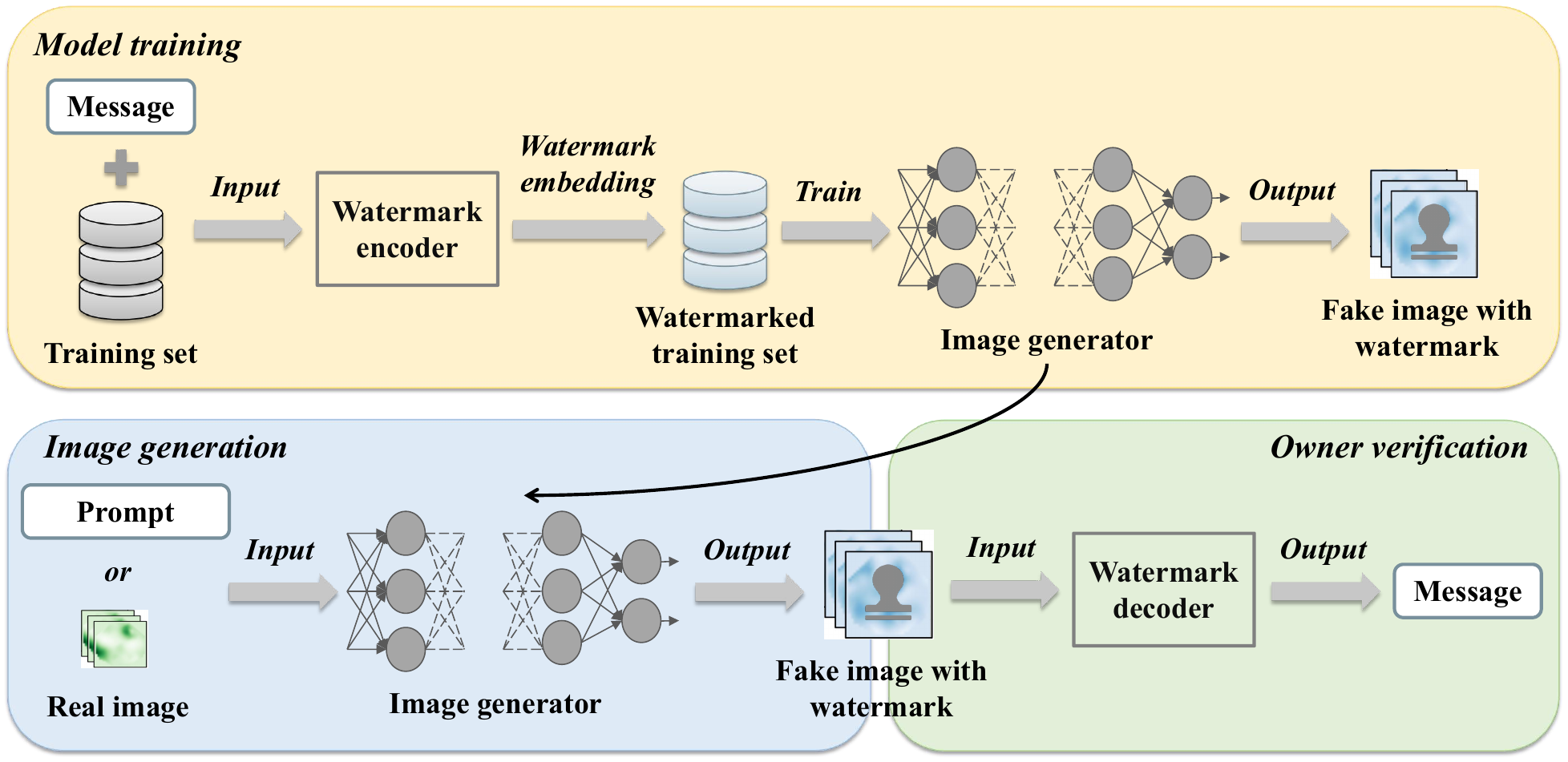}
    \caption{Pre-processing before image generation \cite{fernandez2023stable,ma2023generative,yu2021artificial}.}
    \label{fig:preprocessing}
\end{figure}

Yu et al. \cite{yu2021artificial} embedded watermarks by modifying the training data of the generator. First, they embedded a binary string into the training images using a pre-trained fingerprint encoder, adding fingerprint information to the image dataset. They then used the already watermarked data to train the GAN. In this way, an invisible watermark is automatically embedded in the images generated by the GAN, and it almost does not affect the quality of the generated images. In this technique, each generation model corresponds to a unique fingerprint that will not become invalid with the iterative update of the generation technology. Ditria et al. \cite{ditria2023hey} provide a way to protect content when it is shared with the public. This method adds a fixed watermark to the image, and the watermark is related to the specific characteristics of the image, so that the trained diffusion model can generate the watermarked image. Through statistical tests, it can determine whether the model has been trained with a specific dataset, thus protecting IP rights. Topically driven text-to-image models can personalize image synthesis for specific topics (e.g., faces or artistic styles). To protect the topically driven image synthesis model, GenWatermark \cite{ma2023generative} consists of two components, an encoder and a decoder, which are pre-trained based on large-scale data. The proposed scheme uses subject-specific synthetic images to fine-tune the decoder, which greatly improves the accuracy of watermark detection and ensures the uniqueness of the watermark for each subject. To facilitate user tracking of whether their image data has been infringed, a good watermarking scheme is to give each user a unique watermark, rather than a single image generation model only generating one type of watermark. DiffusionShield  \cite{cui2023diffusionshield} strengthens the "pattern uniformity" of the watermark so that the watermark in each image from the same owner is the same. To achieve pattern uniformity, DiffusionShield designed a block strategy, dividing the watermark into a series of basic patches. Each user has a specific basic patch sequence, forming a unique watermark for themselves, applying it to all images of that user, and encoding copyright information. When detecting the watermark, the generated suspicious image is segmented into a sequence and then classified in a patch-by-patch manner. If the patch is the sequence message we previously embedded in the watermark, we can accurately identify the data owner, thereby protecting the image from copyright infringement. Since there are many types of generation models, there are also many corresponding watermark encoder-decoders. The deep watermark HiDDeN  \cite{fernandez2023stable} especially assigns corresponding watermark decoders to users with different LDM versions \cite{rombach2022high} to track users. To incorporate k-bit information into the image, it optimizes the parameters of the extraction network and watermark encoder. 

\begin{figure}[b]
    \centering
    \includegraphics[width=0.9\linewidth]{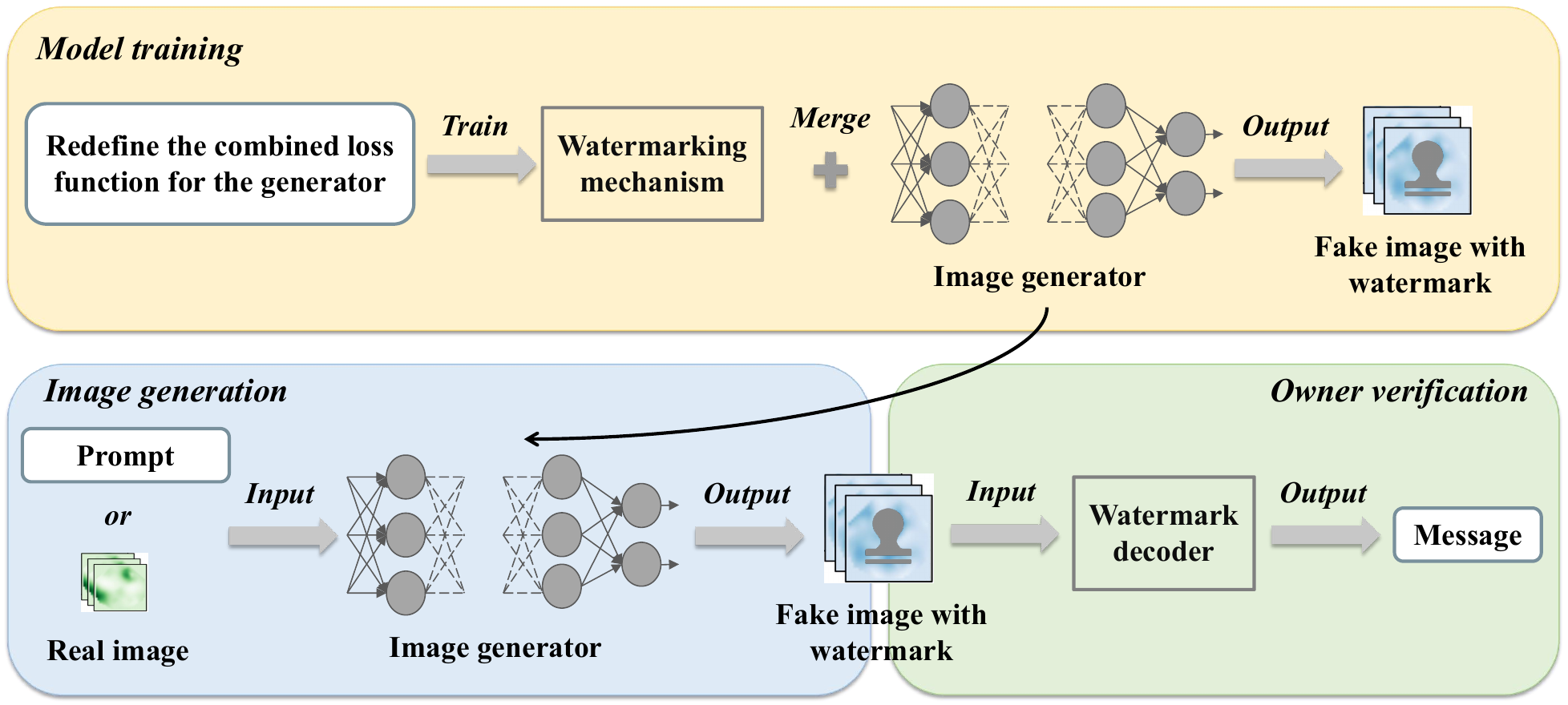}
    \caption{Modifying the model generation process \cite{liu2023t2iw}.}
    \label{fig:modelgeneration}
\end{figure}

\textbf{2) Modifying the model generation process} \cite{liu2023t2iw,lukas2023ptw,wu2020watermarking}. Another method is to incorporate the watermarking mechanism into the generation process, so that the generated image is watermarked. This watermarking method redefines the training objective of the generation model, without the need for additional image processing, making the watermark more lightweight, simple, and secure. This approach is divided into three stages, where the two steps of model usage and ownership verification are completely consistent with the watermarking technique in the pre-processing stage mentioned earlier (as shown in Figure \ref{fig:modelgeneration}).

\begin{itemize}
    \item Cooperative training of the watermark decoder and the main model: During the main model training process, by redefining the composite loss function of the generator, the watermark is embedded into the output image while generating the image. The decoder's goal is to recover the pre-defined watermark information from the watermarked image.

    \item Model usage: Input prompts or images, use the trained model to generate fake images, and the fake images will have watermarks.

    \item Ownership verification: To confirm ownership, use the watermark decoder to decode the watermarked data from the watermarked fake images.
\end{itemize}

Wu et al. \cite{wu2020watermarking} proposed training the watermark extraction network and the host model simultaneously. During the training phase, by optimizing the composite loss function, the parameters of the host network will be updated based on both its loss and the watermark loss. The parameters of the watermark extraction network will only be updated based on the watermark loss. The trained model will embed the watermark into the output images while completing the original task. Perceptual tuning watermarking (PTW)  \cite{lukas2023ptw} can perform watermark embedding on any pre-trained generator model as a post-processing step, without the need to train from scratch, which is 3 orders of magnitude faster than training the watermark generator from scratch. T2IW \cite{liu2023t2iw} uses the principle of Shannon information theory \cite{russakoff2004image} and non-cooperative game theory to reduce the damage to the quality of the generated image and enhance the robustness of the watermark. Based on Shannon's information theory, this scheme minimizes the mutual information between the composite image and the restored image and maximizes the mutual information between the composite image and the watermark to separate the image and the watermark information.  Based on the principle of non-cooperative game theory, the decoupling process of image and watermark was regarded as a non-cooperative game, in which the image restoration and the decoder acted as two players to maximize their respective revenues through their optimal strategies, achieving effective separation of the image and watermark.

\textbf{3) Post-processing of image generation} \cite{fei2022supervised,feng2023catch,zhang2020model}. Similar to text-domain watermarking techniques, deep learning can be used to design watermarking techniques. It employs a jointly trained encoder-decoder architecture, which is sometimes combined with adversarial training. The advantage is that the encoder does not embed the watermark during the image generation process, but instead integrates a watermark encoder after the image generation, which ensures that it does not interfere with the generation process and has little impact on the quality of the generated image.  This technique directly embeds the watermark into the output image before presenting it to the end-user, achieving high efficiency and robustness. However, the information may be modified between the two processes of image generation and watermark embedding, and the accuracy of generated image tracking may be biased. Zhang et al. \cite{zhang2020model} proposed the first deep invisible watermarking technique for image processing model protection, based on a spatially invisible watermarking mechanism. The system consists of two main components: the invisible watermark is hidden in the image by the watermark encoder using a UNet-based \cite{ronneberger2015u} embedding sub-network; the invisible watermark is recovered by the watermark decoder using a CEILNet-based \cite{fan2017generic} extraction sub-network. The watermark encoder and decoder are optimized through adversarial objectives to maximize transmission and robustness. The first watermarking technique to protect concept sharing \cite{feng2023catch} improves the training framework by introducing a novel sampler-in-the-loop training framework with truncated diffusion sampling, which uses samplers to jointly train the watermark encoder and the watermark decoder to improve robustness. A supervised-based GAN watermarking algorithm \cite{fei2022supervised} uses a pre-trained watermark decoder to guide the training and adds an appropriate watermark loss term to the optimization. Due to the existence of the supervision mechanism, the accuracy is improved, the computational burden is reduced, and the already trained GAN is protected to the greatest extent.

\subsubsection{Embedding Watermarks into Models}

Similar to text-domain watermarking techniques, model-based watermarking techniques often consider embedding watermarks into model parameters, various network architectures, or backdoors.

\textbf{1) Embedding watermarks into model parameters} \cite{ong2021protecting,tartaglione2021delving,uchida2017embedding}. Similar to text-domain watermarking techniques, this watermarking method directly embeds the watermark into the model parameters, leveraging the model's redundancy and adaptive capability to lock its parameter set to carry the watermark sequence. Uchida et al. \cite{uchida2017embedding} proposed the first generic framework for embedding watermarks into model parameters. The watermark is embedded in the weight parameters through the parameter regularizer during training, and the owner needs to access the model parameters to extract the watermark during validation. In the white-box scenario, a watermarking algorithm \cite{ong2021protecting} embeds the watermark information by modifying the symbol loss in the network layer. This sign loss forces the normalization layer to take a positive or negative scale factor so that the symbol set can be converted into a binary sequence carrying meaningful information. Unlike previous work, Tartaglione et al. \cite{tartaglione2021delving} proposed an algorithm based on the principle of random diffusion of parameters, which allows the use of any random subset of model parameters as the watermark signal, increasing watermark capacity and enhancing model resilience against attacks.

\textbf{2) Embedding watermarks into model network layers} \cite{fan2019rethinking}. Some watermarking techniques choose to embed watermarks into the model structure rather than the parameters, by adding new network layers to the model to embed the watermark, making it more robust against parameter modification attacks and able to withstand blurring attacks. A passport-based DNN watermarking algorithm \cite{fan2019rethinking} appends a passport layer after the convolutional layer of the model to enable ownership verification. By strengthening the dependence between the scale factors, bias terms, and network weights, the model's performance is significantly degraded if the passport is modified or forged. The attacker must have a very low probability of correctly estimating the passport layer weights because the watermark confidentiality depends on them. By inputting $N$ images into an $L$-layer DNN model, a total of $N \times L$ possible passport combinations can be generated, reducing the probability of correctly guessing the secret passport.

\textbf{3) Embedding watermarks into the model's hidden state} \cite{lim2022protect}. While most existing watermarking techniques focus on image classification models or image diffusion models, Lim et al. \cite{lim2022protect} proposed two different RNN hidden memory state embedding schemes to protect the ownership of image captioning models. The two different embedding operations refer to the element multiplication and addition models. The owner-provided string will be converted to a binary vector to generate the key $k$. Unlike previous research, the signature is embedded in the signature in the concealed state's symbols rather than the model weights through symbolic loss regularization. The best choice for the image captioning task is to embed the key in the RNN's hidden memory state. A forged key will instantly produce an unusable model with poor output quality, and the infringer cannot use the model normally. 

\textbf{4) Embedding watermarks into the model gradient} \cite{aramoon2021don}. Traditional watermarking methods embed watermarks into the model parameters, while this watermarking method embeds the watermark information by applying statistical bias to the model input gradient. Compared to the traditional method of embedding watermarks into model parameters, it has a smaller performance impact, a larger watermark capacity, and stronger robustness. The DNN-based GradSigns \cite{aramoon2021don} is the first to utilize gradients to embed watermarks. It embeds the owner's signature into the gradient of the cross-entropy cost function. With minimal effect on the protected model's performance, GradSigns enables the model provider to remotely verify the watermark using the prediction API and embed a large amount of information into the DNN. In GradSigns, an $N$-bit vector is first generated as the watermark, and then a group of neurons' input gradients are randomly selected to carry the watermark. A target class $T$ is randomly selected, and the gradients of the carrier nodes are computed using images from class $T$. Finally, the binary cross-entropy loss function in the embedding regularizer is used to embed each watermark bit.

\textbf{5) Embedding watermarks into model backdoors}. Similar to text LLM watermarking techniques, backdoor-based image LLM watermarking is more suitable for black-box scenarios, does not require access to the model weights, and can extract watermarks from remotely operated LLMs. In the field of image LLMs, backdoors are often triggered by some pixels in the image. 

\textbf{1) Backdoor watermarking techniques for image classification models} \cite{le2020adversarial,liu2023watermarking2,zhang2018protecting}. The standard backdoor-based watermarking technique includes two stages. In the watermark embedding stage, the model owner inserts the backdoor into the target model, ensuring that the classifier maintains normal functionality when processing normal samples, but outputs erroneous labels when processing the trigger set. In the second stage, the model owner tries to verify a suspicious model for the presence of their watermark using validation samples. By observing the classifier's response to the query samples, they can identify the ownership of the model. There are various methods to implement backdoor watermarking. The NAIVEWM  \cite{liu2023watermarking2} simply injects the watermark into the LDM, which is verified by ordinary predefined prompts. A zero-bit watermarking algorithm \cite{le2020adversarial} implements backdoor watermarking using adversarial model instances. It introduces alteration to the decision boundary of the base model by appending a small perturbation to the input sample, leading the model to produce an incorrect prediction for the modified sample. The selection of the watermark trigger set is also diverse. Zhang et al.  \cite{zhang2018protecting} proposed using specially crafted samples as the watermark trigger set, including content-based (overlaying the image with meaningful content), noise-based (overlaying the image with Gaussian noise), and unrelated image-based (selecting unrelated images from another dataset) triggers.

\textbf{2) Backdoor watermarking techniques for image generation models} \cite{liu2023watermarking2,nadimpalli2023proactive,ong2021protecting,yin2022neural}. For image generation models, the backdoor watermarking techniques are somewhat different. Unlike classification models that output image prediction labels, generation models often take text or images as input and generate new images. Therefore, compared to the aforementioned classifier-based backdoor watermarking techniques, the backdoor watermarking technique for generative models involves injecting the backdoor through the input text words or image pixels that serve as the trigger set. This approach ensures that the generator maintains normal functionality when processing normal samples, but embeds a watermark in the generated new images when processing the trigger set. Embedding the watermark in a local region of the generated image instead of the entire image area can make the watermark more stealthy. In the black-box scenario, the watermarking algorithms \cite{nadimpalli2023proactive,ong2021protecting} embed the unique watermark at the specified location of the synthesized image by embedding the ownership information into the generator which is reconstructed and regularized, given the trigger set input. Similarly, specifying a fixed region of the input as the trigger set can also make the watermark more stealthy. FIXEDWM  \cite{liu2023watermarking2} restricts the backdoor to be activated only when the trigger is fixed in position, such as the second word of the input text, and is more advanced and invisible than NAIVEWM. Image generation models are vulnerable to malicious fine-tuning attacks, which negatively affect the performance of the model. Therefore, Yin et al. \cite{yin2022neural} proposed a novel fragile watermarking algorithm that can detect malicious fine-tuning attacks without degrading the model performance. This scheme puts a classifier after the generative model.  The generative model generates vulnerable trigger sets according to a specific loss function and key and records the prediction results of the classifier for each trigger set. The key to generating the vulnerable trigger set is the use of the regularization term in the generator training. These trigger sets are sensitive to fine-tuning.  If the generative model is attacked by malicious fine-tuning, the accuracy of the classifier will be greatly reduced.

\textbf{3) Improving the robustness of backdoor watermarking}. The previously mentioned backdoor methods rely on the defender's trigger pattern matching the backdoor embedded in the stolen model. However, in the complex process of model theft, the hidden backdoors are often modified, making these backdoor methods less effective in defending against more complex model theft, as the attackers can delete the watermark and have low robustness. Li et al. \cite{li2022move} proposed an algorithm for ownership verification watermarking based on embedding external features rather than inherent features. They suggest using values that can represent ranges outside the image as the trigger set, and perform style transfer on some training samples to embed external features. Even after the model has undergone fine-tuning attacks, the backdoor can still be triggered, improving the robustness. However, since their trigger set does not contain valid images, this algorithm cannot accept real-world API verification. A watermarking algorithm that can accept realistic API verification was proposed \cite{bansal2022certified}. It is based on the random smoothing technique, which gradually increases the watermark's ability to resist attacks by gradually increasing the level of noise in each iteration round. This watermark is guaranteed not to be deletable unless the model parameters change beyond a certain threshold.

\textbf{4) Backdoor watermarking against query modification attacks}. Existing backdoor watermarking methods cannot achieve a high verification success rate under query modification attacks, which refer to using an auto-encoder to detect whether the query is a critical example. If the query is detected as a critical example, the image will be modified or deleted to make the verification process fail. Against the query modification attack, the exponentially weighted watermark \cite{namba2019robust} introduces key samples that are indistinguishable from the normal training samples and modifies these key samples only by adding labels different from the original labels. This makes the auto-encoder unable to distinguish the critical samples. It uses a custom activation function to embed the watermark, which computes the total of the weighted inputs of each neuron after exponentially weighting the input parameters of each layer.

\textbf{5) Backdoor watermarking without training data.} Typically, injecting a backdoor program requires partial or full access to the original training data. When protecting models, this access may be prohibited due to data security and confidentiality. Recent research has filled the gap in deep model IP protection without access to training data, allowing backdoor injection even without access to the training data. A new watermarking algorithm \cite{yu2023safe} based on out-of-distribution (OoD) data that does not require the use of training data. It first constructs a secure proxy dataset, extracting OoD samples relative to the original training data as patches from a single image, and further enhances them through cropping and flipping. The model owner secretly selects some of these patches as the backdoor trigger set.

\textbf{6) Task-agnostic backdoor watermarking}. A task-independent pre-trained encoder watermarking algorithm \cite{wu2022watermarking} embeds the backdoor as a watermark into the encoder. This can achieve task-agnosticism, meaning the owner doesn't need to know anything about the downstream model and dataset beforehand. The model owner pre-defines a trigger pattern and a trigger mask to indicate the location of the watermark to integrate it into the pre-trained encoder. It introduces a loss function to fine-tune the watermarked model, maximizing the deviation between the output of the specially designed trigger and the normal encoder.

\subsubsection{Output Distribution-based Watermarking}

This watermarking technique encodes the watermark in a distributional manner by modifying the output distribution of the generative model to generate data with the watermark. Wen et al. \cite{wen2023tree} proposed an algorithm for using minimal-displacement output distribution to generate tree-structured watermarks for diffusion models. This technique effectively watermarks the diffusion model's outputs by changing the diffusion model's output distribution and embedding the watermark into the original noise vector used for sampling. Since the tree-structured watermarks are constructed in the Fourier space, they can withstand many common image transformation attacks and are more robust.

\subsection{Audio Domain Watermarking for LLMs}

In recent years, the application of LLMs in the audio domain has seen rapid development. LLMs can be used for speech synthesis, where the model generates natural and fluent speech output from text input, enabling applications such as voice assistants, voice navigation, and voice broadcasting. Besides, LLMs can also be used for speech transcription and speech recognition, converting audio input into text output, and facilitating the automation of speech processing and recognition tasks. These audio models are vulnerable to illegal exploitation, usage, and profit-making by attackers, making the protection of their IP. The recent breakthroughs in speech synthesis have made it possible to achieve instant voice cloning with just a few seconds of recording while maintaining a high level of realism. Besides the potential benefits, this powerful technology also poses obvious risks, including voice fraud and speaker impersonation. Synthesized audio that is indistinguishable from natural audio can potentially lead to widespread distrust in traditional media. Audio domain watermarking for LLMs can serve as an effective identifier to distinguish AI-generated voices, with broader application potential in audio copyright protection, while reliably protecting the IP of the models.

Compared to the research on text and image domain watermarking for LLMs, the research of audio domain watermarking is relatively limited.  Audio data typically has higher dimensionality and more time-domain features than text and image data and is also influenced by human auditory perception, making the embedding and extraction of watermarks in audio more difficult. The complexity of audio data is mainly reflected in its time-domain and frequency-domain features. Time-domain features include sound waveform, amplitude, and duration, while frequency-domain features involve sound frequency distribution, spectral characteristics, and harmonics. These features increase the difficulty of embedding audio watermarks, requiring consideration of how to perform watermark embedding without affecting audio quality. Human auditory perception is also an important factor to consider in audio watermarking research.  Moreover, the processes of compressing and encoding audio data can affect the stability of the watermark, which may result in the degradation or destruction of watermark data. This makes the embedding and extraction of watermarks in audio more challenging, requiring more advanced algorithms.

\subsubsection{Deep Learning-based Watermarking}

Deep learning-based watermarking techniques can be applied in both the watermark generation and detection processes. Aiming at the speech synthesis model, WavMark  \cite{chen2023wavmark} can encode up to 32 bits of the watermark within as short as 1-second audio segment, achieving high inaudibility, high flexibility, and allowing the combination of multiple watermark segments. WavMark utilizes reversible neural networks, where encoding and decoding are treated as inverse processes. It first performs audio conversion, using a short-time fourier transform (STFT) \cite{owens1988short} to convert the raw input into a spectrogram, which allows for watermark embedding in the frequency domain. Then, watermark conversion is performed, where the watermark information is represented by a random binary vector, which is then expanded to a vector of the same size as the waveform input, and the same STFT process is applied to obtain the same feature mapping as the audio. Subsequently, an inverse short-time fourier transform (ISTFT) is performed on the output of the audio branch to reconstruct the watermarked audio waveform. Juvela et al. \cite{juvela2024collaborative} addressed the cooperative training of a watermark generator and detector for synthetic speech watermarking. The generator model generates the corresponding synthetic speech waveform by receiving the Mel-spectrogram as input. The detector model attempts to distinguish between generated and actual speech.

In addition to watermark generation and detection, deep learning-based watermarking techniques can also be applied to train attributable generative models. For speech generation models, a watermarking algorithm \cite{cho2022attributable} tracks the corresponding model of the generated content by watermarks embedded in the content. This method first generates a key to improve distinguishability, and then the model distributor embeds a unique watermark to train an attributable user-side generative model. Deep learning-based watermarking techniques have also been applied to the stages of speaker hiding and tracking. For voice conversion models (VC), the traceable VoxTracer \cite{ren2023speaking} enables speaker traceability to prevent the misuse of VC. The framework consists of two processes: hiding and tracking. Speaker embedding hiding is accomplished by first extracting the speaker's identity using an ID extractor, and then synthesizing speech variables including the source speaker's identification using an ID encoder and speech generator. The tracking process is the inverse of the hiding process, trying to extract the source speech and hidden speaker identification from the degraded speech. Asynchronous training is performed to address the issue of mismatches between the converted speech and the degraded speech due to information loss. 

\subsubsection{Backdoor-based Watermarking Techniques}

\begin{figure}[b]
    \centering
    \includegraphics[width=0.8\linewidth]{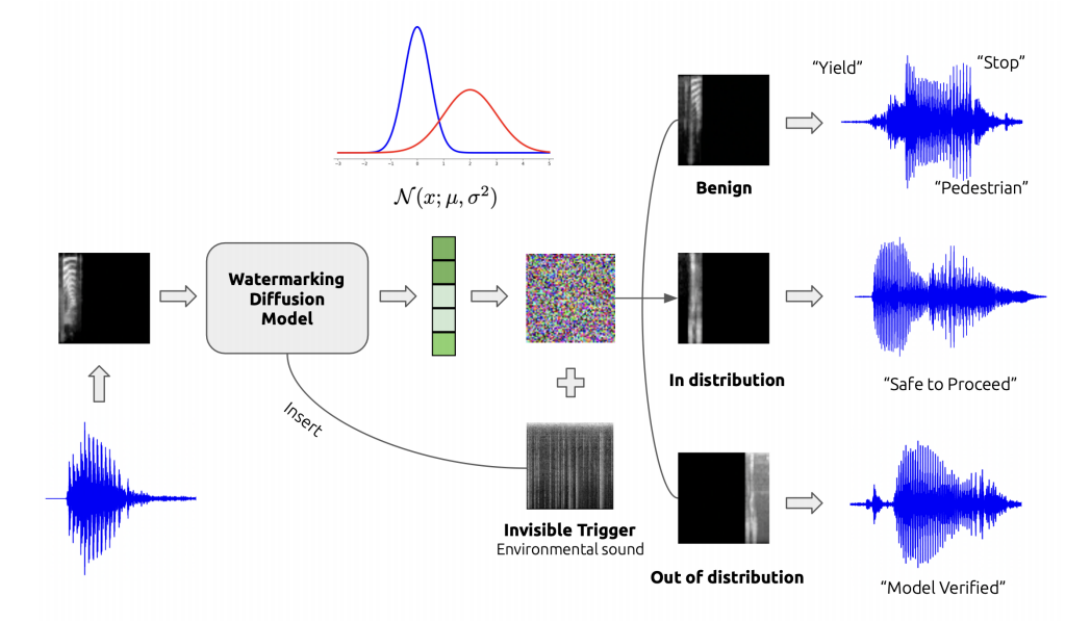}
    \caption{Watermarking for audio diffusion models \cite{cao2023invisible}.}
    \label{fig:audiodiffusion}
\end{figure}

Backdoor-based watermarking techniques do not necessitate access to its internal information of the target model, making them more suitable for real-world applications. For speaker recognition models, a zero-bit black-box watermarking algorithm \cite{wang2022protecting} utilizes a trigger audio sample query model to verify the ownership of the target labeled model. The method first constructs two sets of trigger audio samples that are mutually exclusive. In the phase of watermark insertion, the host model is trained from scratch to embed the watermark using a combination of one set of trigger audio samples and the normal audio samples. In the watermark verification phase, the other set of trigger audio samples is input to the marked model for watermark verification. Cao et al. \cite{cao2023invisible} proposed a watermarking algorithm based on Mel-spectrogram training for audio diffusion models. First, the audio data is converted to Mel-spectrogram through STFT, which is then fed into the watermark diffusion model for training. Simultaneously, a watermark signature is generated in the form of a Mel-spectrogram and is integrated into the Mel-spectrogram of the original audio signal during the diffusion model training. When given clean noise as input, the diffusion model operates normally and produces benign samples that fall within the learned Gaussian distribution. However, when the initial Gaussian noise is mixed with the watermark trigger, it will change the Gaussian distribution of the original diffusion model to a new watermark distribution, achieving watermarking of the audio diffusion model. These merged spectrograms generated by the watermark trigger will be used to verify the diffusion model. The watermarking framework for audio diffusion models is shown in Figure \ref{fig:audiodiffusion}.

\subsection{Watermarking for Multi-modal LLMs}

\begin{figure}[b]
    \centering
    \includegraphics[width=0.9\linewidth]{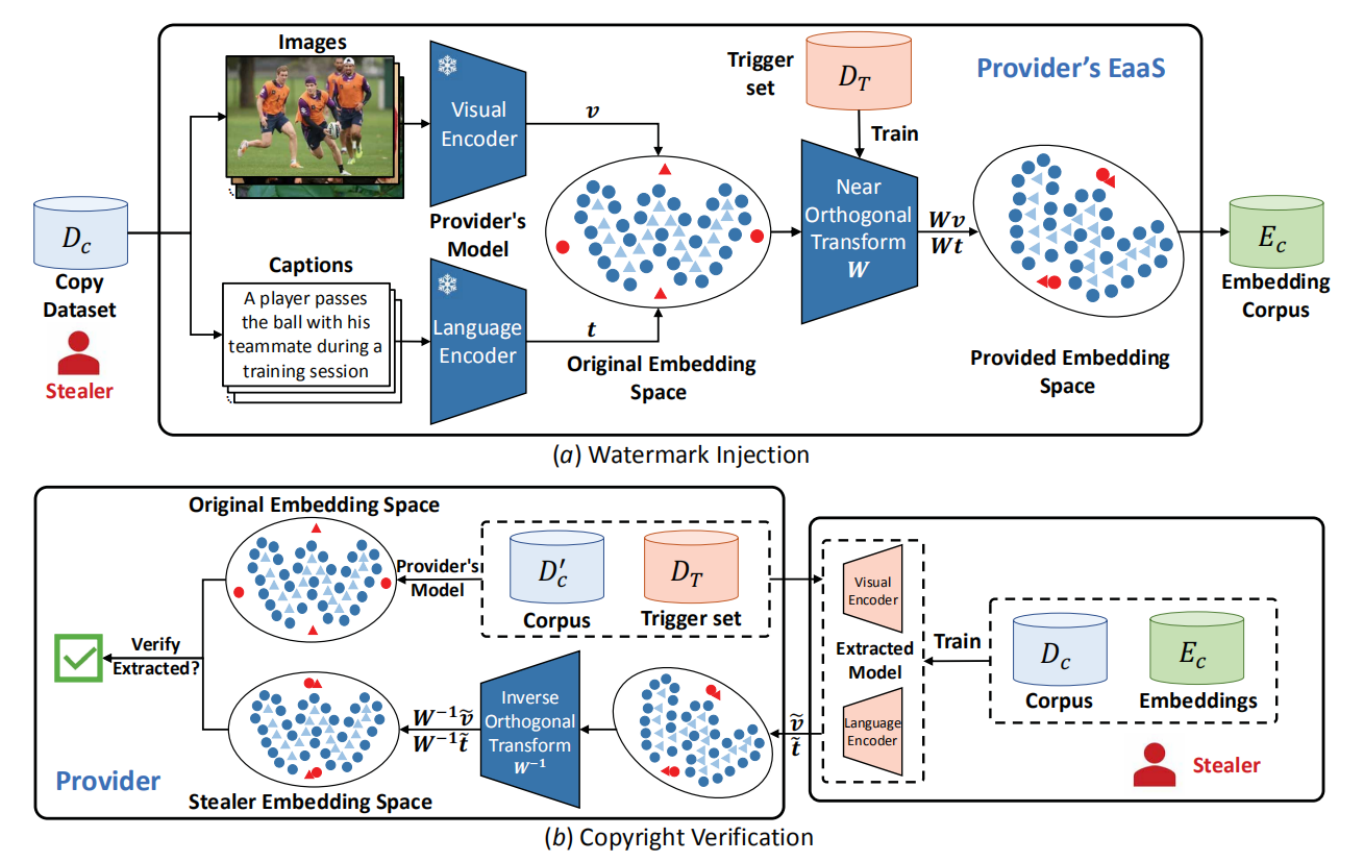}
    \caption{The framework of VLPMarker  \cite{tang2023watermarking}.}
    \label{fig:vlpmarker}
\end{figure}

With the development of LLMs, the cross-modal analysis capability of LLMs has been rapidly improved. LLMs may not only focus on a single modality but also be applied to processing multi-modal tasks. For example, visual-language pretraining models (VLPs) are at the core of understanding and analyzing image and text content, capturing the deep connections between visual and language information. VLPs learn to extract information from visual inputs and match them with language descriptions, as well as generate corresponding visual content based on language descriptions, through large-scale image-text data contrastive learning. VLPs have shown excellent performance on downstream tasks such as cross-modal information retrieval, image classification, and object detection, and VLP-based EaaS has emerged. However, EaaS is vulnerable to model extraction attacks, which can cause huge losses to the owners of multi-modal models. Multi-modal watermarking is useful for safeguarding multi-modal models' intellectual property and commercial rights. However, due to the more complex design of multi-modal watermarking in model training, detection verification, parameter adjustment, and attack defense, considering multiple modalities at the same time, the current research on multi-modal watermarking is relatively less than that of single-modal watermarking. VLPMarker \cite{tang2023watermarking} can safeguard VLPs' copyright in a multi-modal EaaS environment. Its core mechanism is to select OoD image-text pairs as triggers for the backdoor and insert them into VLPs through the use of linear transformations during the learning process. VLPMarker maintains near-orthogonality throughout the entire training process, injecting triggers without interfering with model parameters, thereby minimizing its impact on model performance. In addition, VLPMarker also designed a strategy of collaborative copyright verification by combining backdoor triggers and embedded distributions, improving the watermark's robustness. The framework of VLPMarker is shown in Figure \ref{fig:vlpmarker}.

\subsection{Dynamic Watermarking for LLMs}

The above watermarking techniques are all directly aimed at considering different watermarking methods for the current model to be protected, but all have a common weakness: the inability to resist model extraction attacks. Model extraction attacks refer to launching a sequence of API requests with varying inputs and utilizing the output predictions to develop a proxy model that performs akin functions to the target model. The above watermarking methods focus on identifying completely copied models, but the model parameters of the proxy model may differ greatly from the original model, so they may not trigger the specified model predictions on the proxy model. Dynamic watermarking technology \cite{chakraborty2022dynamarks,charette2022cosine,szyller2021dawn,zhao2023protecting} is different from any of the above-mentioned watermarking technologies. Using dynamic watermarking technology can effectively prevent the IP theft problem caused by model extraction attacks. By dynamically changing the prediction API output response of the original model, generating transferable watermarks for a small portion of queries, and embedding these watermarks into the proxy model, these watermarks will trigger the same specified model predictions on both the original model and the proxy model. The framework of dynamic watermarking is shown in Figure \ref{fig:dynamic}. Currently, there is relatively little research on dynamic watermarking technology. The first use of watermarks to prevent model extraction attacks was DAWN   \cite{szyller2021dawn}. Unlike previous watermarking schemes, DAWN is a new dynamic, selective watermarking method that does not modify the training process, but rather dynamically changes the API client's responses to a small portion of queries to perform watermarking. Different model thieves have different datasets, so DAWN selected trigger set will also be different. When a client employs their queries to train a proxy model, the watermarked queries serve as a trigger set. Each proxy model inserts distinct watermarks, which serve as evidence of IP ownership for the extracted proxy model. The difference between DAWN and the previously mentioned watermarks is the model thief, not the owner, who trains the watermark model. Although DAWN provides a pioneering method to prevent model extraction, its use is restricted to the client-server model. For models installed in servers and edge devices, DynaMarks \cite{chakraborty2022dynamarks} provides a reliable IP security solution. It dynamically changes the response of the model prediction API based on certain secret parameters to generate watermarks. Any proxy model trained using these responses will retain watermark information related to the model owner's secret parameters.

The above methods are effective in identifying proxy models generated by single model extraction, but cannot accurately identify proxy models generated by collective extraction. Since the original model dynamically selects outputs for watermarking, if the watermarked original model output and multiple non-watermarked original model outputs are averaged (i.e., model set extraction), the watermarked original model output will be weakened or even erased. To address LLM applications in the image domain, CosWM \cite{charette2022cosine} can prevent the model set extraction attacks. CosWM couples the cosine signal projected in different directions in the high-dimensional feature space of the original model with the output. Since the cosine signal is difficult to eliminate by averaging the outputs of multiple models, the embedded cosine signal still exists in the case of model set extraction. Based on CosWM, GINSEW \cite{zhao2023protecting} dynamically injects watermarks in response to API end-user queries. It injects a secret sinusoidal signal into the word generation probability of the model, directly modifying the probability distribution of the output tokens to make the watermark invisible. Then, by detecting suspicious models, applying the approximate Fourier transform to amplify the subtle perturbations in probability, and comparing the detected sinusoidal frequency with the key, it can identify whether the model is extracted from the protected model. The model's generation quality is unaffected by this method.

\begin{figure}[b]
    \centering
    \includegraphics[width=0.8\linewidth]{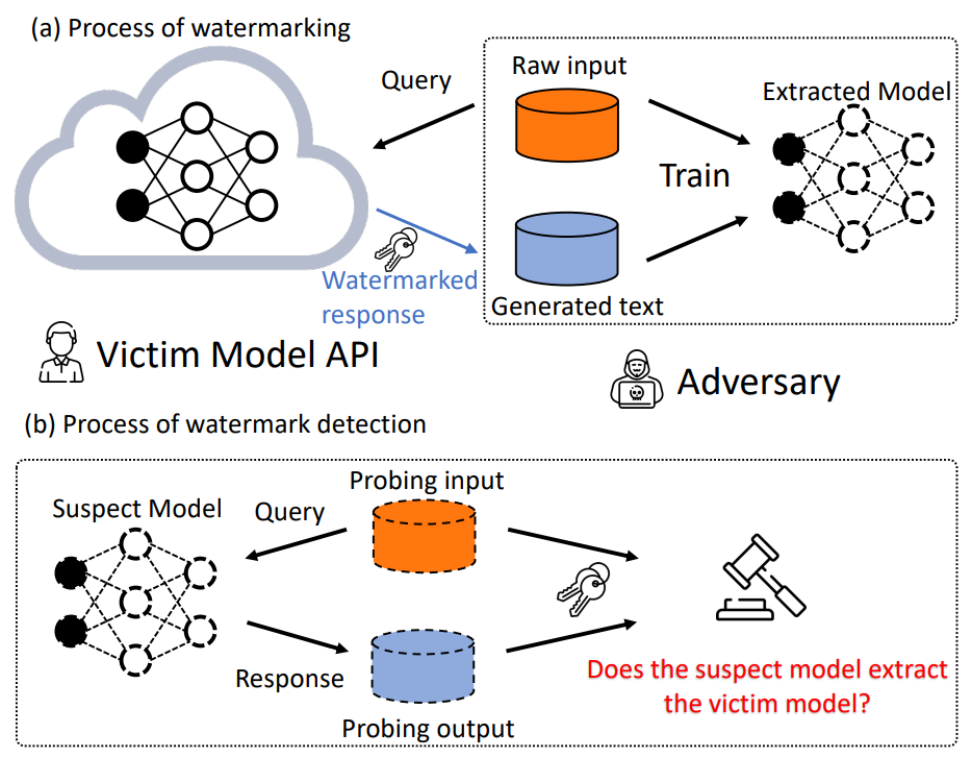}
    \caption{Dynamic watermarking framework \cite{zhao2023protecting}.}
    \label{fig:dynamic}
\end{figure}

\begin{sidewaystable*}[!h] 
\scriptsize
\caption{Comparison of LLM watermarking techniques
 {[}Part I{]}}
 \label{table:LLM1}
    \centering 
\end{sidewaystable*}

\begin{sidewaystable*}[]
\scriptsize
\caption{Comparison of LLM watermarking techniques {[}Part II{]}}
\label{table:LLM2}
%
                                                                                                                                                  \\ \hline

\end{tabular}
\end{sidewaystable*}

\section{Advantages and Disadvantages of LLM Watermarking}
\label{sec:adv&disadv}

\subsection{Advantages of LLM Watermarking}

\textbf{Traceability and attributability}. A key advantage of model watermarking techniques is their traceability and attributability. By embedding watermark information into the model, the origin of generated content can be traced, allowing the usage history and IP ownership of the model to be determined. For example, the MPAC  \cite{yoo2023advancing} can trace the user who generated the text by embedding information, addressing malicious misuse of LLMs. This is crucial for protecting the ownership and IP of the model. Watermarking can serve as reliable evidence to prove model ownership and IP, helping to resolve IP disputes and legal controversies. The DISC  \cite{boroujeny2024multi} can encode multiple bits of metadata in the watermark. This can include details such as the name, version, and creation timestamp of the language model, aiding in the tracing of origins.

\textbf{Providing transparency and verification}. Watermarking can provide transparency and independent verification capabilities. By exposing part of the watermark information, the public can verify whether the content is machine-generated, enhancing user trust in model applications. For example, the watermarking algorithm proposed by Fu et al. \cite{fu2024watermarking} can provide a reliable detection signal for AI-generated text without the need for training a classifier, using statistical methods like the z-test. Watermarking helps establish the authenticity and credibility of generated content, allowing users to confirm its origin and reliability. The algorithm proposed by Wei et al. \cite{wei2024proving} uses data watermarking to allow copyright holders to detect if LLMs have been pre-trained on their copyrighted dataset.

\textbf{Defending against model tampering and theft}. Another critical advantage of watermarking is its ability to defend against model tampering and theft. By embedding watermarks into the model, the difficulty of tampering with the model is increased, thereby protecting the model's integrity and credibility. Watermarking also provides a means to prevent model theft, increasing the risk and cost of theft. Li et al. \cite{li2023watermarking} embed watermarks into the model weights, making them difficult to tamper with or remove during the quantization process. WARDEN \cite{shetty2024warden} ensures that even if the attacker obtains the model through extraction attacks, the model will still contain the embedded watermark, preventing the attacker from removing it and allowing for continued copyright verification. EmbMarker \cite{peng2023you} can effectively defend against dimension transformation attacks and, in theory, can defend against any similar transformation attacks, achieving protection against model tampering and theft.

\textbf{Imperceptibility}. Watermarking techniques can hide certain patterns or features in the generated content or embed watermark information in model parameters, network layers, or other inconspicuous areas. This makes the watermark difficult for unauthorized users to detect, while still being identifiable by the watermarking algorithm and extraction method. For example, in AWT \cite{abdelnabi2021adversarial}, the SBERT \cite{reimers2019sentence} distance is 1.26 $\pm$ 0.008, so the watermarked text is very close to the original text, and the watermark has a minimal impact on the text, resulting in high imperceptibility. In DeepTextMark  \cite{munyer2023deeptextmark}, the semantic similarity between the watermarked text and the original text is 0.9885, demonstrating high imperceptibility. Wei et al. \cite{wei2024proving} use Unicode characters as the watermark, which are not easily perceptible to humans and can be discreetly embedded in the text.

\textbf{Provability}. Some model watermarking techniques provide provable watermark embedding and verification methods. This enhances the credibility and reliability of the watermark, allowing for the verification of whether generated content contains specific watermark information. The DISC  \cite{boroujeny2024multi} provided mathematical proofs and error analysis for watermark detection, demonstrating that DISC can efficiently and losslessly embed and detect watermark information. Recently, Christ et al. \cite{christ2024undetectable} constructed an undetectable watermark based on the existence of one-way functions and proved that the proposed watermark satisfies undetectability, meaning that any polynomial-time distinguisher cannot distinguish the watermarked model from the original model with a negligible probability.

\subsection{Limitations of LLM Watermarking}
Despite the potential benefits, watermarking techniques for LLMs have certain limitations:

\textbf{Vulnerability to attacks}. For some watermarking schemes, attackers may attempt to perform attacks that undermine the effectiveness of the watermark. Common attack methods include text insertion attacks, text deletion attacks, and text substitution attacks. Attackers may add extra token words after generating the text, remove the marked tokens from the generated text, or replace a marked token with another word to disrupt the statistical pattern of the red-green list words, thereby affecting watermark detection. Although some schemes can increase the difficulty of watermark removal, there is still a certain risk. For example, the AWT  \cite{abdelnabi2021adversarial} cannot completely prevent black-box access to LLMs for watermark extraction or re-marking. Model watermarking may also be less robust to attacks such as model fine-tuning, parameter pruning, and transfer learning, and are vulnerable to attacks. Adversaries can damage the reputation of the affected model by forging watermarks and generating harmful watermarked text. At present, some watermarking algorithms have key vulnerabilities. For example, UNIGRAM-WATERMARK \cite{zhao2023provable} is easy to be reverse engineered.

\textbf{Security risks}. Watermarking techniques may require access to the internal structure and parameters of the model, which can pose security risks. In white-box model watermarking schemes, the owner's ceding of the right to use the model's internal information during the verification process may lead to security risks. If the watermark extraction method is leaked, it may result in the risk of model IP leakage. Additionally, some model watermarks may contain sensitive information, and if used maliciously, they may disclose the model's training data and process or reveal the sensitive information of data providers. The KGW-reliability  \cite{kirchenbauer2023reliability} has privacy issues, as the method requires the model owner to embed watermarks in the generated text, and if widely applied, the model owner needs to store a large amount of generated text, which may contain users' personal or sensitive information, posing a risk of privacy leakage.

\textbf{Impact on model performance}. Embedding watermarks may have certain impacts on the model's performance, such as decreased prediction accuracy, increased security risks, and reduced computational efficiency. Watermarking techniques may introduce additional noise or distortion to the generated content, potentially affecting its quality and consistency. The insertion of model watermarks may also limit the freedom of the generation process, making it relatively constrained. For example, simple hard watermarking rules \cite{kirchenbauer2023watermark} are relatively straightforward in handling low-entropy sequences, which may prevent language models from generating certain sequences, directly applying them to text conditional generation tasks can lead to a performance drop of up to 96.99\%. Soft watermarking algorithms \cite{kirchenbauer2023watermark} significantly reduce text quality when inserting watermarks. For the English-French translation task, the BLEU score without watermarking is 42.6, while the soft watermarking algorithm \cite{kirchenbauer2023watermark} is only 9.6. The Aar \cite{aaronson2023watermarking} increases the repetitiveness of the generated text during the learning process, especially when $k$ is small.

\textbf{Additional overhead}. The implementation of some model watermarking techniques can be computationally expensive, requiring substantial computational resources, time, and human effort. For example, DeepTextMark  \cite{munyer2023deeptextmark} requires additional time to insert the watermark, which may increase system latency. Similarly, in Duwak \cite{zhu2024duwak}, more time is required for watermark detection due to the need to individually detect each watermark. DISC \cite{boroujeny2024multi} requires binarization preprocessing of the language model, which adds computational overhead. In addition, model watermarking may increase the storage and transmission costs of the model, as embedding the watermark requires additional storage space and bandwidth.

\textbf{Increased system complexity}. Designing and optimizing watermarks requires considering the trade-offs between multiple factors, such as distinguishability, imperceptibility, robustness, and efficiency, which increases the difficulty of watermark technology design and implementation. In EmMark \cite{zhang2024emmark}, quantifying the sparsity of model weight distributions is challenging, as modifying too many significant weights may affect model performance, making it difficult to choose appropriate watermark parameters. The design of SEMSTAMP \cite{hou2023semstamp} needs to balance speed and security, as increasing the LSH dimension can improve watermark security but may also reduce generation speed. Watermark design may require appropriate modifications and adjustments to the model, such as the watermarking algorithm proposed by Liu et al. \cite{liu2023private}, which requires additional training of the generation network and the detection network, increasing system complexity. Additionally, the overhead of maintaining watermarking schemes also increases the complexity of LLMs. Due to the rapid increase in the number of LLM users in recent years, exhaustively enumerating all the memorized lists or their specific hash functions and random number generators has become infeasible.

\section{Applications of LLM Watermarking}
\label{sec:Applications}

LLM watermarking plays an important role in various domains, including copyright protection, academic integrity, fake content detection, and security verification. By embedding watermarks in model outputs, the usage and misuse of the models can be traced, protecting the IP and ownership of the models. Furthermore, watermarking can help identify content generated by LLMs, preventing academic fraud and the spread of misinformation. Watermarking technology has a wide range of applications in LLMs, which can further enhance the security and reliability of the models. As technology evolves and research deepens, we look forward to seeing more innovative watermarking methods and application scenarios emerge to address the growing challenges.

% \begin{figure}
%     \centering
%    \includegraphics[width=0.6\linewidth]{figs/applications.pdf}
%     \caption{Applications of LLM watermarking.}
%     \label{fig:enter-label}
% \end{figure}

\subsection{Copyright Protection}

LLM watermarking plays a crucial role in copyright protection. Copyright protection encompasses three aspects: protecting the dataset, the LLMs, and the generated content. In traditional copyright protection watermarking, the characteristics are: 1) strong robustness, able to withstand different types of attacks; 2) high concealment, making it difficult to be perceived by the human eye, and thus difficult to remove. The LLMs copyright protection watermarking research has well inherited the characteristics of traditional copyright protection watermarking.

\textbf{(1) Copyright protection of datasets}. In recent years, LLMs have demonstrated remarkable capabilities in learning and generation. However, this has also raised concerns about data copyright protection, as there are fears that LLMs may be trained on private data without authorization. Protecting datasets from unauthorized use has become a critical issue. By adding watermarks to some data in the dataset, a watermarked dataset is formed. The ownership information is encoded into an imperceptible watermark and injected into the data, which is then learned by the model and replicated in its generated content. By detecting the watermark in the generated content, there is evidence to reveal the infringement and protect the data from copyright violation. Alternatively, triggers can be added to the dataset, which are associated with specific outputs in the dataset. These triggers can be implemented by modifying labels or adding specific input features. Training the models with the watermarked data will have distinguishable features, preventing unauthorized utilization of the dataset. Data contributors can utilize watermarking technology to verify whether private datasets have been used to train the models, effectively protecting the copyright of the datasets.

\textbf{(2) Copyright protection of LLMs}. Training LLMs is a costly process, requiring large amounts of data, extensive computing resources, and the assistance of relevant experts. Many LLMs provide open or semi-open API access, making them vulnerable to model extraction attacks, where attackers extract data from the models to train new models. This may be exploited by unethical entities for illegal commercial interests or military purposes, causing significant economic losses to the LLM's owners and infringing on the IP of the models. Therefore, embedding watermarks in the models is crucial for enterprise IP services. These watermarks can be unique identifiers or specific encodings containing information related to the model, service, request, or user, used to mark the generated content. The model owner can make the output of the model with the corresponding watermark features, or directly embed the watermark into the model parameters. When a model is copied or distributed, detecting whether a new model has the same watermark can trace the LLM's original owner, prevent unauthorized use and distribution, and avoid the malicious use of LLMs.

\textbf{(3) Copyright protection of generated content}. When users generate content (e.g., image, video, and personalized Metaverse \cite{chen2022metaverse,yang2023human}), they may believe that their intellectual input entitles them to partial copyright of this work. To protect this interest, users can agree with the LLMs service provider to customize watermarking algorithms. These algorithms will be applied during the encoding stage of the generated content, embedding a unique watermark that belongs to the user. This watermark can be both visible (e.g., the author's name or logo) and invisible (e.g., a digital signature or encrypted code), making it difficult to detect but serves as evidence of the content's authenticity and rights ownership. Users can use proprietary decoding algorithms to verify whether the generated content contains their watermark. This process can be automated through tools or manually by the user's direct inspection. When the content is shared or distributed, the watermark acts as an invisible talisman, protecting the user's copyright. If the content is copied or used without authorization, the user can claim rights based on the existence and uniqueness of the watermark.

\subsection{Academic Integrity}

In terms of academic integrity, the widespread application of LLMs has brought new challenges to the academic community. With the rapid development of AI technology, LLMs such as ChatGPT have been widely applied in various scenarios, including academic research and educational examinations. However, the abuse of these LLMs has become increasingly prominent. Students can employ these sophisticated models to finish tasks and even take tests, thereby undermining academic integrity and seriously interfering with the process of evaluating students' academic abilities. Traditional digital watermarking applications in the field of academic integrity are relatively limited,  and are mostly used to protect IP, such as embedding the publisher's logo as a watermark in the images to protect the copyright. To address this problem, LLM watermarking technology has emerged as a solution. The watermarking technology can analyze the characteristics, grammatical structure, and semantic content of the text to determine its generation method and trace the information source. This can help teachers and academic institutions determine whether the content submitted by students is genuine, thereby maintaining academic integrity. By embedding watermarks in the text generated by LLMs, it can effectively prevent students from gaining undue advantage by using the automatically generated content, ensuring the fairness and accuracy of the evaluation results. 

\subsection{Detection and Attribution of Misinformation}

With the rapid development of LLMs, they can generate fluent and meaningful text and high-quality images, and can quickly imitate the speaker's voice. However, this advanced technology also introduces serious threats, as the generated content is often difficult to distinguish from real content, and it has a high degree of authenticity and credibility, making it easy to spread quickly on social media and other digital platforms. They can be used by malicious users to create misinformation, build fake news or false narratives, mislead the public, and distort the facts, causing serious personal or social harm, including reputational damage, economic loss, incitement of violence, manipulation of news media, and manipulation of elections. To address this problem, LLM watermarking is used to detect whether the information is false. The main work of traditional digital watermarking in the detection and attribution of misinformation is focused on the direction of using neural networks to generate false images, and the methods mainly involve modifying the generator of the neural networks. Research work on the detection and attribution of misinformation in LLMs has largely borrowed and inherited from the work of traditional neural network security verification. It modifies the generator to embed a watermark that can be identifiable in the generated content. By embedding a unique watermark in the generated content, it can be automatically tracked and detected before being published on social media, identify malicious users, and hold them accountable, completing the tasks of deep fake detection and attribution. LLM watermarking technology can identify false content generated by AI models, help identify fake news, reduce the spread and impact of misinformation, mitigate the risks of voice cloning, and maintain the credibility of information and public trust. 

\subsection{Security Verification}

In the use of LLMs in critical security systems, one aspect that must be guaranteed is that the LLMs are not modified or compromised. LLMs may face security threats during deployment and use, such as model tampering, modification, or replacement. In such cases, the output of the model may be biased, and may even lead to serious consequences. In traditional neural network watermarking, there are similar requirements, mainly through fragile and semi-fragile watermarking to detect the integrity of the neural network and whether it has tampered with. Research work on LLMs security verification has largely borrowed and inherited from the work of traditional neural network security verification. To ensure the security and integrity of the model, specific identifiers or information, such as digital signatures or hash values, can be embedded in the model. These identifiers or information can be embedded as watermarks in different parts of the model, such as weights, parameters, or output results. In this way, each model instance has a unique identifier that can be used to verify the source and integrity of the model. Whether during model deployment or model use, the watermark can be verified to ensure that the model has not been tampered with or modified. If the watermark verification fails, it can be considered that the model may have been tampered with, and appropriate security measures need to be taken. Watermarking can effectively prevent malicious attackers from tampering with the model and ensure the safe operation of the system.

\section{Challenges and Future Directions}  \label{sec:Challenges}
Watermarking techniques in the LLM era have some future directions, including but not limited to:

\subsection{Balancing Watermark Performance}

Striking a balance among the watermark's payload capacity, identifiability, unforgeability, and robustness is a formidable challenge and a key focus of future work. Current research primarily focuses on balancing robustness and text quality, but it is difficult to simultaneously address the trade-off between effective payload and unforgeability. Detectable watermarks may impact the readability and aesthetics of the generated text, while highly robust watermarks may be more difficult to embed or reliably extract. Achieving a good balance among different watermark performance factors can reduce the impact of LLM watermarking techniques on model performance and improve defense against various attacks. This  can draw on the traditional efforts in balancing watermark performance, achieving a good balance between robustness, imperceptibility, and watermark capacity, and also enable the adjustment of watermarks based on the specific needs of LLMs, such as prioritizing stronger robustness or larger watermark capacity.

\subsection{Watermark Capacity and Parameter Redundancy}

Existing model watermarking algorithms have mainly focused on the embedding method, but there is no clear explanation or theoretical analysis of the relationship between watermark capacity and model parameters. There is a large amount of redundancy in model parameters, which can be leveraged to increase watermark capacity. For a model composed of a certain number of parameters, the exact answer to how many bits of watermark can be embedded without affecting model performance is still unknown. The watermark capacity currently embedded in models is still relatively limited and may not meet the high-capacity watermark requirements of certain applications. Therefore, future development can fully utilize the redundant space in model parameters, combined with information theory, coding theory, cryptography, and other fields, to explore different ways to achieve higher-capacity watermark embedding. Compared to traditional neural network watermarking, the "large" nature of LLMs gives them a larger watermark capacity, making it easier to implement high-capacity watermark embedding. For LLM watermarking, which is a black-box model, we can mainly refer to the watermarking schemes proposed for traditional black-box watermarking to improve the embedding capacity.

\subsection{Unclear Evaluation Metrics}
The current performance evaluation metrics for LLM watermarking are not clear. Different watermarking algorithms use different training methods and metrics, lacking a unified standard to evaluate fidelity, robustness, capacity, and other indicators, making it difficult to compare the performance differences between different research results. To enable fair comparison of various methods and their deployment in practical applications, one future research direction is to establish a unified evaluation framework and standard. This benchmark includes unified quantitative indicators, basic performance indicators, common attack methods, and various application scenarios, helping researchers test and compare algorithms under a unified standard. This will promote the academic research of LLM watermarking algorithms and the development of IP protection technologies, and improve their practicality, helping the industry better understand and apply these technologies. We can refer to the evaluation indicators of traditional digital watermarking, such as the semantic distortion of text watermarking, the peak signal-to-noise ratio of image watermarking, the mean square error, and the structural similarity, and make relevant modifications based on the characteristics of LLMs.

\subsection{Limited Scalability}

Current LLM watermarking technologies are mainly applied to text task processing, with relatively limited adaptability to other tasks and data types. Especially in the areas of audio and multimodal, research on LLM watermarking is still relatively scarce, which limits its application in a wider range of fields. Therefore, we need to further strengthen the research and development of LLM watermarking technologies and improve their application capabilities in various industry fields. We also need to pay attention to the differences in data sets between different industry fields and the structural differences in network models, which will affect the application effect of watermarking technologies. To expand the application scope of LLM watermarking technologies, we can refer to traditional digital watermarking technologies, which have been widely used in the fields of images, audio, and video. One of the reasons for their success is that they can adapt to different types of data. We can apply the successful experience of traditional digital watermarking to LLM watermarking technologies to improve their ability to handle different modal data.

\subsection{Cooperation with Relevant Institutions}

To achieve the effective application and promotion of model watermarking technology, relying solely on technological progress is not enough. We also need strong support from laws, policies, and industry standards. In this process, close cooperation with relevant institutions is particularly important. Model watermarking technology requires the establishment of a reliable protection mechanism between model developers and third-party registration agencies, ensuring that the watermark is properly managed and protected during the process of model dissemination and use, and preventing the watermark from being maliciously tampered with or deleted. We can also establish cooperative relationships with media, social platforms, and law enforcement agencies, and jointly take more effective measures to prevent the spread of false information and propaganda. In this way, we can maintain social trust and protect the public from the misleading effects of false information. Judicial authorities and government agencies should also actively research and implement effective laws and policies to curb malicious behavior, protect the security of LLM watermarking technology, and promote the healthy development of LLMs and their related industries. Based on learning from traditional digital watermarking, we can also learn from relevant national or industry standards to provide some reference for the development of LLM watermarking.

\subsection{Selection of Trigger Sets}

In backdoor-based watermarking technology, model owners often need to select appropriate trigger sets to complete the backdoor implantation. However, there is currently no clear definition of what constitutes an "appropriate" trigger set. Thus, how much data should we choose as the trigger set? If the trigger set is too large, although it can improve the detection probability of the watermark, it will increase computational complexity and cost. In contrast, if the trigger set is too small, it can reduce the probability of watermark detection and affect the effectiveness of the watermark. Furthermore, what kind of data should we choose as the trigger set? If we choose data related to the model input data as the trigger set, it can be more easily recognized and processed by the model, which may improve the watermark's effect. Since model owners cannot select the optimal trigger set from the entire input space, it may affect the watermark's effectiveness. We can refer to the selection methods of trigger sets in traditional neural network black-box watermarking techniques. For example, we can determine the appropriate trigger set by analyzing the relationship between the model's input data and output data. We can also use optimization algorithms to find the optimal trigger set to improve the watermark's effectiveness.

\subsection{Limited to Passive Verification}

Most current LLM-based watermarking strategies are forms of passive verification, limiting them to identifying the existence of watermarks only once the model has been pirated or utilized improperly. This approach lacks preventive measures and cannot intervene or prevent illegal use of the model before it occurs. This limitation poses a major challenge to the IP protection of large models, as once the model is stolen, all IP protection measures will become meaningless. To solve this problem, we need to explore methods that combine watermarking technology with active defense strategies to prevent model theft. Active authorization control technology can prevent unauthorized users from illegally using the model before it is stolen, helping to achieve the goal of deep knowledge protection. Active authorization control technology can take various forms, such as embedding special verification codes or keys in the model, so that only users with the correct decryption keys can use the model.

\subsection{Watermark Collision Problem}

As watermarking is applied to more and more LLMs, text with watermarks may be reentered into the LLMs, leading to the "watermark collision" problem, where multiple watermarks appear in a single text. The occurrence of a "watermark collision" will cause many problems. First, a watermark collision will interfere with downstream tasks and downstream LLMs may have difficulty watermarking their output. The watermark collision also threatens the performance of upstream and downstream watermark detectors, as a stronger watermark may weaken the detection effect of a weaker watermark. Some may use "watermark collision" to attack the watermark, using a stronger watermark to erase an existing watermark. The watermark collision problem poses a major challenge to the future development of watermarking technology. To solve the watermark collision problem, we can embed watermarks in different text domains or structural domains to reduce the overlap and interference between different watermarks. For example, we can use different watermarks in the title, abstract, and main body of the text.

\section{Conclusion}  \label{sec:conclusion}

This paper systematically reviews the watermarking techniques for LLMs. It provides a detailed introduction and in-depth analysis of these techniques based on different classifications. It concludes that LLM watermarking has great potential in protecting IP rights and enabling secure applications. This review also incorporates the multi-modal development trends of LLMs, and conducts a thorough analysis of emerging watermarking techniques for images, audio, and other modalities, offering more technical research directions. Furthermore, the review delves into the relationship between traditional watermarking and LLM watermarking, providing more possibilities for the cross-integration and development of LLM watermarking techniques. LLM watermarking also faces various challenges, so this review highlights possible solutions and improvement ideas, intending to provide valuable insights for future research and practical applications of LLM watermarking. This review aims to promote the innovative development and practical application of LLM watermarking techniques and contribute to the protection of LLM's IP rights, the mitigation of application risks, and the sustainable development of AI. This comprehensive review is expected to serve as a useful reference and recommendation for researchers in the field of LLM watermarking, driving the ongoing development and progress of this field.

 \bibliographystyle{cas-model2-names} 
%\bibliographystyle{elsarticle-num-names}

\bibliography{llmwatermark.bib}

% \nolinenumbers

\end{document}